\documentclass[preprint,review,12pt]{elsarticle}
\makeatletter
\def\ps@pprintTitle{%
 \let\@oddhead\@empty
 \let\@evenhead\@empty
 \def\@oddfoot{\centerline{\thepage}}%
 \let\@evenfoot\@oddfoot}
\makeatother 

\usepackage{lineno,hyperref}
\modulolinenumbers[5]

\usepackage{color,soul}
\usepackage{amsmath}
\usepackage{amssymb}
\usepackage{graphicx}
\usepackage{subcaption}
\usepackage{lipsum}
\usepackage{float}

\usepackage[percent]{overpic}
\usepackage{enumerate}

\begin{document}

\begin{frontmatter}



\title{A phase-field study of elastic stress effects on phase separation in ternary alloys}



\author[mymainaddress]{Sandeep Sugathan}
\ead{ms14resch01001@iith.ac.in}

\author[mymainaddress]{Saswata Bhattacharya\corref{mycorrespondingauthor}}
\cortext[mycorrespondingauthor]{Corresponding author}
\ead{saswata@iith.ac.in}

\address[mymainaddress]{Indian Institute of
Technology, Department of Material Science and Metallurgical Engineering, Hyderabad, 502285, India}

\begin{abstract}
Most of the commercially important alloys are multicomponent,
producing multiphase microstructures as a result of 
processing. When the coexisting phases are elastically coherent, 
the elastic interactions between these phases play a major role in the development of 
microstructures. To elucidate the key effects of elastic
stress on microstructural evolution when more than two misfitting phases 
are present in the microstructure, we have developed a microelastic phase-field 
model in two dimensions to study phase separation in
ternary alloy system.
Numerical solutions of a set of coupled Cahn-Hilliard equations for the composition 
fields govern the spatiotemporal evolution of the three-phase microstructure. 
The model incorporates coherency strain interactions between the phases using
Khachaturyan's microelasticity theory. 
We systematically vary the misfit strains (magnitude and sign) between the phases along 
with the bulk alloy composition
to study their effects on the morphological development of the phases and the resulting phase separation kinetics. 
We also vary the ratio of interfacial 
energies between the phases to understand the interplay between elastic and interfacial energies
on morphological evolution. The sign and degree of misfit affect
strain partitioning between the phases during spinodal decomposition, 
thereby affecting their compositional history and morphology.
Moreover, strain partitioning affects solute partitioning and alters the kinetics of coarsening of the phases. 
The phases associated with higher misfit strain appear coarser and exhibit wider 
size distribution compared to those having lower misfit.
When the interfacial energies satisfy complete wetting condition, phase separation
leads to development of stable core-shell morphology depending on
the misfit between the core (wetted) and the shell (wetting) phases.
\end{abstract}


\end{frontmatter}


\section{Introduction}
\label{Introduction}

Technologically important alloys with multiple components 
often exhibit microstructures containing multiple coexisting phases. 
To enhance the properties of these alloys, we require a fundamental
understanding of the multiphase microstructures and 
their kinetics of evolution as a function of key processing parameters 
to establish processing-microstructure-property relations in these alloys. 
In a multicomponent alloy containing multiple coherent phases, 
an interplay of several factors, particularly, 
alloy chemistry, relative interfacial energies between the phases, coherency strains arising due to 
lattice parameter mismatch between the phases, and kinetic parameters such as composition dependent 
diffusivities of components, leads to modification in the pathways of morphological evolution. 

Elastic stresses developed in the microstructure due to the formation of coherent interfaces 
among multiple phases induce important morphological changes and affect strain partitioning and as a result
solute partitioning~\cite{voorhees2004thermodynamics,kostorz1995metals,doi1996elasticity}. The elastic 
interactions between the coherent domains in the microstructure
depend on the magnitude and sign of misfit between them,
as well as  the anisotropy and inhomogeneity in elastic moduli~\cite{fratzl1999modeling,abinandanan1997computer}.

There are several examples in the existing literature elucidating the significance of coherency strain effects 
on the development of multiphase microstructures. Here we cite some important examples relevant to our study of
the development of coherent three-phase microstructures as a result of solid to solid phase trasformations.
Long term aging of alloys like \textit{Cu-Ni-Cr}~\cite{findik2002modulated},  
\textit{Fe-Al-Mn-Cr}~\cite{liu1985alpha} and 
\textit{Fe-Ni-Mn-Al}~\cite{hanna2005new} results in the evolution of
coherent three-phase microstructures.
Moreover, ternary alloys (e.g., \textit{Al-Li-Zr}, \textit{Al-Li-Sc}, \textit{Al-Sc-Zr}, \textit{Al-Li-Cu}, 
Inconel, alloy steels) develop microstructures with three structurally distinct phases  sharing 
elastically coherent interfaces during multi-step heat treatment~\cite{Cozar_1973,radmilovic2011highly,KumarMakineni2017}, although development of such phases do not happen through spinodal decomposition (SD).

With the advent of `high entropy' alloys (HEA),experimental studies have reported the development of coherent
three-phase microstructures via SD~\cite{pickering2015fine,santodonato2015deviation,xiao2017microstructure}.
In a recent paper, Morral and Chen~\cite{morral2017high} conjectured the existence of three-phase miscibility gap in multicomponent HEA systems.  

Although there are extensive studies of microstructural evolution during phase separation in binary
alloy systems with coherent elastic
misfit~\cite{johnson1989effects,abinandanan1993coarsening,Lee1995,nishimori1990pattern,onuki1991anomalously}, 
there are very few studies focusing on the effects of elastic stress when more than two phases are present 
in the microstructure ~\cite{zhou2014computer,shi2019growth}.

De Fontaine~\cite{de1972analysis} and Morral and Cahn~\cite{morral1971spinodal} analyzed the 
compositional stability, fluctuations and early stage kinetics of ternary SD and derived 
Cahn-Morral equations to describe phase separation. Hoyt~\cite{hoyt1989spinodal} developed 
a master equation for ternary SD and derived linearized diffusion 
equations and three independent partial structure functions. 
The mean field coarsening theory of Lifshitz, Slyozov and Wagner~\cite{lifshitz1961kinetics,wagner1961theorie} 
was extended to study coarsening in ternary alloys~\cite{bhattacharyya1972activation,hoyt1998coarsening} 
and multicomponent, multiphase systems~\cite{philippe2013ostwald,wang2017phase}.

Chen~\cite{chen1993computer,chen1994computer} and Eyre~\cite{eyre1993systems} studied the influence of alloy chemistry 
on morphological evolution during SD in ternary alloys using computer simulations. 
Later, Bhattacharyya and Abinandanan~\cite{BHATTACHARYYA2009646} studied the effect of relative 
interfacial energies between the coexisting phases on ternary SD. In a recent study, 
Ghosh \textit{et al.}~\cite{ghosh2017particles} extended the model to investigate 
the influence of immobile particles with selective wetting present in the microstructure on 
SD in ternary polymer blends. The studies mentioned above do not include the elastic stress 
effects arising due to lattice parameter mismatch between the coexisting phases.
Since the elastic stresses developed in the microstructure may play a crucial role in the morphological evolution of
ternary alloys, we have extended the diffuse interface model to study the effect of coherency strains between the 
phases on the evolution of microstructure during ternary SD. We aim to investigate the effects of degree and
sign of relative misfit between the phases on ternary phase separation.

The framework of the paper is as follows: in Section~\ref{formulation}, we describe the formulation of the 
phase field model with a regular solution bulk free energy and the technique of incorporating elasticity. 
We present the results of our simulations in Section~\ref{results} and discuss the effect of relative misfit strains 
and interfacial energies between coexisting phases on decomposition pathways, solute partitioning, microstructural evolution 
and coarsening kinetics. We summarize the major conclusions from our study in Section~\ref{conclusions}. 

\section{Formulation and numerical implementation of the model}
\label{formulation}

\subsection{Energetics}
\label{Energetics}

In the phase-field model discussed here, we consider a ternary substitutional alloy containing three atomic species A, B and C. 
The concentration of $i'$th species $c_i(\textbf{r},t)$ $(i=A,B,C)$ is a function of position \textbf{r} and time $t$.
The bulk energetics presented here is an extension of an existing diffuse-interface 
formulation~\cite{eyre1993systems,bhattacharyya2003study,ghosh2017particles} 
of ternary Cahn-Hilliard model incorporating additional terms describing elastic interactions.

The total free energy for the coherent, isotropic, compositionally inhomogeneous system
is expressed as a function of composition field variables $c_B(\textbf{r},t)$ and $c_C(\textbf{r},t)$ ($\because c_A+c_B+c_C=1$):

\begin{equation}
F = N_v\int_v\Big(f_0(c_A,c_B,c_C) + \sum_{i=A,B,C}\kappa_i|\nabla c_i|^2+f_{el}\Big)dV,
\label{eq:energy}
\end{equation}

where $N_v$ is the number of molecules per unit volume (assumed to be independent of composition and position) and
$\kappa_i(i=A,B,C)$ are the gradient energy coefficients associated with compositions fields.

The bulk chemical free energy per atom $f_0(c_A,c_B,c_C)$ of the homogeneous alloy is given by the regular solution expression:
\begin{equation}
f_0(c_A,c_B,c_C) = \frac{1}{2}\sum_{i\neq j}\chi_{ij}c_ic_j+\sum_{i}c_i\ln{c_i},
\end{equation}
where $\chi_{AB}$, $\chi_{AC}$ and $\chi_{BC}$ are the pair-wise interaction parameters.
The elastic strain energy density $f_{el}$ is given by

\begin{equation}
f_{el}=\frac{1}{2}\sigma_{ij}(\textbf{r})\epsilon_{ij}^{el}(\textbf{r}),
\label{eq:elastenergy}
\end{equation}

where $\sigma_{ij}(\textbf{r})$ denotes the stress field calculated 
from mechanical equilibrium equation and
$\epsilon_{ij}^{el}(\textbf{r})$ indicates the elastic strain field arising due to the coherency strain field,
$\epsilon_{ij}^0(\textbf{r})$. We represent all the vector and tensor fields using Einstein summation convention.

$\epsilon_{ij}^{el}(\textbf{r}) = \delta\epsilon_{ij}(\textbf{r}) + E_{ij} - \epsilon_{ij}^{0}(\textbf{r})$ and 
$\sigma_{ij}^{el}(r) = \lambda_{ijkl}\epsilon_{kl}^{el}(r)$. $\lambda_{ijkl}$ indicates the elastic stiffness tensor 
(homogeneous across three phases), $E_{ij}$ is the homogeneous strain and $\delta\epsilon_{ij}(\textbf{r})$ is the 
periodic strain tensor, defined as the gradient of the displacement field vector. $\epsilon_{ij}^0(\textbf{r})$ is
expressed as a function of composition fields:
\begin{equation}
\epsilon_{ij}^{0}(\textbf{r}) = \sum_{p=0}^1 \epsilon_{ij}^{(p)} \theta_p(\textbf{r}), 
\end{equation}
where $\theta_0(\textbf{r}) = c_B(\textbf{r})-c_B^0$ and $\theta_1(\textbf{r}) = c_C(\textbf{r})-c_C^0$,
and $\epsilon_{ij}^{(p)} $ is the position independent part of the eigenstrain field associated with $\theta_p(\textbf{r})$. 
($c_B^0,c_C^0$) is the homogeneous alloy composition and $\delta_{ij}$ is the Kronecker delta. 
We assume the eigenstrains in the system to be dialataional: $\epsilon_{ij}^{(0)}=\epsilon_{\alpha\beta}\delta_{ij}$ 
and $\epsilon_{ij}^{(1)}=\epsilon_{\alpha\gamma}\delta_{ij}$.
$\epsilon_{\alpha\beta}$ and $\epsilon_{\alpha\gamma}$ denote the lattice expansion coefficients 
associated with $c_B$ and $c_C$, respectively. $\epsilon_{\beta\gamma}$ represents the difference between
$\epsilon_{\alpha\beta}$  and $\epsilon_{\alpha\gamma}$ 
($\epsilon_{\beta\gamma}=\epsilon_{\alpha\beta}-\epsilon_{\alpha\gamma}$).

We consider the alloy system to be a linear elastic solid. Solution of mechanical equilibrium equation 
($\frac{\partial \sigma_{ij}}{\partial x_j}=0$) in Fourier space (assuming the local displacement field to be 
periodic) gives the local elastic strain and stress fields. We obtain an expression for elastic energy in
reciprocal space using Khachaturyan’s microelastic theory \cite{khachaturyan2013theory} as follows:
\begin{equation}
F_{el}=\frac{1}{2}\sum_{p,q=0}^1\int\frac{d^3\textbf{k}}{(2\pi)^3}B_{pq}(\textbf{n})\tilde{\theta_p}(\textbf{k})\tilde{\theta_q}^*(\textbf{k}),
\label{eq: elastenk}
\end{equation}

where \textbf{k} denotes the Fourier wave vector, $\textbf{n}=\frac{\textbf{k}}{|\textbf{k}|}$ is the unit
vector in reciprocal space, $\tilde{\theta_p}(\textbf{k})$ represents the Fourier transform of 
$\theta_p(\textbf{r})$. $B_{pq}(\textbf{n})=\lambda_{ijkl}\epsilon_{ij}^{(p)}\epsilon_{kl}^{(q)}
-n_i\hat{\sigma_{ij}}^{(p)}\omega_{jk}(\textbf{n})\hat{\sigma_{kl}}^{(q)}n_l$
is the elastic interaction energy between $\theta_p$ and $\theta_q$
and $\omega_{il}^{-1}(\textbf{n})=C_{ijkl}n_jn_k$ is the inverse Green tensor.
$\tilde{\theta^*}$ denotes the complex conjugate of $\tilde{\theta}$. 

\subsection{Kinetics}
\label{Kinetics}

The temporal evolution of the conserved field variables are governed by a set of Cahn-Hilliard equations. 
We solve the equations for both $c_B$ and $c_C$ to study the microstructural evolution.
The temporal evolution of the composition fields is described by the continuity
equations:
\begin{equation}
\frac{\partial c_i}{\partial t}=-\nabla\cdot\bar{J_i},
\end{equation}

where $\bar{J_i}$ represents the net flux of component i ($i=B,C$). The kinetic equations for 
microstructural evolution are obtained by following the results of 
Kramer \cite{kramer1984interdiffusion} and combining them with Gibbs-Duhem equations and Onsager relations
for the effective mobilities: 

\begin{equation}
\begin{split}
\frac{\partial c_B}{\partial t}=M_{BB} & \nabla^2\Bigg(g_B-2\kappa_{BB}\nabla^2c_B-2\kappa_{BC}\nabla^2c_C+\frac{\delta F_{el}}{\delta c_B}\Bigg)+ \\ M_{BC} & \nabla^2\Bigg(g_C-2\kappa_{BC}\nabla^2c_B-2\kappa_{CC}\nabla^2c_C+\frac{\delta F_{el}}{\delta c_C}\Bigg),
\end{split}
\label{eq: dCbdt}
\end{equation}

\begin{equation}
\begin{split}
\frac{\partial c_C}{\partial t}=M_{BC} & \nabla^2\Bigg(g_B-2\kappa_{BB}\nabla^2c_B-2\kappa_{BC}\nabla^2c_C+\frac{\delta F_{el}}{\delta c_B}\Bigg)+ \\ M_{CC} & \nabla^2\Bigg(g_C-2\kappa_{BC}\nabla^2c_B-2\kappa_{CC}\nabla^2c_C+\frac{\delta F_{el}}{\delta c_C}\Bigg),
\end{split}
\label{eq: dCcdt}
\end{equation}

where $\frac{\delta F_{el}}{\delta c_B}$ and $\frac{\delta F_{el}}{\delta c_C}$ are variational derivatives of elastic energy with respect to composition order parameters, $g_B=\frac{\partial f_0}{\partial c_B},g_C=\frac{\partial f_0}{\partial c_C}$, $\kappa_{BB}=\kappa_A+\kappa_B, \kappa_{BC}=\kappa_{CB}=\kappa_A, \kappa_{CC}=\kappa_A+\kappa_C$. $M_{BB}$, $M_{BC}$ and $M_{CC}$ are effective mobilities defined as follows \cite{allnatt_lidiard_1993}:
\begin{equation}
M_{BB}=(1-C_B)^2M_{B}+C_B^2(M_A+M_C),
\nonumber
\end{equation}
\begin{equation}
M_{CC}=(1-C_C)^2M_{C}+C_C^2(M_A+M_B),
\nonumber
\end{equation}
\begin{equation}
M_{BC}=M_{CB}=C_BC_CM_A-C_B(1-C_C)M_{C}-C_C(1-C_B)M_{B}.
\end{equation}

\subsection{Numerical implementation}
\label{Numerical implementation}

We use a semi implicit Fourier spectral method \cite{chen1998applications,PhysRevE.60.3564} for 
solving the evolution equations (Eqns.~\ref{eq: dCbdt} and~\ref{eq: dCcdt}).

\begin{equation}
\begin{split}
\frac{\partial\tilde{c_B}}{\partial t}=-M_{BB} & k^2\Bigg(\tilde{g_B}+2\kappa_{BB}k^2\tilde{c_B}+2\kappa_{BC}k^2\tilde{c_C}+\Bigg[\frac{\delta F_{el}}{\delta c_B}\Bigg]_k\Bigg)- \\
 M_{BC} & k^2\Bigg(\tilde{g_C}+2\kappa_{BC}k^2\tilde{c_B}+2\kappa_{CC}k^2\tilde{c_C}+\Bigg[\frac{\delta F_{el}}{\delta c_C}\Bigg]_k\Bigg),
\end{split}
\label{eq: Cb_tilde}
\end{equation}

\begin{equation}
\begin{split}
\frac{\partial\tilde{c_C}}{\partial t}=-M_{BC} & k^2\Bigg(\tilde{g_B}+2\kappa_{BB}k^2\tilde{c_B}+2\kappa_{BC}k^2\tilde{c_C}+\Bigg[\frac{\delta F_{el}}{\delta c_B}\Bigg]_k\Bigg)- \\
M_{CC} & k^2\Bigg(\tilde{g_C}+2\kappa_{BC}k^2\tilde{c_B}+2\kappa_{CC}k^2\tilde{c_C}+\Bigg[\frac{\delta F_{el}}{\delta c_C}\Bigg]_k\Bigg),
\end{split}
\label{eq: Cc_tilde}
\end{equation}

where $\tilde{c_B}$ and $\tilde{c_C}$ are the Fourier transforms of the respective compositions in the real space,
$\tilde{g_B}$ and $\tilde{g_C}$ are the Fourier transforms of the bulk driving force terms.
The symbol $[]_k$ denotes a Fourier transform of the quantity within the brackets.
The variational derivatives of elastic energy with respect to composition field variables in reciprocal space are 
derived from the expression of elastic energy in Fourier space (Eqn. \ref{eq: elastenk}), considering a change in 
the functional relative to the change in functions, $c_B$ and $c_C$:

\begin{equation}
\Bigg[\frac{\delta F_{el}}{\delta c_B}\Bigg]_k = \Bigg[\frac{\delta F_{el}}{\delta\theta_0}\Bigg]_k = B_{00}\tilde{\theta_0} + B_{01}\tilde{\theta_1},
\end{equation}

\begin{equation}
\Bigg[\frac{\delta F_{el}}{\delta c_C}\Bigg]_k = \Bigg[\frac{\delta F_{el}}{\delta\theta_1}\Bigg]_k = B_{01}\tilde{\theta_0} + B_{11}\tilde{\theta_1},
\end{equation}

where $\tilde{\theta_0}$ and $\tilde{\theta_1}$ are the Fourier transforms of the $\theta_0$ and $\theta_1$, respectively
and $B_{pq}$ is the elastic interaction energy between $\theta_p$ and $\theta_q$ ($p,q=0,1$).

\subsection{Simulation details}
\label{Simulation details}
All the simulation parameters in the model are presented in 
non-dimensional form using characteristic length, time and energy values. 
We perform simulations of microstructural evolution on a two-dimensional 
square grid of size $2048\times 2048$ with dimensionless grid spacing 
$\Delta x =\Delta y = 1.0$. 
For approximating a bulk system (without surfaces), 
we use periodic boundary conditions. 

The simulations start with a homogeneous alloy of a prescribed concentration. 
Initially, we add a conserved Gaussian noise of $0.1\%$ strength to each grid point 
in the simulation box to mimic thermal fluctuations. 
We choose a non-dimensional time step $\Delta t=0.1$ for the
evolution of composition field variables
to ensure temporal accuracy of our numerical scheme.

We consider a system ($\chi_{AB}=\chi_{AC}=\chi_{BC}=3.5$, $\kappa_A=\kappa_B=\kappa_C=4$) 
with a symmetric ternary miscibility gap with equal interfacial energies 
between the equilibrium phases 
($\Gamma_{\alpha\beta}=\Gamma_{\alpha\gamma}=\Gamma_{\beta\gamma}$). 
Moreover, we vary the alloy composition and the degree and sign of misfit between the phases 
to study their effects on microstructural evolution during the early stages of phase separation. 
We have chosen nine different alloy systems listed in Table~\ref{cases_C4} for the simulations.
We have used the following naming convention to specify an alloy system:
an uppercase letter denotes the composition of the alloy, an integer subscript
represents the state of misfit strain.
Integer $`1'$ represents the case where the lattice expansion coefficients, 
$\epsilon_{\alpha\beta}$ and $\epsilon_{\alpha\gamma}$, associated with 
concentrations of B and C, respectively, have equal magnitude and same sign.
On the other hand, integer $`2'$ represents the case where 
$\epsilon_{\alpha\beta}=-\epsilon_{\alpha\gamma}$, and integer $`3'$ represents the case where 
$|\epsilon_{\alpha\beta}|\neq|\epsilon_{\alpha\gamma}|$.
\begin{table}[H]
\caption{Alloy systems used for simulations  
(prescribed alloy composition, magnitude and sign of misfit strains).}
\label{cases_C4}
\begin{center}
\begin{tabular}{ |c|c|c|c|c|c| } 
 \hline
 System & $(c_B^0,c_C^0)$ & $\epsilon_{\alpha\beta}$  & $\epsilon_{\alpha\gamma}$  \\
 \hline
      $P_1$ & (0.33,0.33) & 0.01 & 0.01  \\
      $P_2$ & (0.33,0.33) & 0.01 & -0.01  \\
      $P_3$ & (0.33,0.33) & -0.0139 & 0.0028  \\
      $Q_1$ & (0.25,0.25) & 0.01 & 0.01 \\
      $Q_2$ & (0.25,0.25) & 0.01 & -0.01  \\
      $Q_3$ & (0.25,0.25) & -0.0139 & 0.0028   \\
      $R_1$ & (0.45,0.45) & 0.01 & 0.01  \\
      $R_2$ & (0.45,0.45) & 0.01 & -0.01  \\
      $R_3$ & (0.45,0.45) & -0.0139 & 0.0028  \\
 \hline
\end{tabular}
\end{center}
\end{table}
De Fontaine developed a criterion for the stability
of ternary alloys with respect to bulk, incoherent SD~\cite{de1972analysis}. 
According to his analysis, the chemical spinodal surface of a ternary alloy shows three regions of stability
based on the eigenvalues and eigenvectors of a stability matrix $\mathcal S$ 
formed by the second order derivatives of the bulk free energy
$f_0$ with respect to the composition fields, $c_B$ and $c_C$:
\begin{itemize}
        \item
        Region I: central region of absolute instability, where both eigenvalues are negative
        ($\mathcal S$ is negative definite) and composition fluctuations grow along all directions for the alloys,
        \item
        Region II: intermediate region of conditional stability, 
        where one of the eigenvalues is negative
        ($\mathcal S$ is indefinite) and the eigenvector associated with the negative eigenvalue gives the direction of SD for the alloys,
        \item
        Region III: region of absolute stability at the three corners, 
        where both eigenvalues are positive
        ($\mathcal S$ is positive definite) and alloys remain stable with respect to composition fluctuations in any direction,.
\end{itemize} 
\begin{figure}[H]
 \centering
 \includegraphics[width=0.5\linewidth]{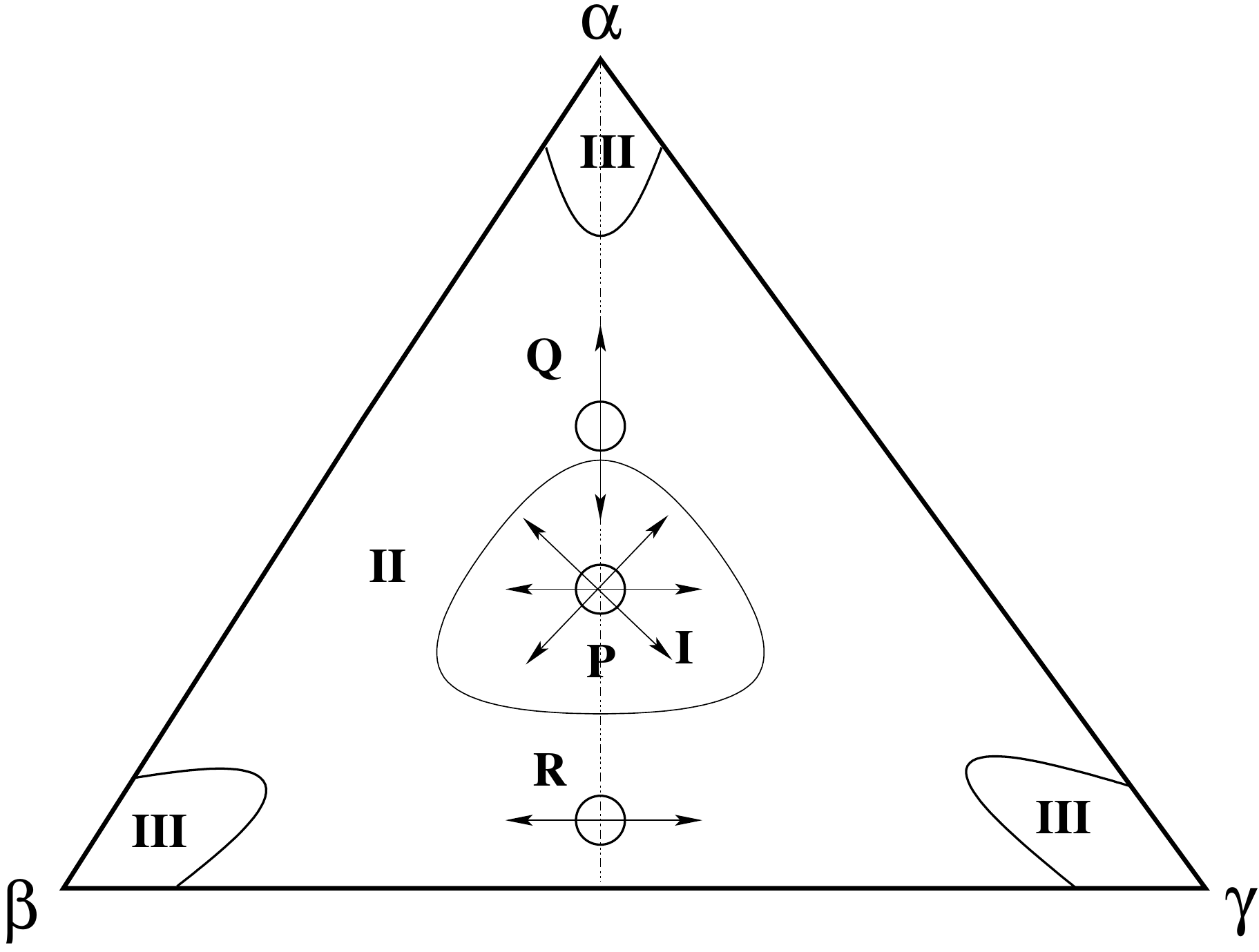}
\caption{Isothermal section of the chemical spinodal surface indicating directions 
of instability with respect to fluctuations in composition (schematic).}
\label{fig:dF_spino}
\end{figure} 
The points $P$, $Q$ and $R$ on the isothermal section of the chemical spinodal
surface ($\mid{\mathcal S}\mid = 0$) indicate the prescribed alloy compositions used in our study (Fig.~\ref{fig:dF_spino}).
The arrows indicate the directions of instability according to the stability matrix $\mathcal S$.

We present the time snapshots of microstructures  
using an RGB (red-green-blue) map representing the local compositions of components $A$, $B$ and $C$. According to the color map, blue hue indicates $A-$rich $\alpha$ phase,
green indicates $B-$rich $\beta$ phase, red indicates $C-$rich $\gamma$
phase and the intermediate shades (linear combination of the terminal colors) 
represent the interfacial compositions. We also estimate the particle size distribution of different 
domains in each alloy system using an open source program developed by Fialkowski 
\textit{et al.}~\cite{fialkowski2001scaling,aksimentiev2002morphology,fialkowski2002morphology} for 2D morphological analysis.    


\section{Results and discussion}
\label{results}

\subsection{Microstructural evolution}
\label{microstr_evolution}

\begin{figure}[H]
  \centering
\begin{subfigure}{.3\textwidth}
  \centering
  \includegraphics[width=0.9\linewidth]{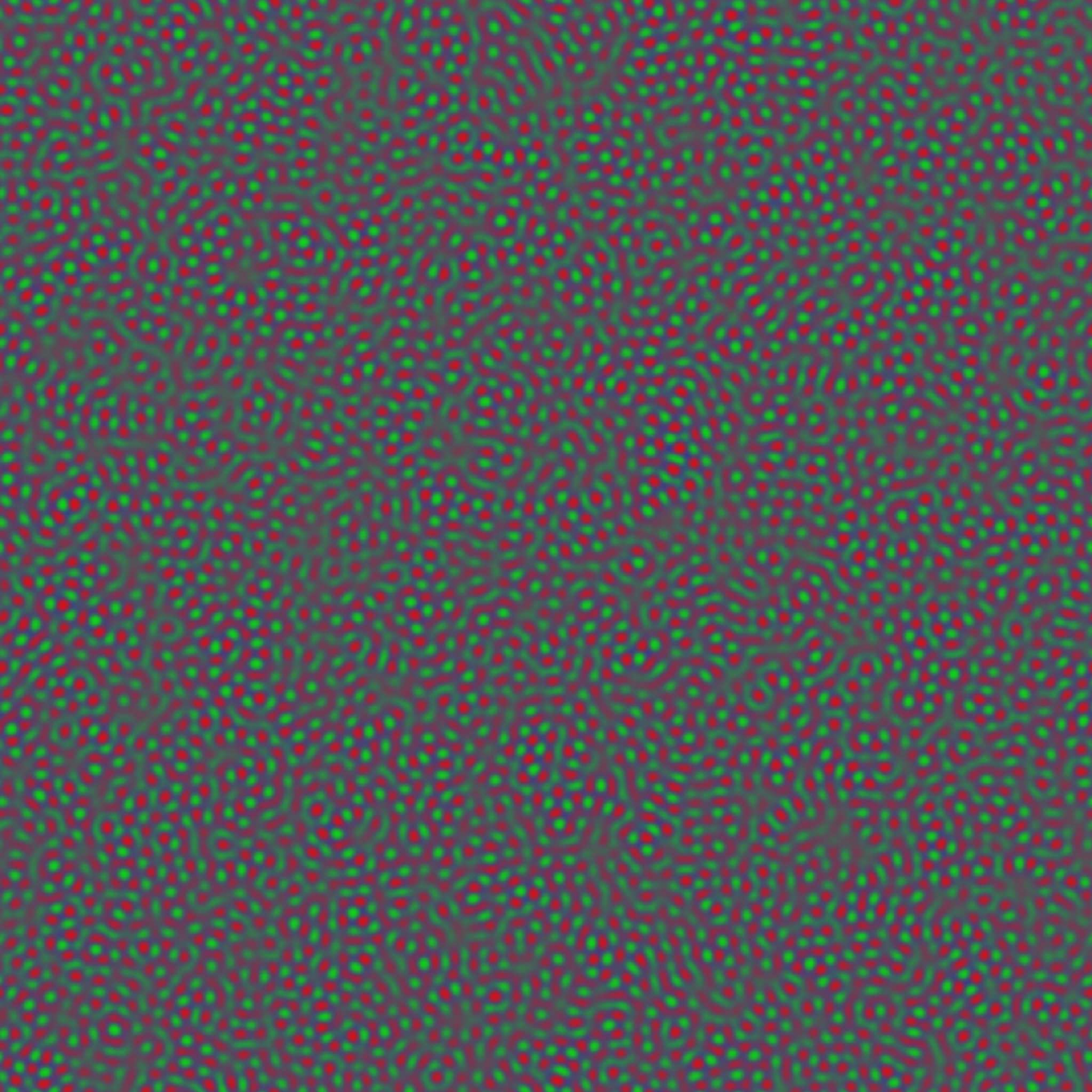}
  \caption{$P_1$}
  \label{P1_1}
\end{subfigure}
\begin{subfigure}{.3\textwidth}
  \centering
  \includegraphics[width=0.9\linewidth]{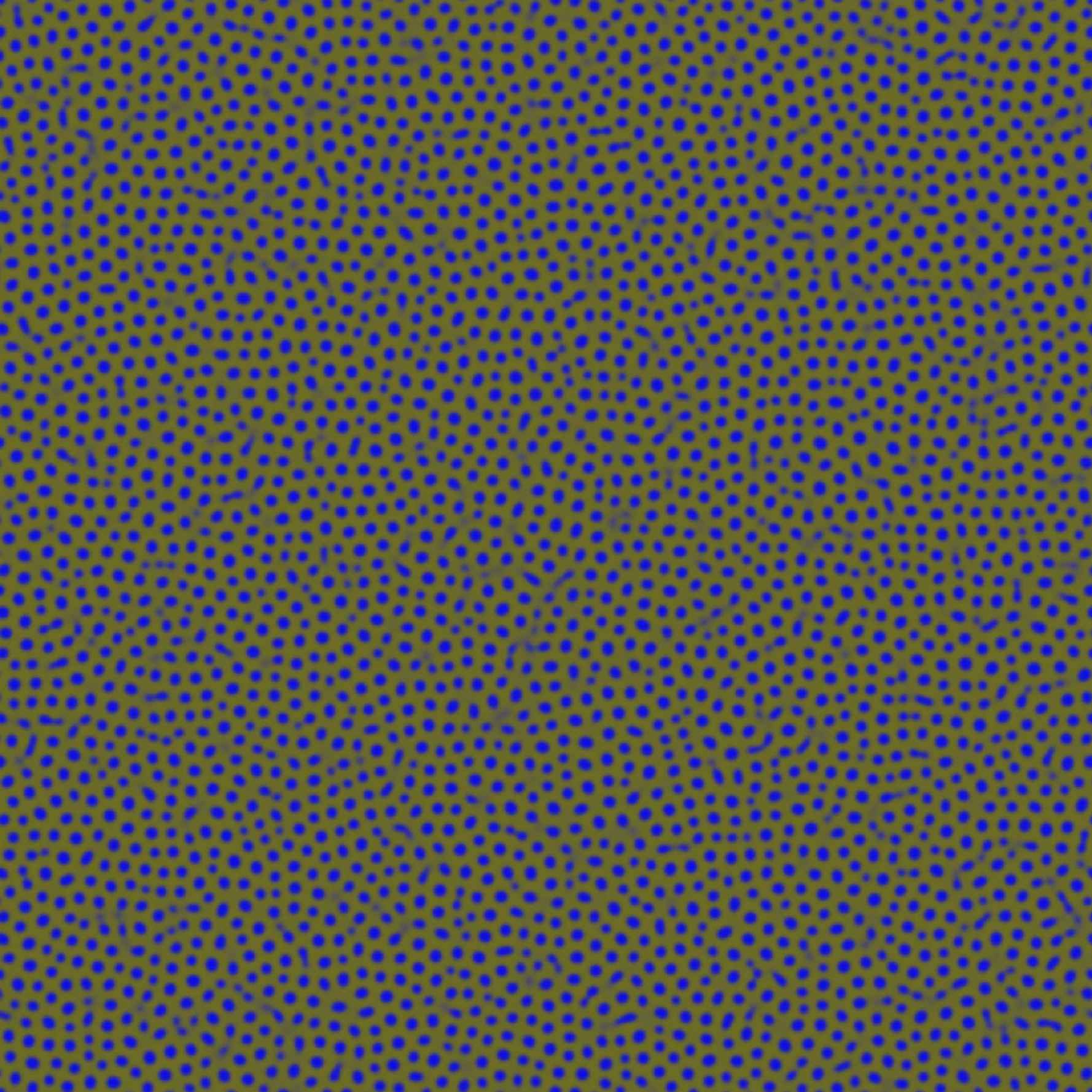}
  \caption{$P_2$}
  \label{P2_1}
\end{subfigure}
\begin{subfigure}{.3\textwidth}
  \centering
  \includegraphics[width=0.9\linewidth]{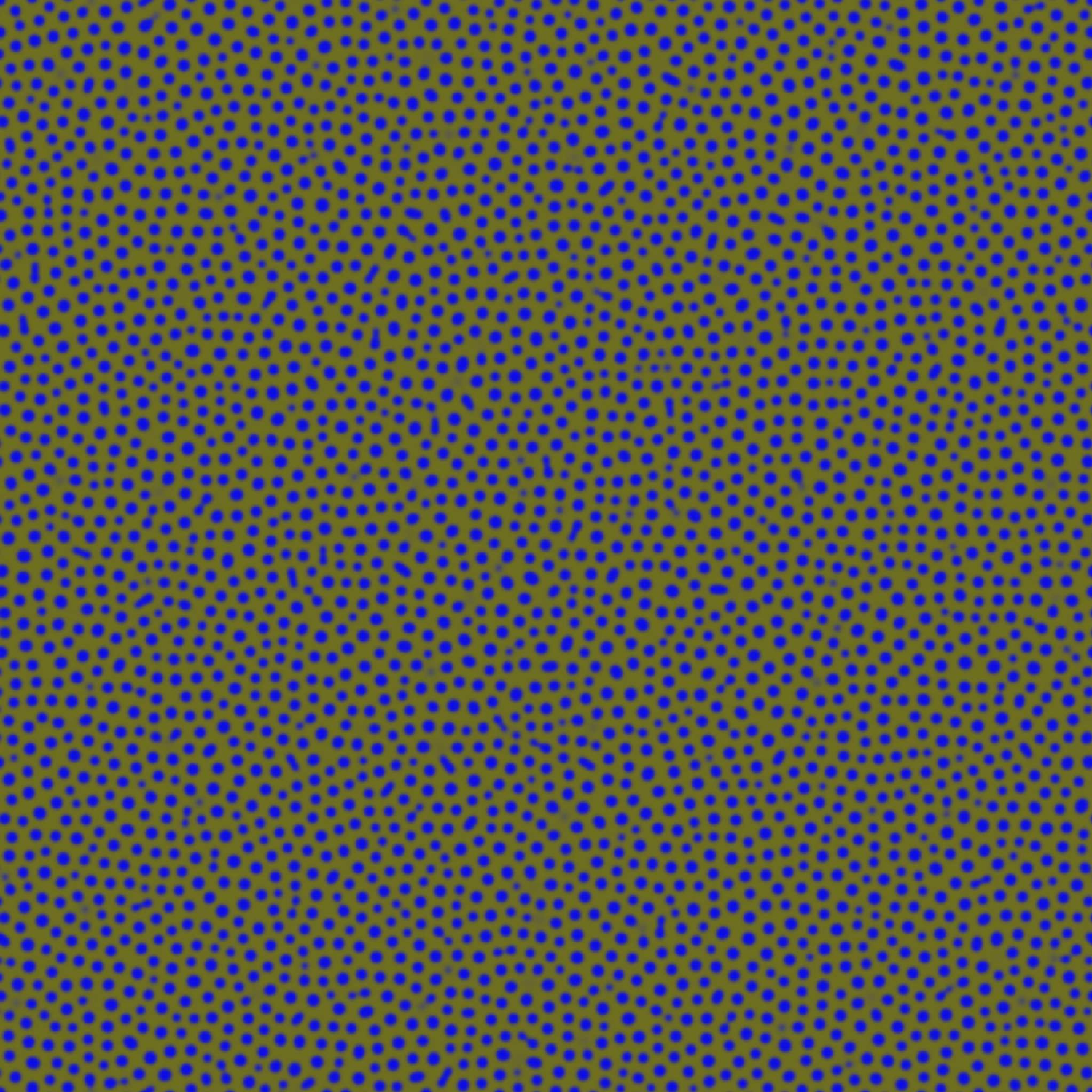}
  \caption{$P_3$}
  \label{P3_1}
\end{subfigure}\\
\begin{subfigure}{.3\textwidth}
  \centering
  \includegraphics[width=0.9\linewidth]{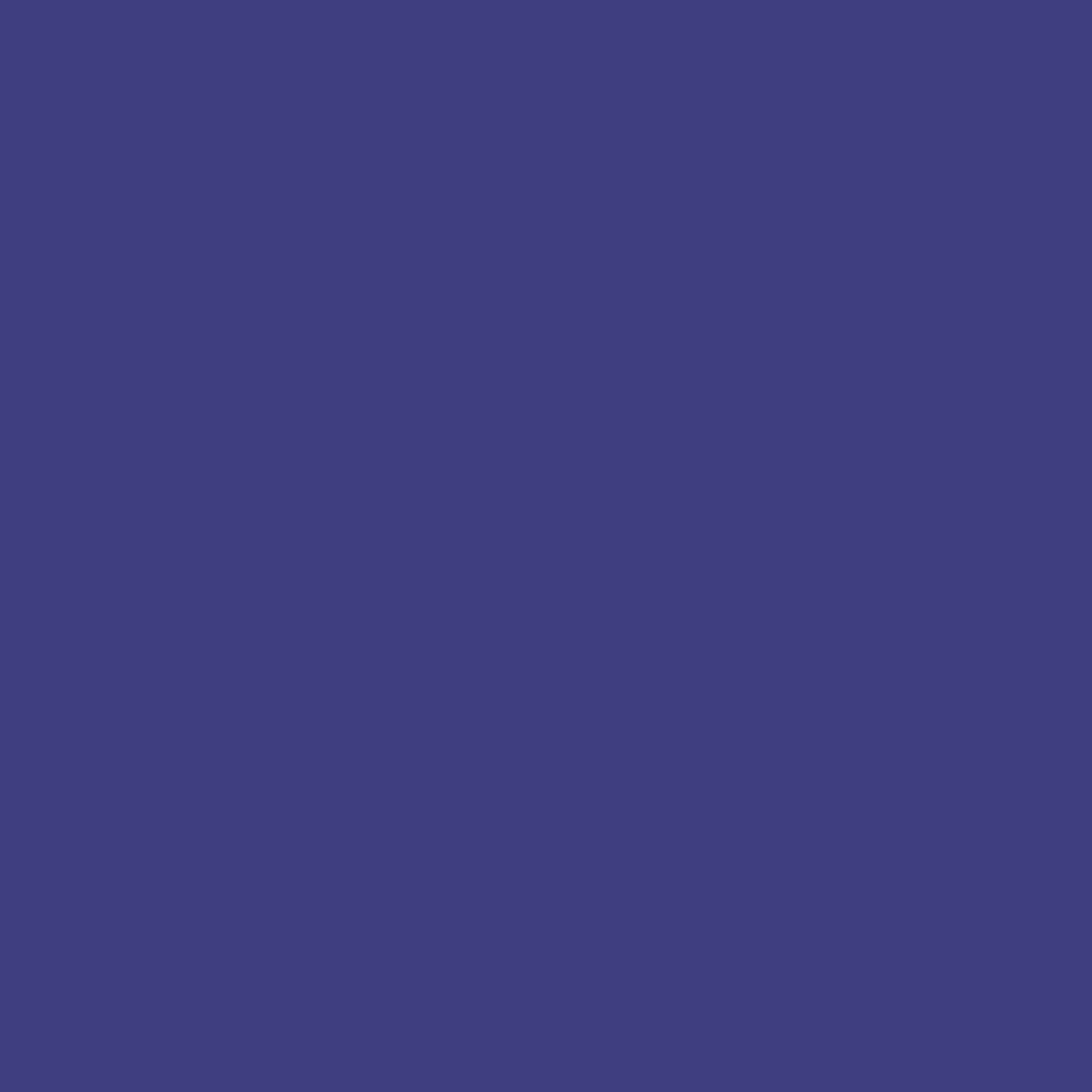}
  \caption{$Q_1$}
  \label{Q1_1}
\end{subfigure}
\begin{subfigure}{.3\textwidth}
  \centering
  \includegraphics[width=0.9\linewidth]{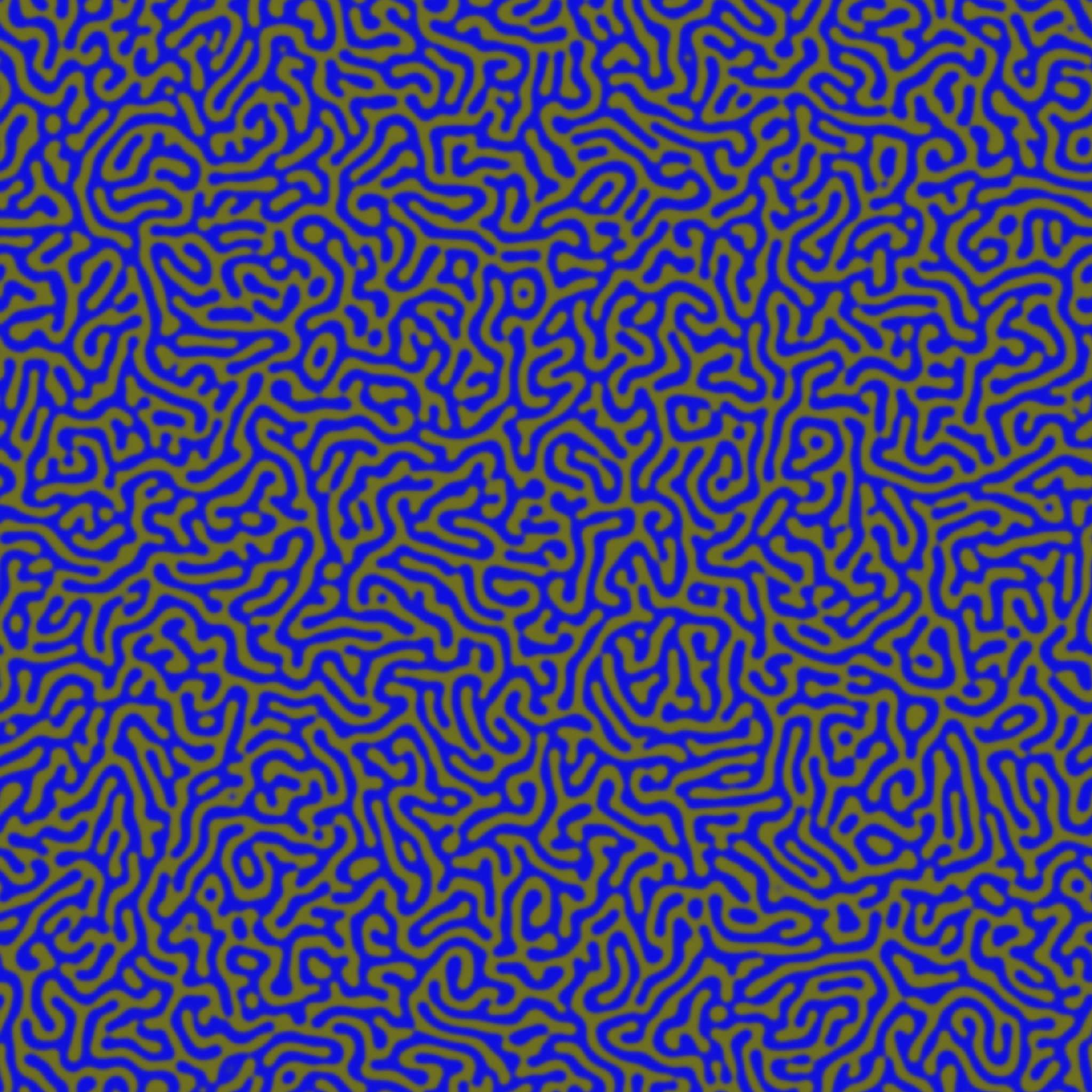}
  \caption{$Q_2$}
  \label{Q2_1}
\end{subfigure}
\begin{subfigure}{.3\textwidth}
  \centering
  \includegraphics[width=0.9\linewidth]{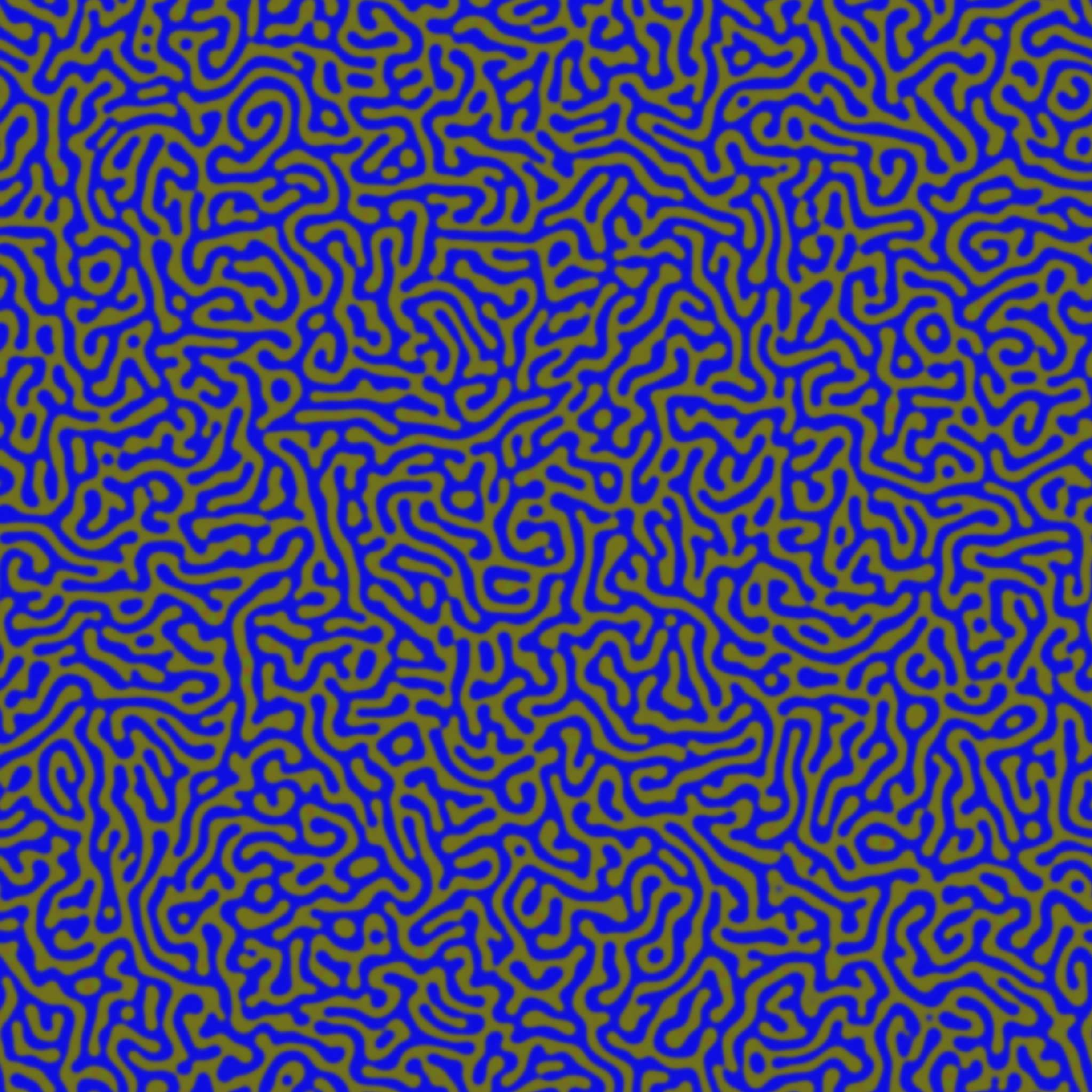}
  \caption{$Q_3$}
  \label{Q3_1}
\end{subfigure}\\
\begin{subfigure}{.3\textwidth}
  \centering
  \includegraphics[width=0.9\linewidth]{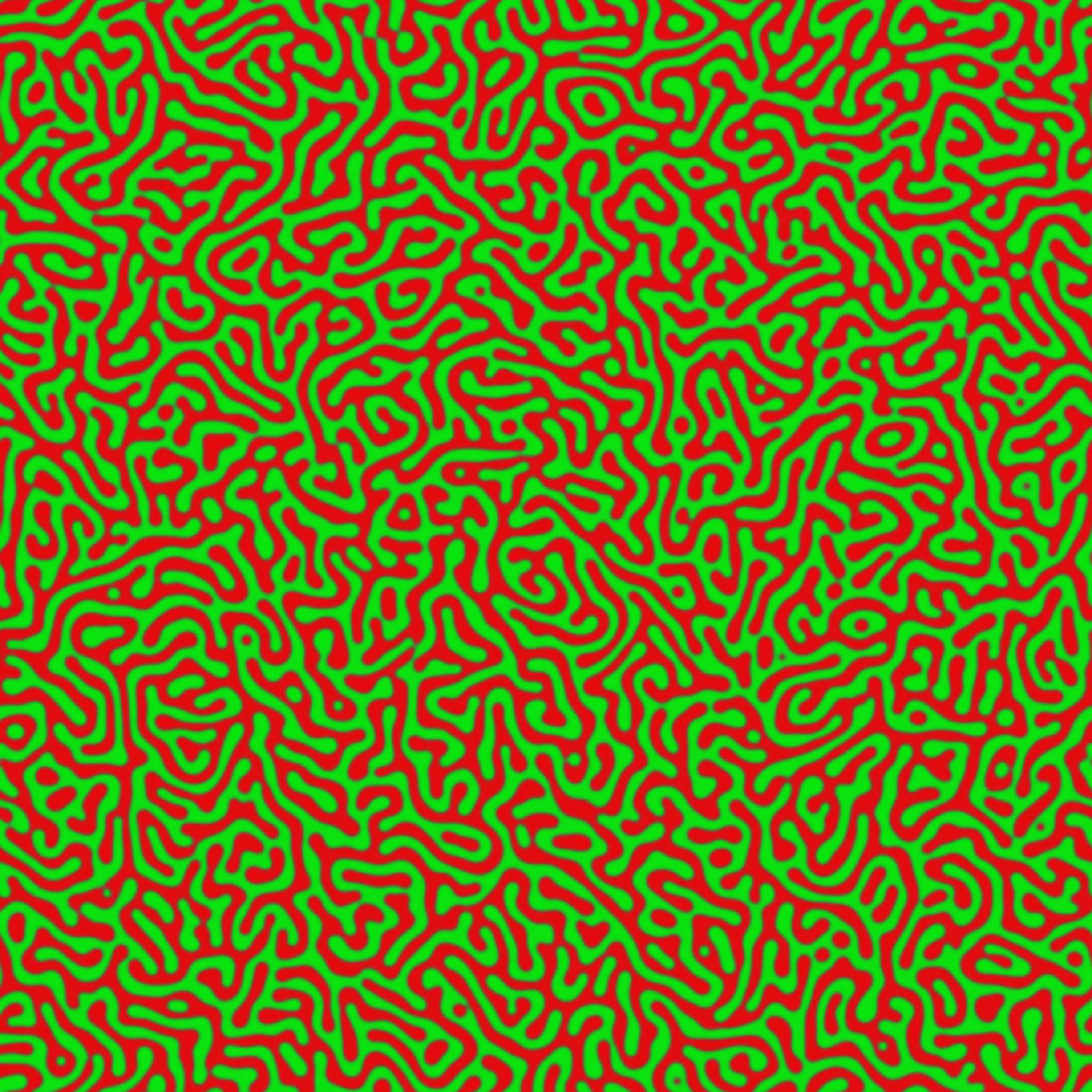}
  \caption{$R_1$}
  \label{R1_1}
\end{subfigure}
\begin{subfigure}{.3\textwidth}
  \centering
  \includegraphics[width=0.9\linewidth]{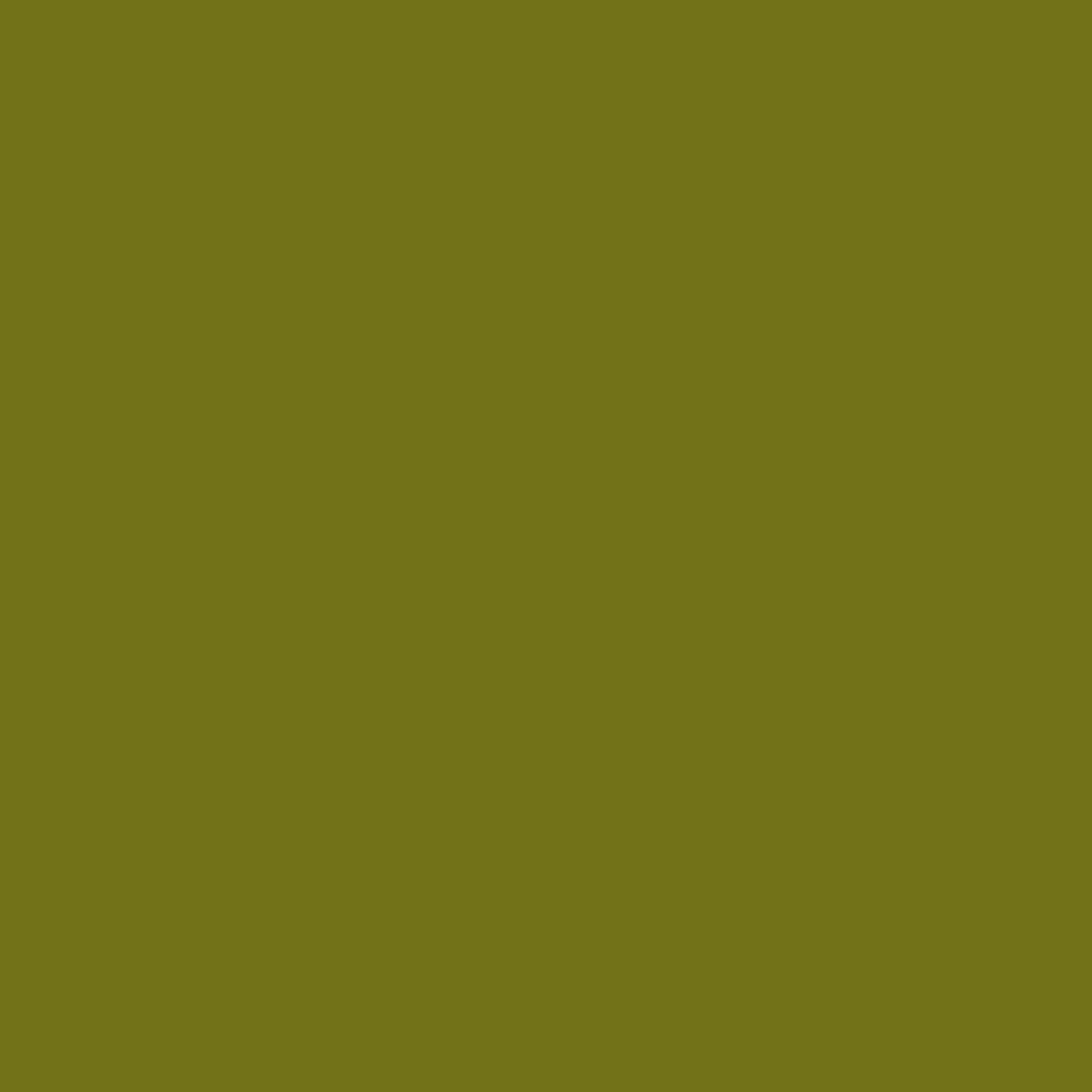}
  \caption{$R_2$}
  \label{R2_1}
\end{subfigure}
\begin{subfigure}{.3\textwidth}
  \centering
  \includegraphics[width=0.9\linewidth]{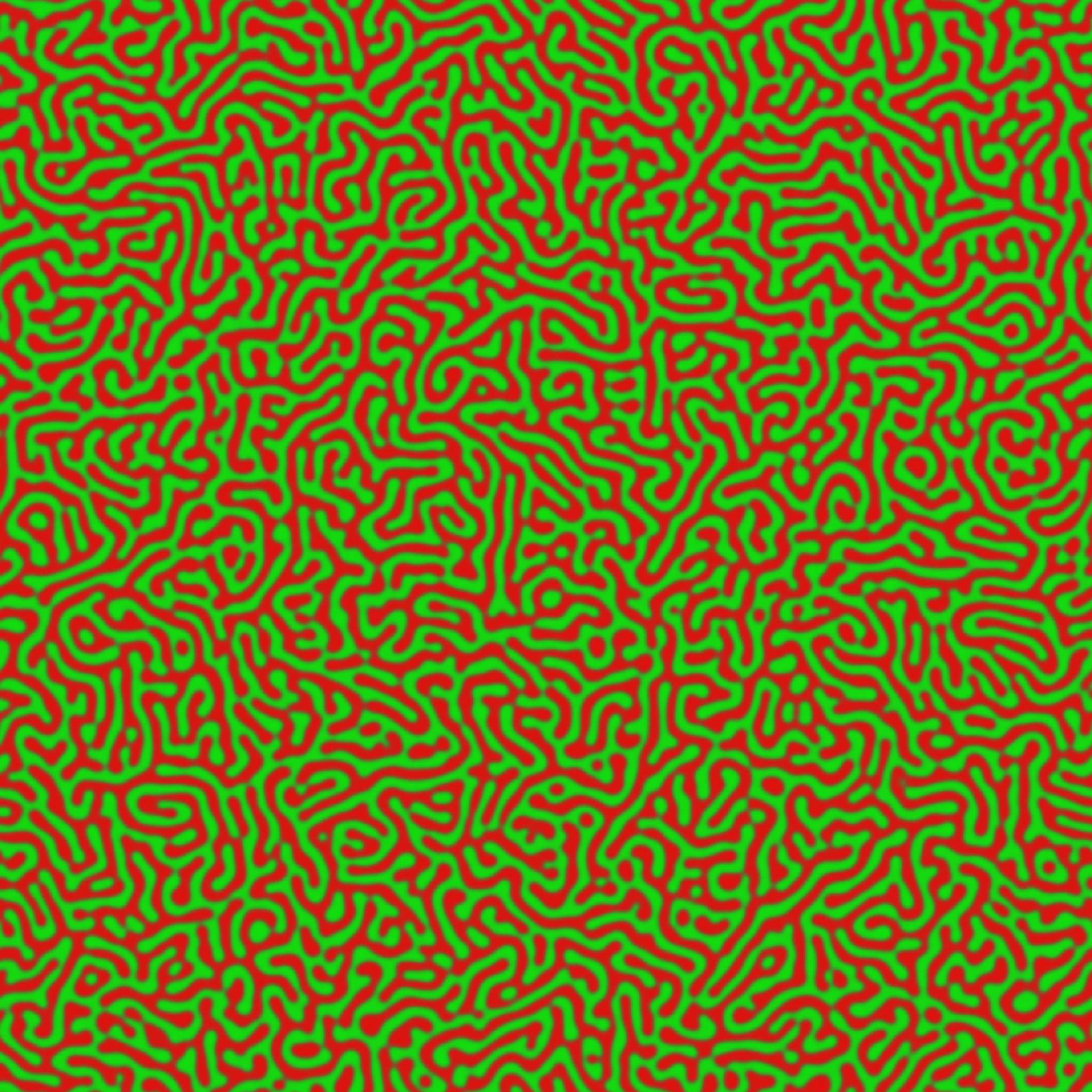}
  \caption{$R_3$}
  \label{R3_1}
\end{subfigure}\\
\caption{Time snapshots of microstructure of alloy systems $P$, $Q$, and $R$ at $t=1000$ (early stage).
The subscripts 1, 2, and 3 indicate different sets of misfit strains $\epsilon_{\alpha\beta}=\epsilon_{\alpha\gamma}$,
$\epsilon_{\alpha\beta}=-\epsilon_{\alpha\gamma}$, and $|\epsilon_{\alpha\beta}|\neq|\epsilon_{\alpha\gamma}|$,
respectively, as given in Table~\ref{cases_C4}.}
\label{fig:early}
\end{figure}

\begin{figure}[H]
  \centering
\begin{subfigure}{.3\textwidth}
  \centering
  \includegraphics[width=0.9\linewidth]{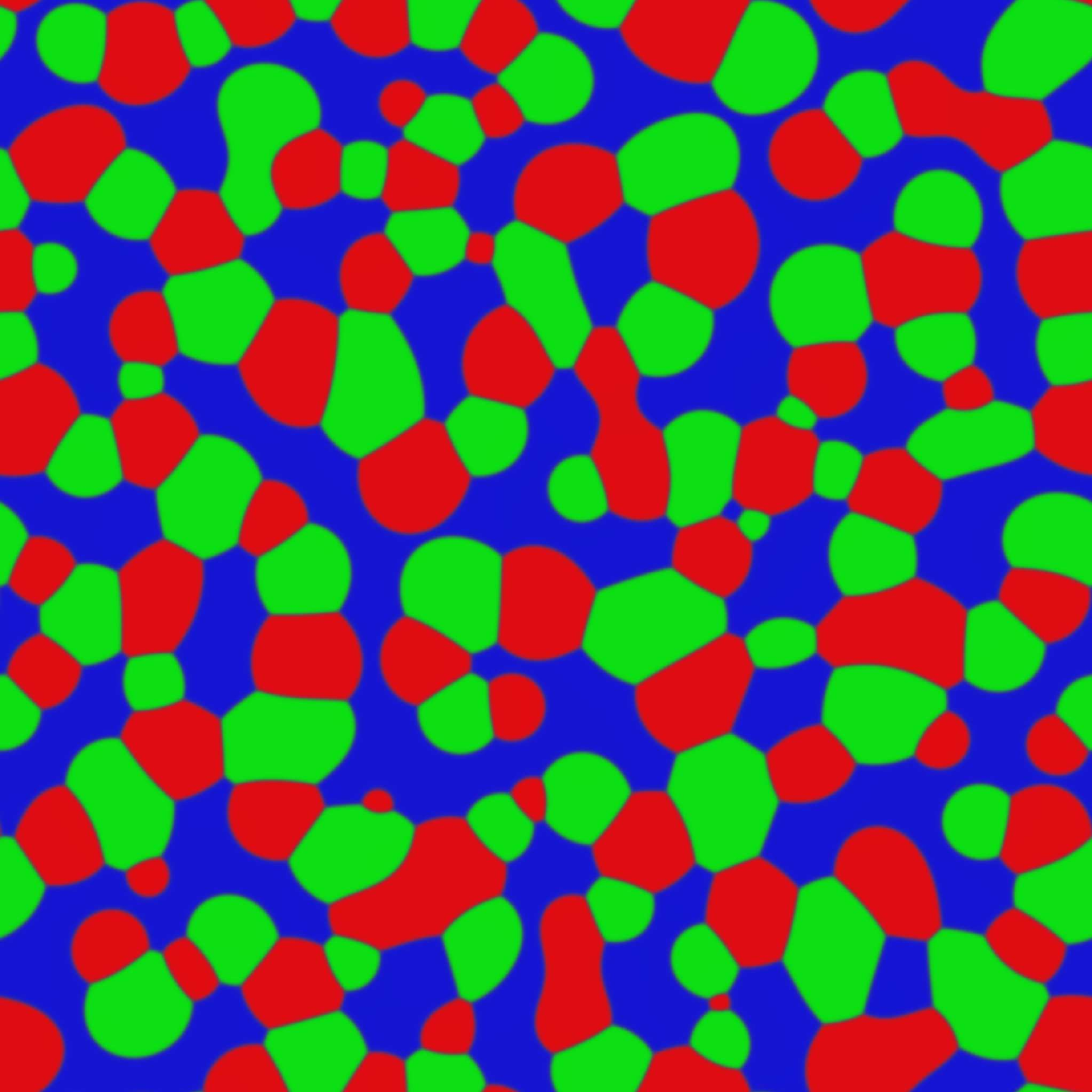}
  \caption{$P_1$}
  \label{P1_2}
\end{subfigure}
\begin{subfigure}{.3\textwidth}
  \centering
  \includegraphics[width=0.9\linewidth]{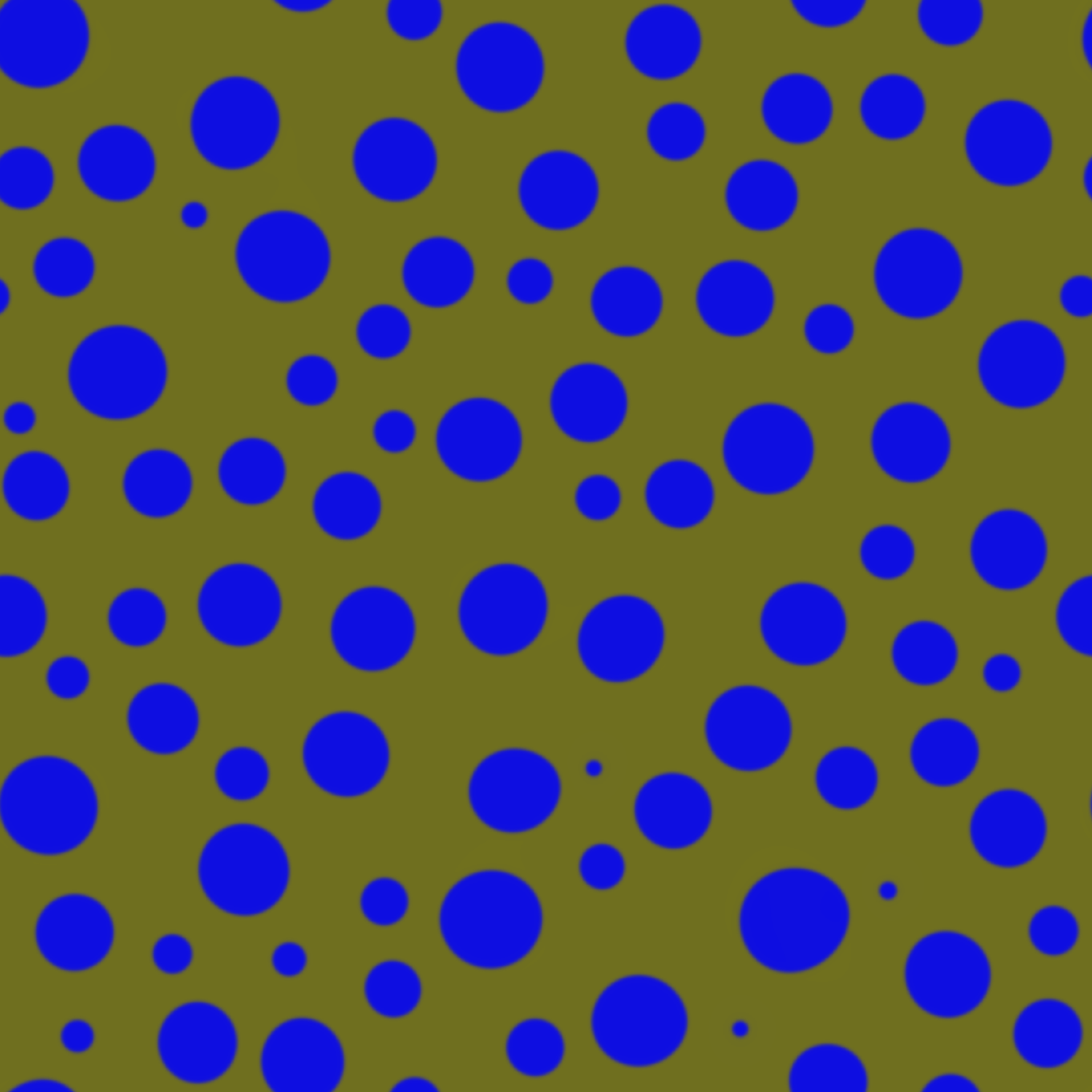}
  \caption{$P_2$}
  \label{P2_2}
\end{subfigure}
\begin{subfigure}{.3\textwidth}
  \centering
  \includegraphics[width=0.9\linewidth]{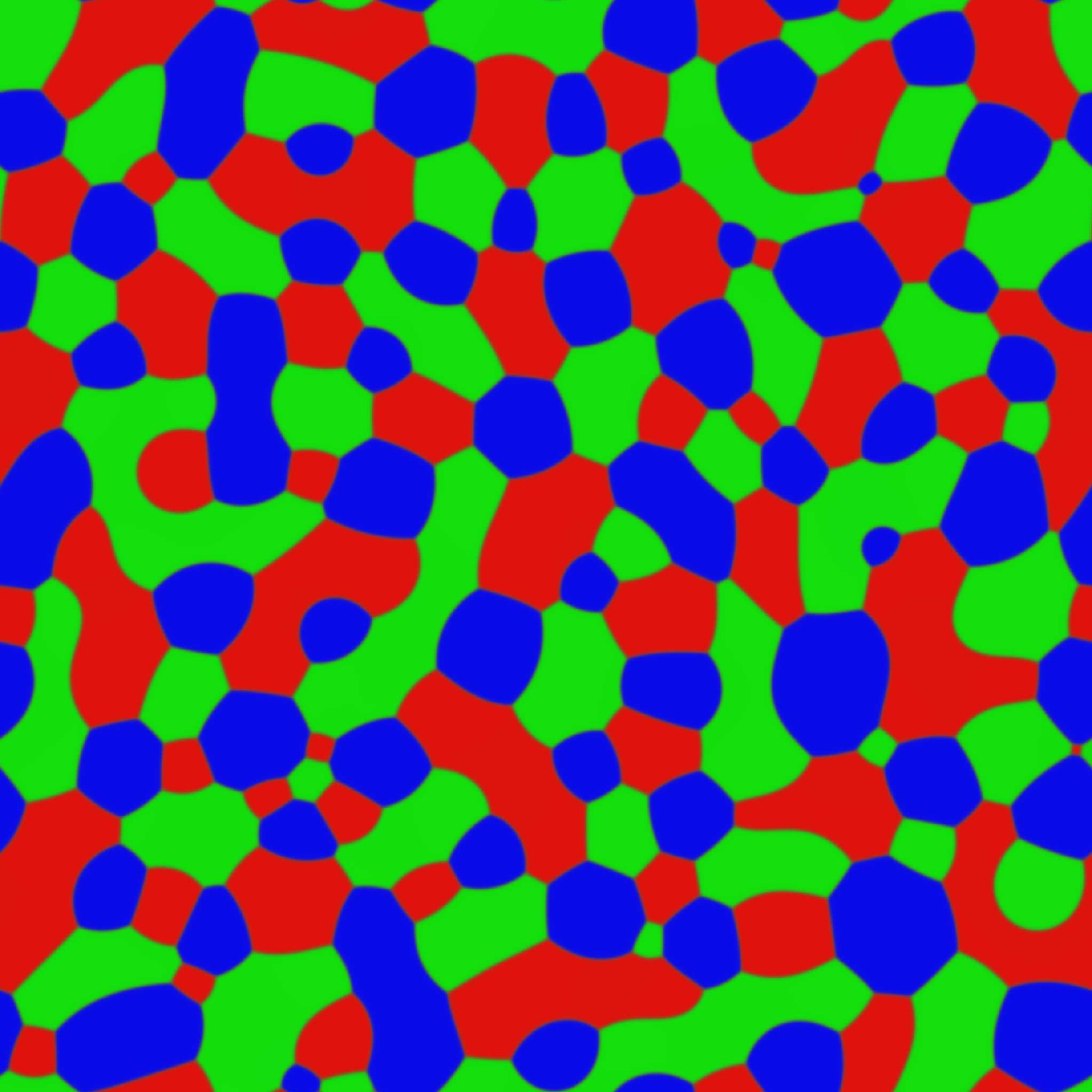}
  \caption{$P_3$}
  \label{P3_2}
\end{subfigure}\\
\begin{subfigure}{.3\textwidth}
  \centering
  \includegraphics[width=0.9\linewidth]{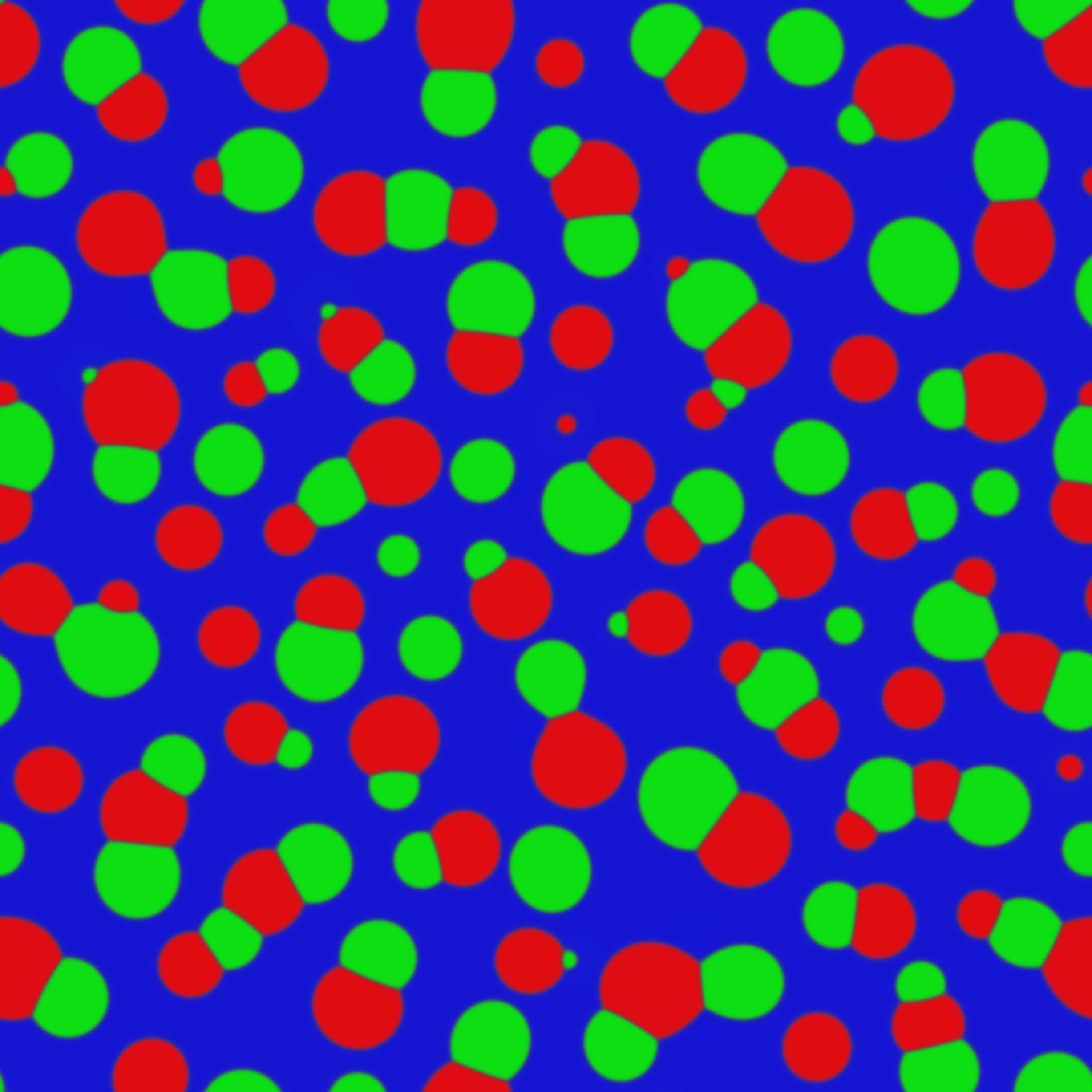}
  \caption{$Q_1$}
  \label{Q1_2}
\end{subfigure}
\begin{subfigure}{.3\textwidth}
  \centering
  \includegraphics[width=0.9\linewidth]{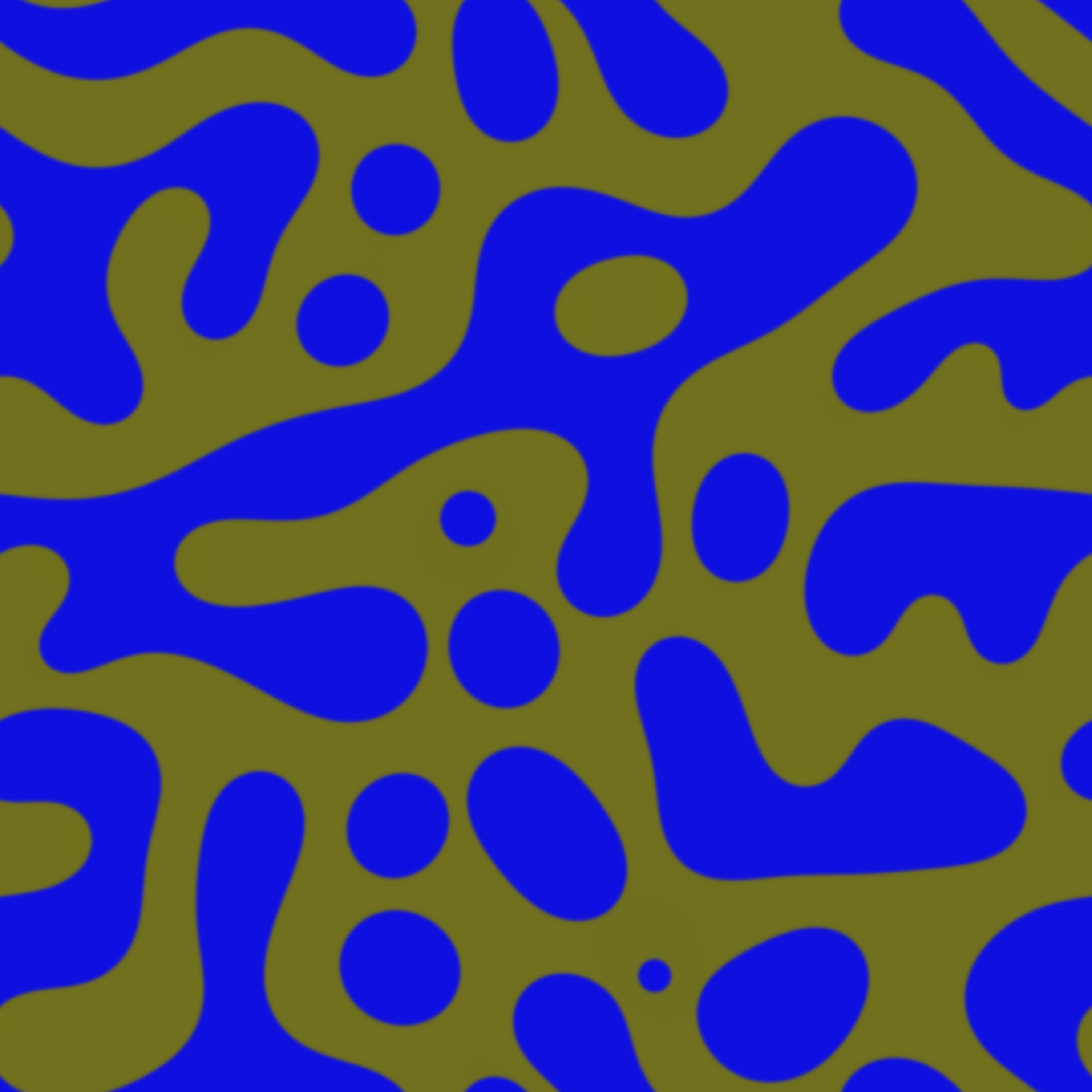}
  \caption{$Q_2$}
  \label{Q2_2}
\end{subfigure}
\begin{subfigure}{.3\textwidth}
  \centering
  \includegraphics[width=0.9\linewidth]{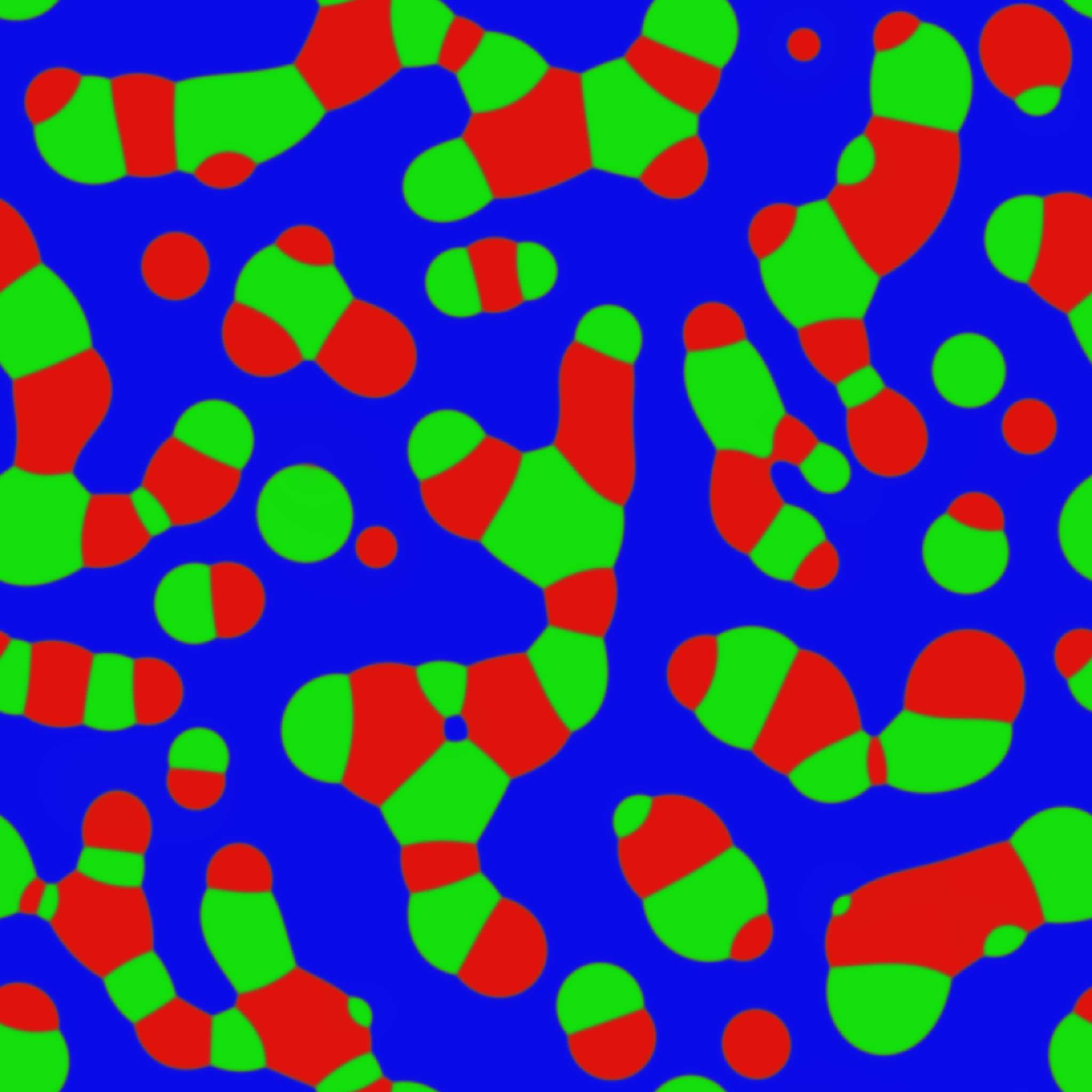}
  \caption{$Q_3$}
  \label{Q3_2}
\end{subfigure}\\
\begin{subfigure}{.3\textwidth}
  \centering
  \includegraphics[width=0.9\linewidth]{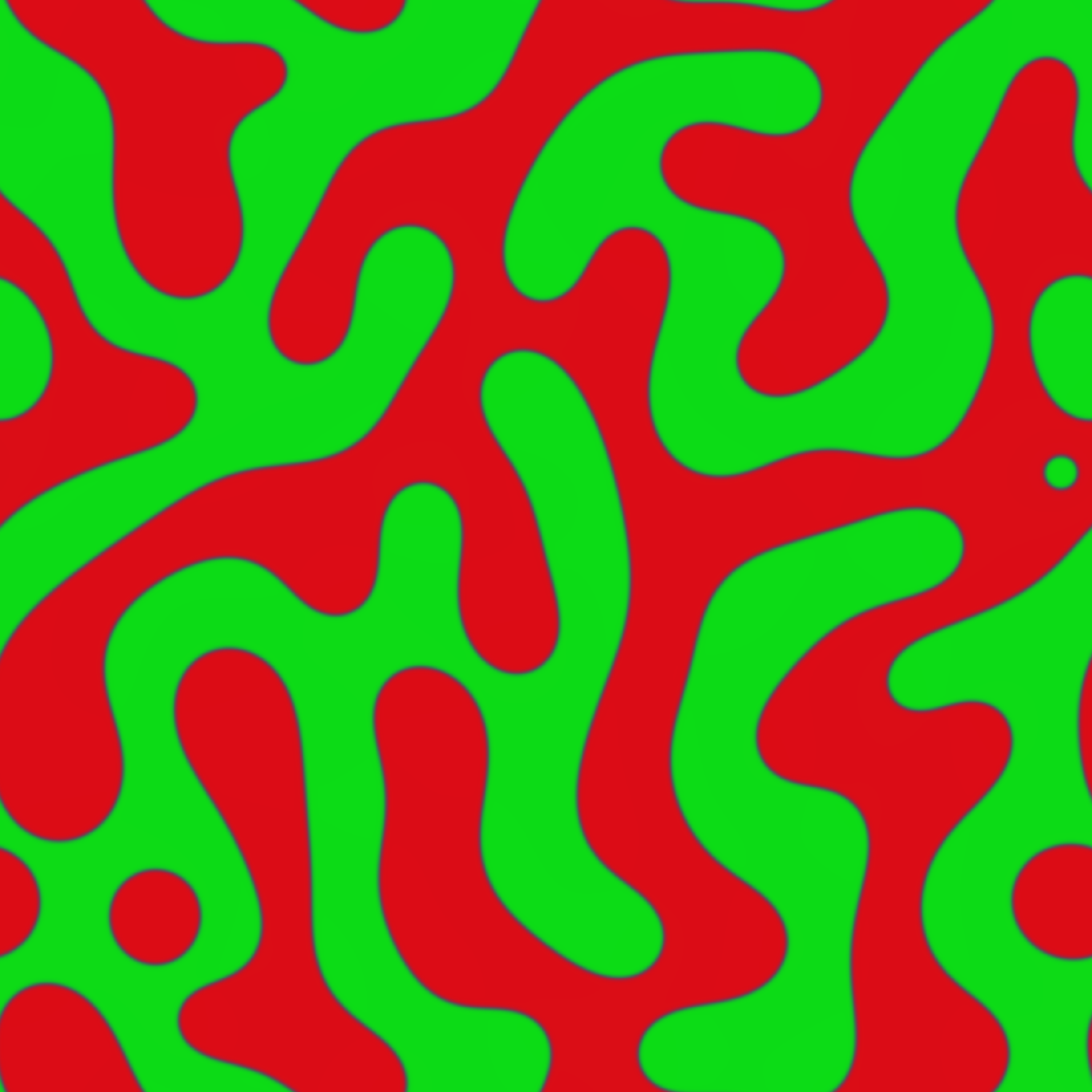}
  \caption{$R_1$}
  \label{R1_2}
\end{subfigure}
\begin{subfigure}{.3\textwidth}
  \centering
  \includegraphics[width=0.9\linewidth]{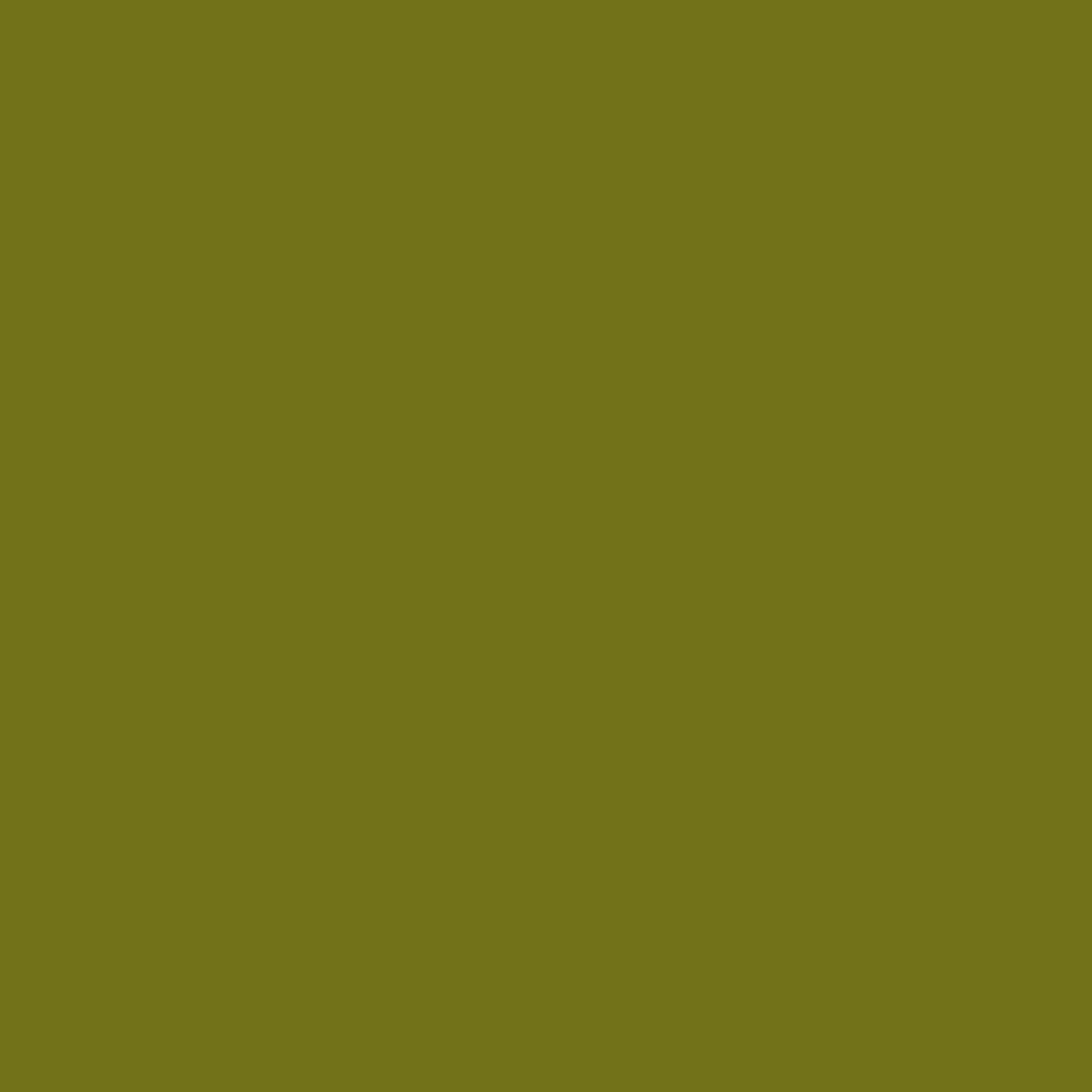}
  \caption{$R_2$}
  \label{R2_2}
\end{subfigure}
\begin{subfigure}{.3\textwidth}
  \centering
  \includegraphics[width=0.9\linewidth]{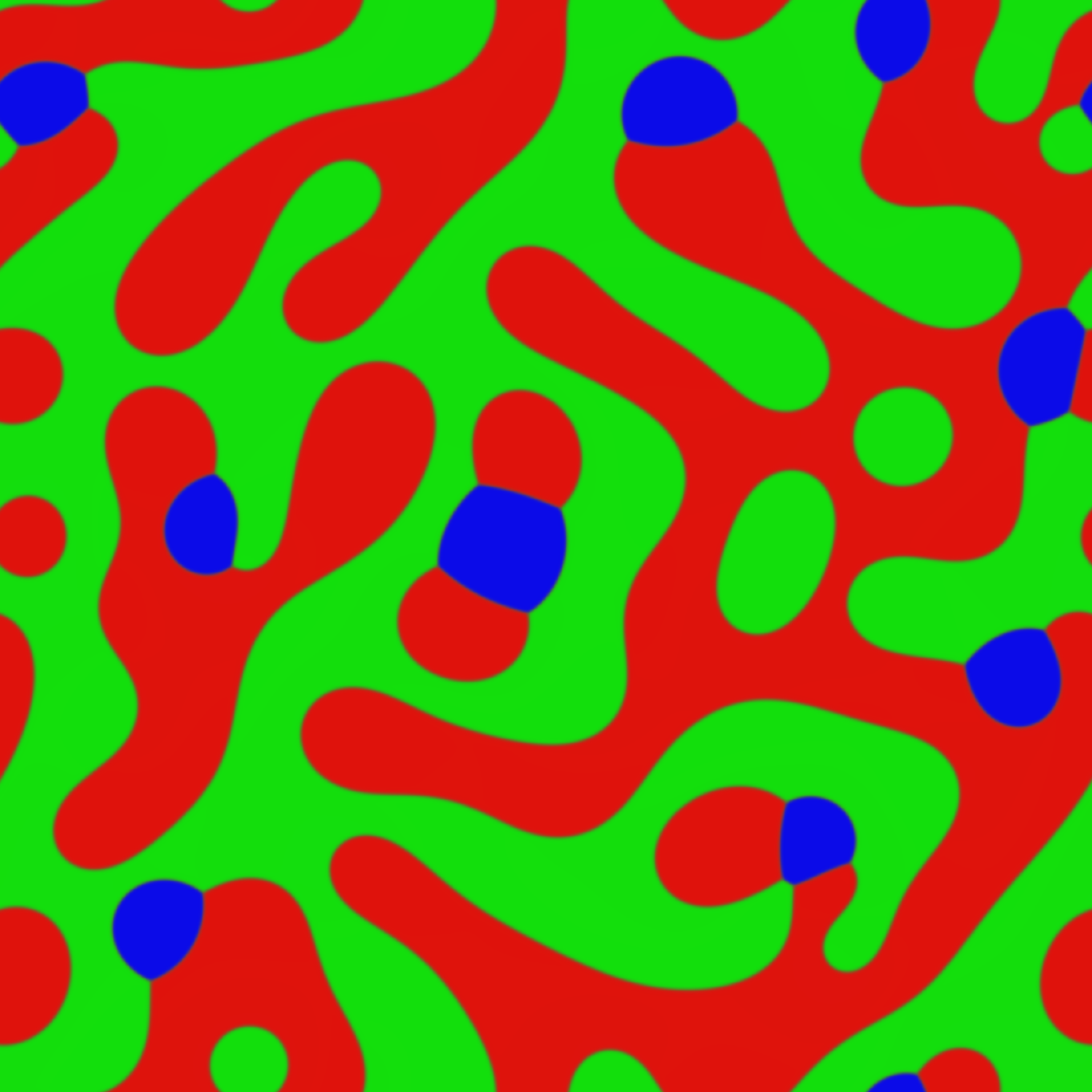}
  \caption{$R_3$}
  \label{R3_2}
\end{subfigure}\\
\caption{Time snapshots of microstructure of alloy systems $P$, $Q$, and $R$ at $t=300000$ (late stage).
The subscripts 1, 2, and 3 indicate different sets of misfit strains $\epsilon_{\alpha\beta}=\epsilon_{\alpha\gamma}$,
$\epsilon_{\alpha\beta}=-\epsilon_{\alpha\gamma}$, and $|\epsilon_{\alpha\beta}|\neq|\epsilon_{\alpha\gamma}|$,
respectively, as given in Table~\ref{cases_C4}.}
\label{fig:late}
\end{figure}

The time snapshots of the evolution of microstructure for equiatomic $P$ alloys,
$A-$rich $Q$ alloys and $A-$poor $R$ alloys at early and late stages of SD are shown in 
Figs.~\ref{fig:early} and~\ref{fig:late}, respectively.
In alloy $P_1$, early stages of SD produce $\beta$ and $\gamma$ phases in the microstructure (Fig.~\ref{P1_1}). 
Large misfit of $\alpha$ phase with $\beta$ and $\gamma$ 
($\epsilon_{\alpha\beta}=\epsilon_{\alpha\gamma}=0.01$) delays $\alpha$ phase separation. Since 
$\epsilon_{\beta\gamma}=0$, the system prefers more $\beta\gamma$ interfaces to minimize 
coherency strain energy. Thus, the microstructure at the late stages of SD (Fig.~\ref{P1_2})  
contains an interconnected network of $B-$rich and $C-$rich domains. 
The $\alpha$ particles appear coarser compared to $\beta$ and $\gamma$. 
Since the magnitudes of mismatch between $\beta$ and $\gamma$ ($\epsilon_{\beta\gamma}$) are large 
in both $P_2$ and $P_3$, B and C species do not partition during the early stages of SD.
As a result, these alloys undergo pseudo-binary SD during the early stages
to form a two-phase microstructure containing $A-$rich domains in an $A-$poor matrix (Figs.~\ref{P2_1},~\ref{P3_1}).
However, the misfit between $\beta$ and $\gamma$ phases in $P_2$ is larger  
than that in $P_3$. Therefore, secondary decomposition of $A-$poor domain 
to $B-$rich and $C-$rich phases is suppressed in $P_2$. 
The $\alpha$ domains remain nearly circular throughout the evolution (Fig.~\ref{P2_2}). 
During the early stages of SD,
$\alpha$ domains coalesce to reduce interfacial energy. Coarsening of $\alpha$ domains  
sets in when $t>10000$.   

In system $P_3$, the $A-$poor matrix undergoes secondary phase separation resulting in the formation of 
a network of $\beta$ and $\gamma$ domains. 
Since $\mid\epsilon_{\alpha\gamma}\mid<\mid\epsilon_{\alpha\beta}\mid$, the late stage 
microstructures contain more $\alpha-\gamma$ interfaces than $\alpha-\beta$ interfaces.
During late stages, $\alpha$ domains lose their circular shapes due to the impingement of diffusion and strain fields.(Fig.~\ref{P3_2}).    

According to de Fontaine's stability analysis, primary SD
of $A-$rich $Q$ alloy leads to composition fluctuations along the median perpendicular to line
$BC$, producing $A-$rich and $A-$poor domains. Subsequent SD of $A-$poor domains along
line $BC$ would yield $B-$rich and $C-$rich phases (Fig~\ref{fig:dF_spino}).
In alloy $Q_1$, since $\epsilon_{\alpha\beta}$ and $\epsilon_{\alpha\gamma}$ have same magnitude and sign, there is an increase in the
effective misfit between the $A-$rich and $A-$poor regions. As a result, there is a delay in
primary phase separation in this alloy system (Fig~\ref{Q1_1}). But, when the $A-$poor regions finally appear, they
rapidly decompose yielding $\beta$ and $\gamma$ phases. 
The $\beta$ and $\gamma$ domains share a strain-free interface between them ($\epsilon_{\beta\gamma}=0$)
and predominantly emerge in the form of networks of conjoined circular particles. 
They maintain the composite morphology and near circular shapes even at later stages of SD (Fig~\ref{Q1_2}). 

Alloys $Q_2$ and $Q_3$ also undergo primary SD forming a bicontinuous microstructure consisting of
$A-$rich and $A-$poor domains. Since the misfits between $\beta$ and $\gamma$
are larger in both alloys, the secondary decomposition of $A-$poor domains is delayed (Figs.~\ref{Q2_1},~\ref{Q3_1}).

However, the magnitude of mismatch between $\beta$ and $\gamma$ phases in $Q_2$ is higher than that in
$Q_3$. Hence, secondary decomposition of $A-$poor domains to $\beta$ and
$\gamma$ phases is inhibited in $Q_2$. The microstructure remains bicontinuous throughout
the evolution (Fig.~\ref{Q2_2}). Whereas, in system $Q_3$, the $A-$poor domains undergo secondary 
phase separation forming dual semicontinuous
networks of alternating $\beta$ and $\gamma$ domains embedded in $\alpha$ matrix(Fig.~\ref{Q3_2}).

For an $A-$poor $R$ alloy with composition in the region of conditional stability (Region II), 
de Fontaine's analysis predicts decomposition along $BC$ line producing $B-$rich and $C-$rich domains.
In alloy $R_1$, since $\epsilon_{\beta\gamma}=0$, the primary SD occurs rapidly forming a bicontinuous
microstructure consisting of $B-$rich and $C-$rich domains (Fig.~\ref{R1_1}). At later stages, the 
two-phase microstructure is stable and $\alpha$ phase does not form due to its high degree of misfits 
associated with $\beta$ and $\gamma$ phases (Fig.~\ref{R1_2}). 

The large magnitude of mismatch between $\beta$ and $\gamma$ phases ($\epsilon_{\beta\gamma}=0.02$)
completely suppresses primary SD in alloy $R_2$ (Figs.~\ref{R2_1},~\ref{R2_2}). The high degree of
misfits associated with the coexisting phases $\alpha$, $\beta$ and $\gamma$ appears to shift 
the alloy from region of conditional stability to region of absolute stability. 

Alloy $R_3$ undergoes initial phase separation forming a bicontinuous microstructure containing 
$B-$rich and $C-$rich domains (Fig.~\ref{R3_1}). Since the lattice mismatch between $\alpha$ and $\gamma$ 
is very low ($\epsilon_{\alpha\gamma}=0.0028$), this alloy goes through next stage of decomposition leading 
to the formation of $\alpha$ particles at $\beta$-$\gamma$ boundaries (Fig.~\ref{R3_2}).  

\subsection{Kinetic paths of $\alpha$, $\beta$ and $\gamma$ phases in equiatomic $P$ alloys}
\label{kin_paths:P123}

We follow the compositional history of six representative $A-$ rich, $B-$ rich, and 
$C-$ rich regions during microstructural evolution and plot best fitted curves 
on a Gibbs triangle to show the kinetic paths of evolution of $\alpha$, $\beta$ and $\gamma$
phases during SD in equiatomic alloys $P_1$, $P_2$ and $P_3$. We compare the sequences of SD 
obtained from kinetic paths with the directions of decomposition
predicted by de Fontaine's stability analysis~\cite{de1972analysis}. 

According to de Fontaine's stability analysis, when an equiatomic ternary alloy (characterized by a symmetric miscibility gap)
undergoes bulk, incoherent SD producing three phases with equal interfacial energy, 
the compositional history of each phase should be symmetric and should 
lie along the medians of Gibbs triangle (see Fig.~\ref{fig:dF_spino}).   
Later, Bhattacharyya has shown that the difference in the relative interfacial energies 
between the phases may alter the kinetic paths of decomposition~\cite{bhattacharyathesis}.   
Here we aim to show the influence of strain partitioning between the coherent phases on the kinetic paths 
even when the phases have equal interfacial energies between them. 
\begin{figure}[H]
  \centering
\begin{subfigure}{.32\textwidth}
  \centering
\begin{overpic}[width=1.05\textwidth]{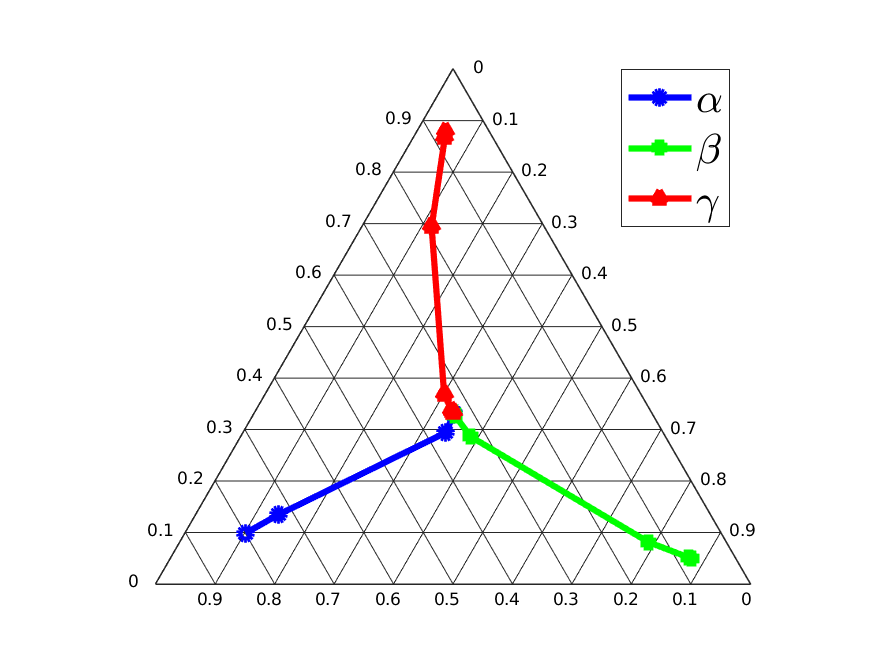}
 \put (7,5) {A}
 \put (90,5) {B}
 \put (48,70) {C}
\end{overpic}
  \caption{$P_1$}
  \label{fig: P1_kPath}
\end{subfigure} 
\begin{subfigure}{.32\textwidth}
  \centering
\begin{overpic}[width=1.05\textwidth]{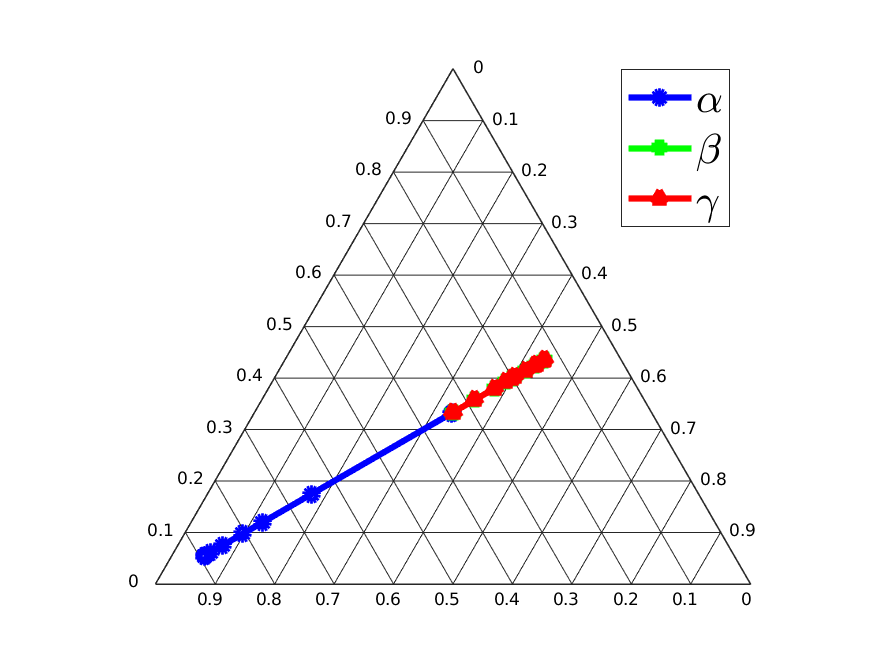}
 \put (7,5) {A}
 \put (90,5) {B}
 \put (48,70) {C}
\end{overpic}  \caption{$P_2$}
  \label{fig: P2_kPath}
\end{subfigure} 
\begin{subfigure}{.32\textwidth}
  \centering
\begin{overpic}[width=1.05\textwidth]{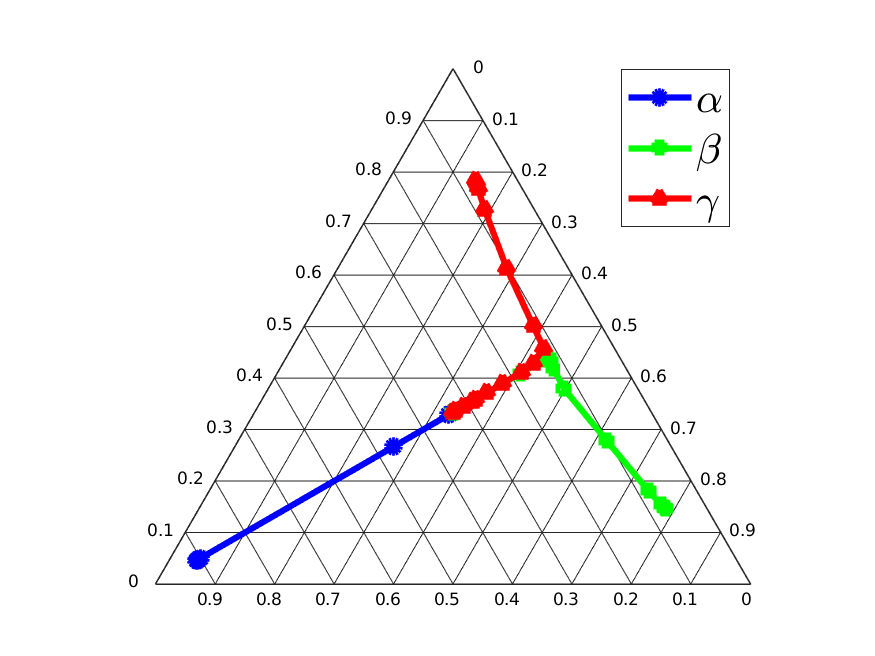}
 \put (7,5) {A}
 \put (90,5) {B}
 \put (48,70) {C}
\end{overpic}  \caption{$P_3$}
  \label{fig: P3_kPath}
\end{subfigure}
\caption{Kinetic paths of $\alpha$, $\beta$ and $\gamma$ phases for (a) $P_1$, (b) $P_2$ and 
(c) $P_3$. The lines represent best fitted curves through the data points.}
\label{fig:P_kPath}
\end{figure}
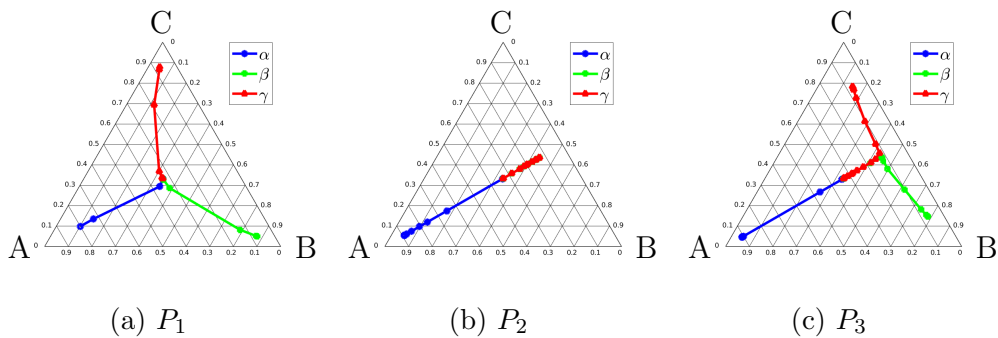 
As shown in Figs.~\ref{fig: P1_kPath},~\ref{fig: P2_kPath},~\ref{fig: P3_kPath}, 
the kinetic paths of formation of $\alpha$, $\beta$ and $\gamma$ phases
in $P_1$, $P_2$ and $P_3$ differ from the predictions 
based on de Fontaine's stability analysis. 
The delay in the formation of  
of $\alpha$ in alloy system $P_1$ is evident from the asymmetry of the 
kinetic paths (see Fig.~\ref{fig: P1_kPath}). 
The initial SD takes place along a line parallel to BC while the $\alpha$
phase appears later.   
On the other hand, in system $P_2$, absence of kinetic paths associated with 
$\beta$ and $\gamma$ phases indicate a pseudo-binary SD leading to $A-$rich and
$A-$ poor regions along a line perpendicular to BC (see Fig.~\ref{fig: P2_kPath}).
System $P_3$ follows a similar trend at the initial stages (Fig.~\ref{fig: P3_kPath}), 
where primary decomposition occurs perpendicular to BC leading to the formation of $A-$rich and $A-$poor 
domains. However, further partitioning of $B$ and $C$ leads to secondary SD 
along a line parallel to BC direction in alloy $P_3$ forming $\beta$ and $\gamma$ phases 
in the microstructure. 

\subsection{Structure functions ($S_{ii}(\mathbf{k},t),i = A,B,C$) for equiatomic $P$ alloys}
\label{struc_functions:P123}

We also characterize the three-phase microstructures using three time-dependent 
and linearly independent structure functions $S_{ii}(\textbf{k},t)$~\cite{PhysRevE.60.3564,chakrabarti1993late}: 
\begin{equation}
S_{ii}(\mathbf{k},t) = \frac{1}{N}\Big \langle \sum_{r}\sum_{r^\prime}e^{-i\mathbf{k}.\mathbf{r}}
[c_i(\mathbf{r}+\mathbf{r}^\prime,t)-\bar{c_i}][c_i(\mathbf{r}^\prime,t)-\bar{c_i}] \Big \rangle,
\end{equation}
where $i = A,B,C$, and $c_i$ denotes the average concentration of the species $i$.
$\mathbf k$ denotes the scattering vector whose magnitude is equal to the 
reciprocal of the wavelength of modulation of the composition field. 

$S_{ii}(\mathbf{k},t)$ provides a measure for the spatial correlations in 
composition of component $i$ in the reciprocal space. 
A linear combination of $S_{ii}(\mathbf{k},t)$ mathematically 
represents the scattered intensity obtained from small 
angle X-ray and neutron scattering (SAXS,SANS) studies of a ternary solid solution.
For isotropic systems, we perform a circular averaging 
of the structure functions and 
plot them as a function of radial distance in reciprocal space. 
\begin{figure}[H]
  \centering
\begin{subfigure}{.32\textwidth}
  \centering
  \includegraphics[width=1.0\linewidth]{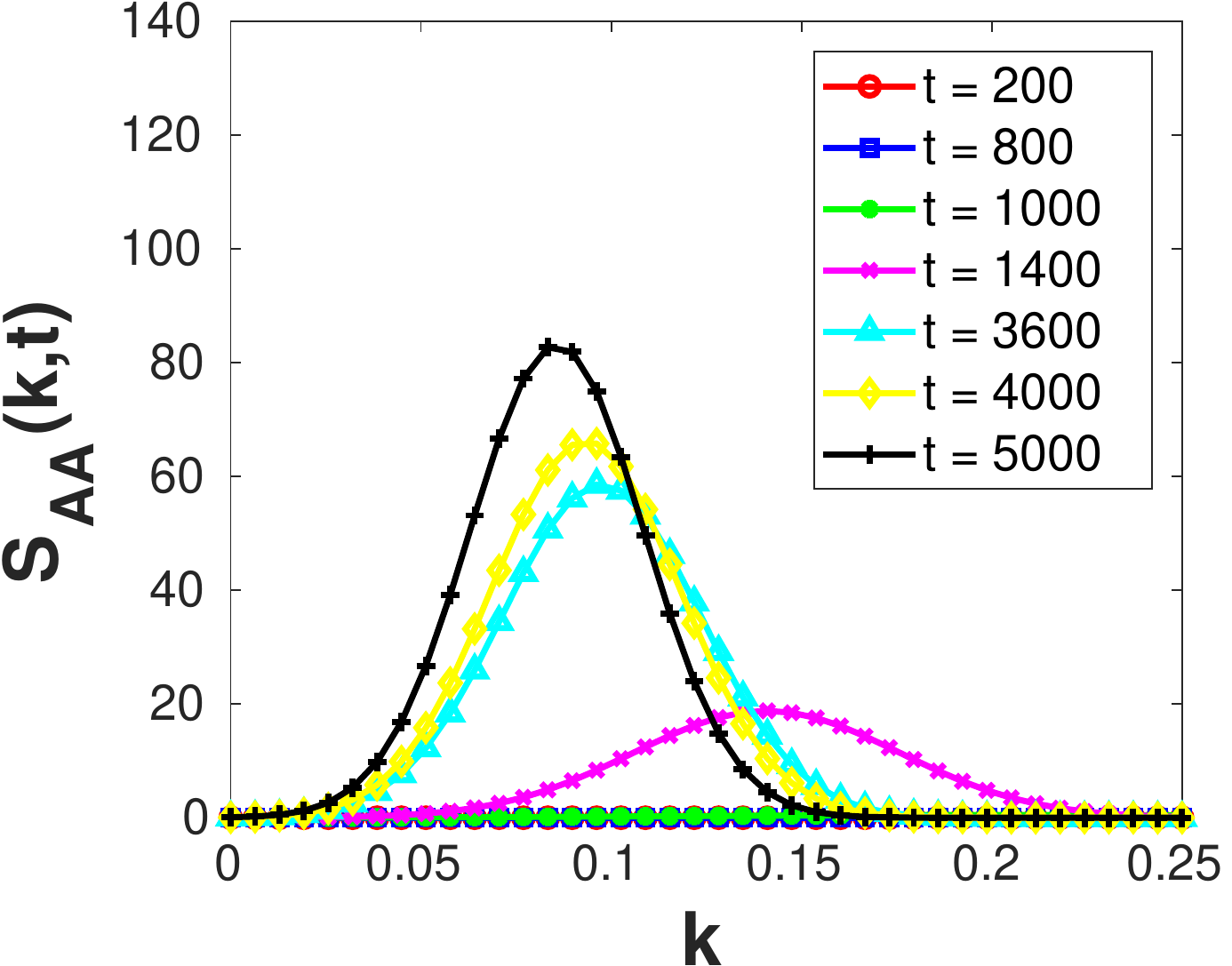}
  \caption{}
  \label{fig: P1_Saa}
\end{subfigure} 
\begin{subfigure}{.32\textwidth}
  \centering
  \includegraphics[width=1.0\linewidth]{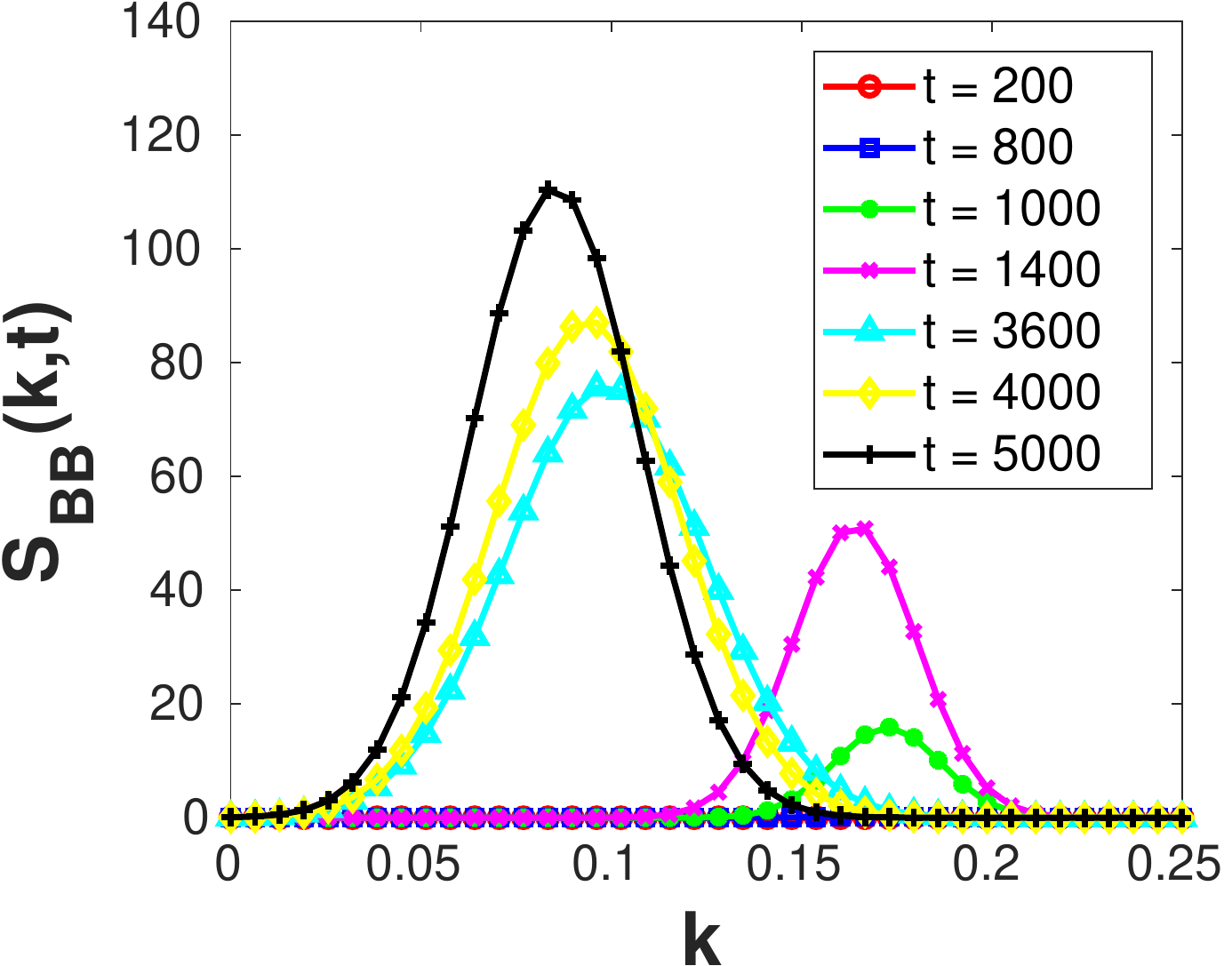}
  \caption{}
  \label{fig: P1_Sbb}
\end{subfigure} 
\begin{subfigure}{.32\textwidth}
  \centering
  \includegraphics[width=1.0\linewidth]{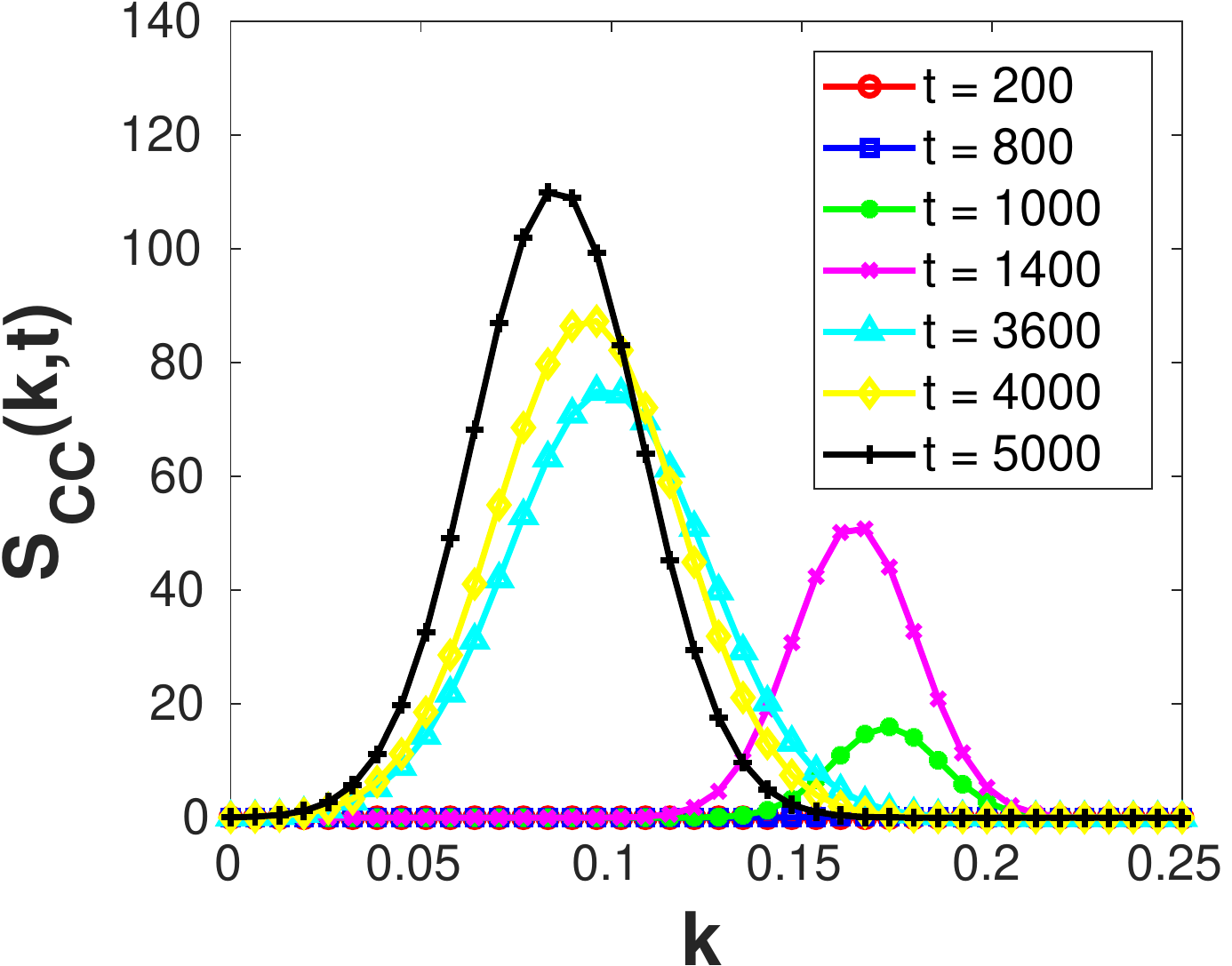}
  \caption{}
  \label{fig: P1_Scc}
\end{subfigure} \\
\begin{subfigure}{.32\textwidth}
  \centering
  \includegraphics[width=1.0\linewidth]{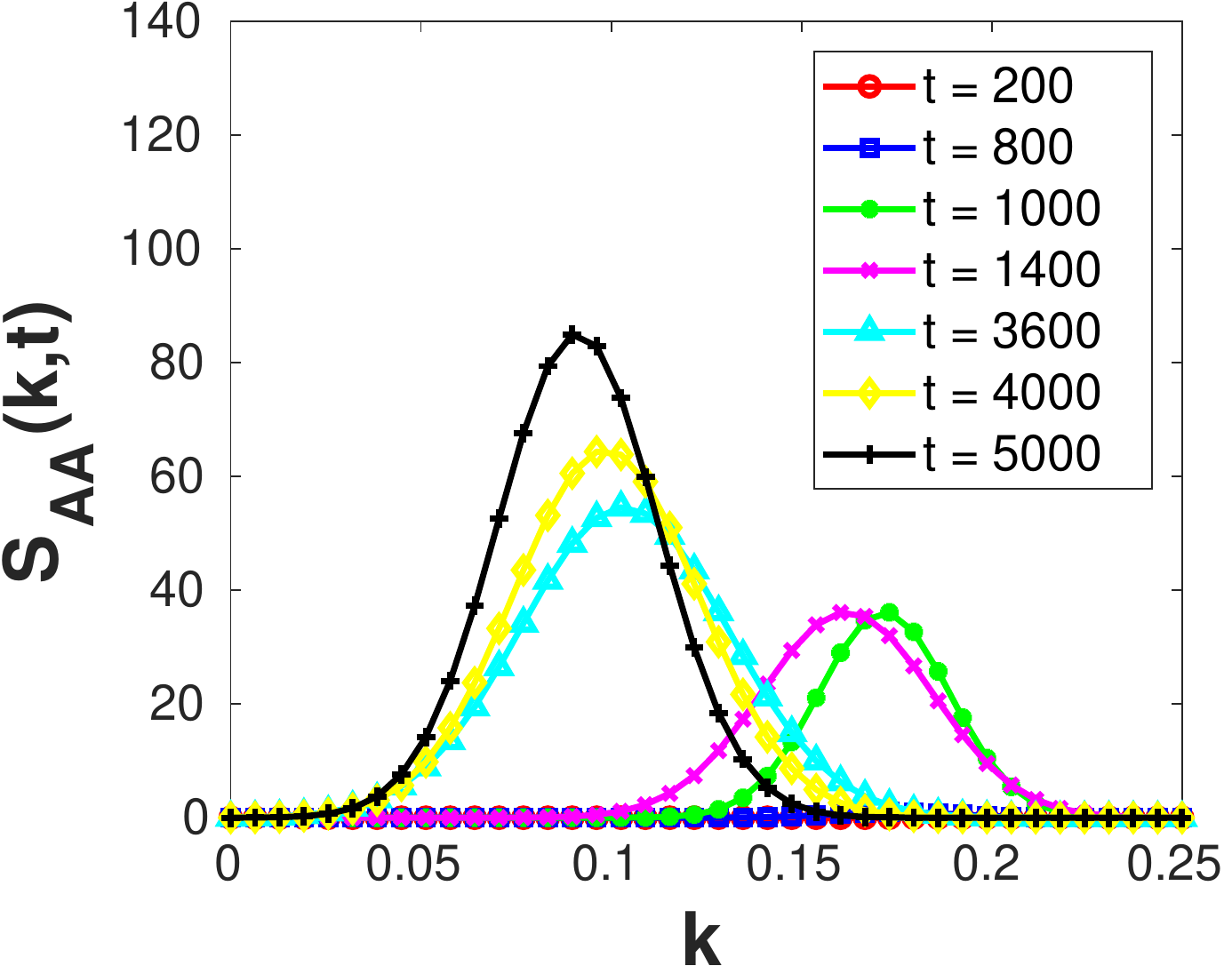}
  \caption{}
  \label{fig: P2_Saa}
\end{subfigure} 
\begin{subfigure}{.32\textwidth}
  \centering
  \includegraphics[width=1.0\linewidth]{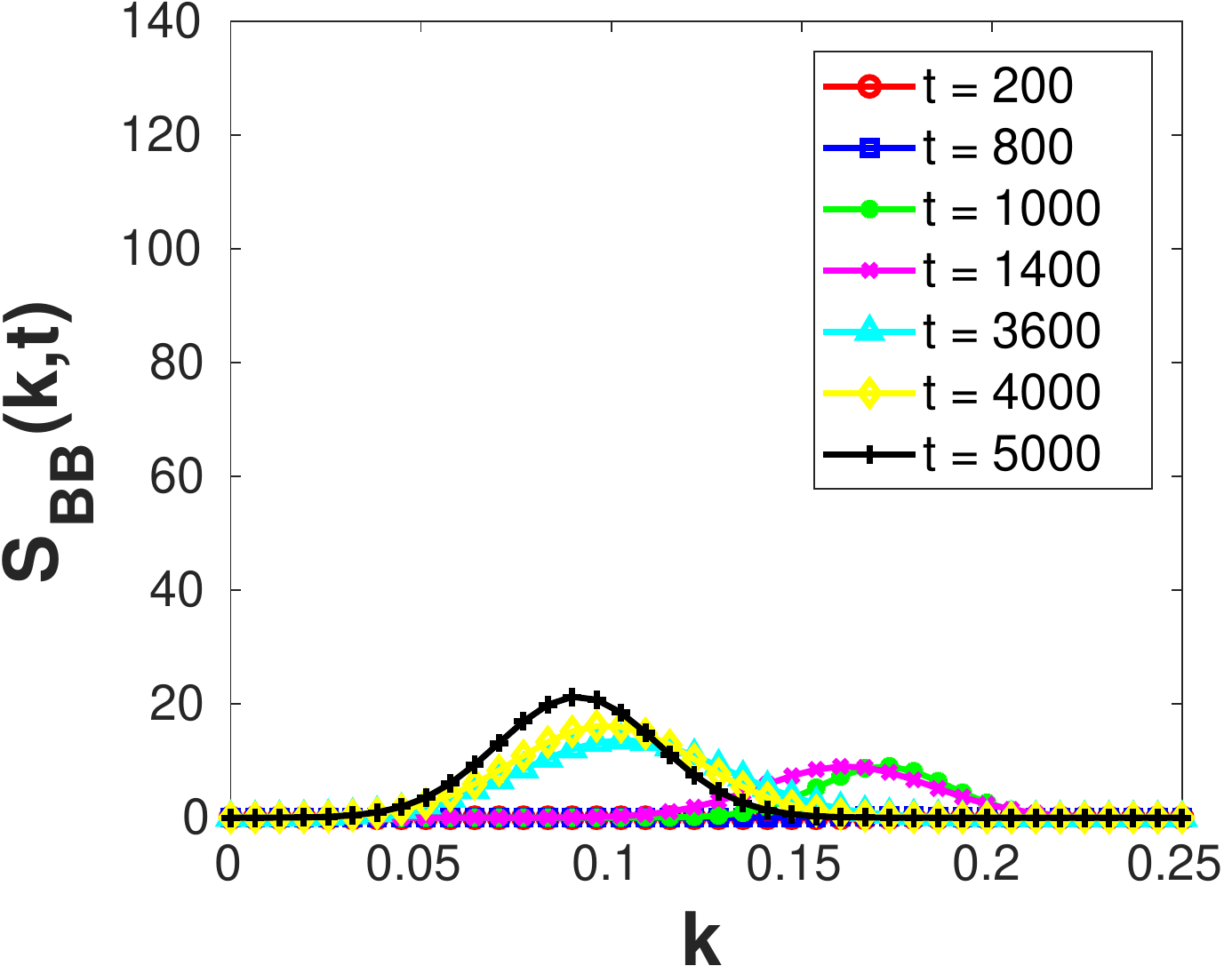}
  \caption{}
  \label{fig: P2_Sbb}
\end{subfigure} 
\begin{subfigure}{.32\textwidth}
  \centering
  \includegraphics[width=1.0\linewidth]{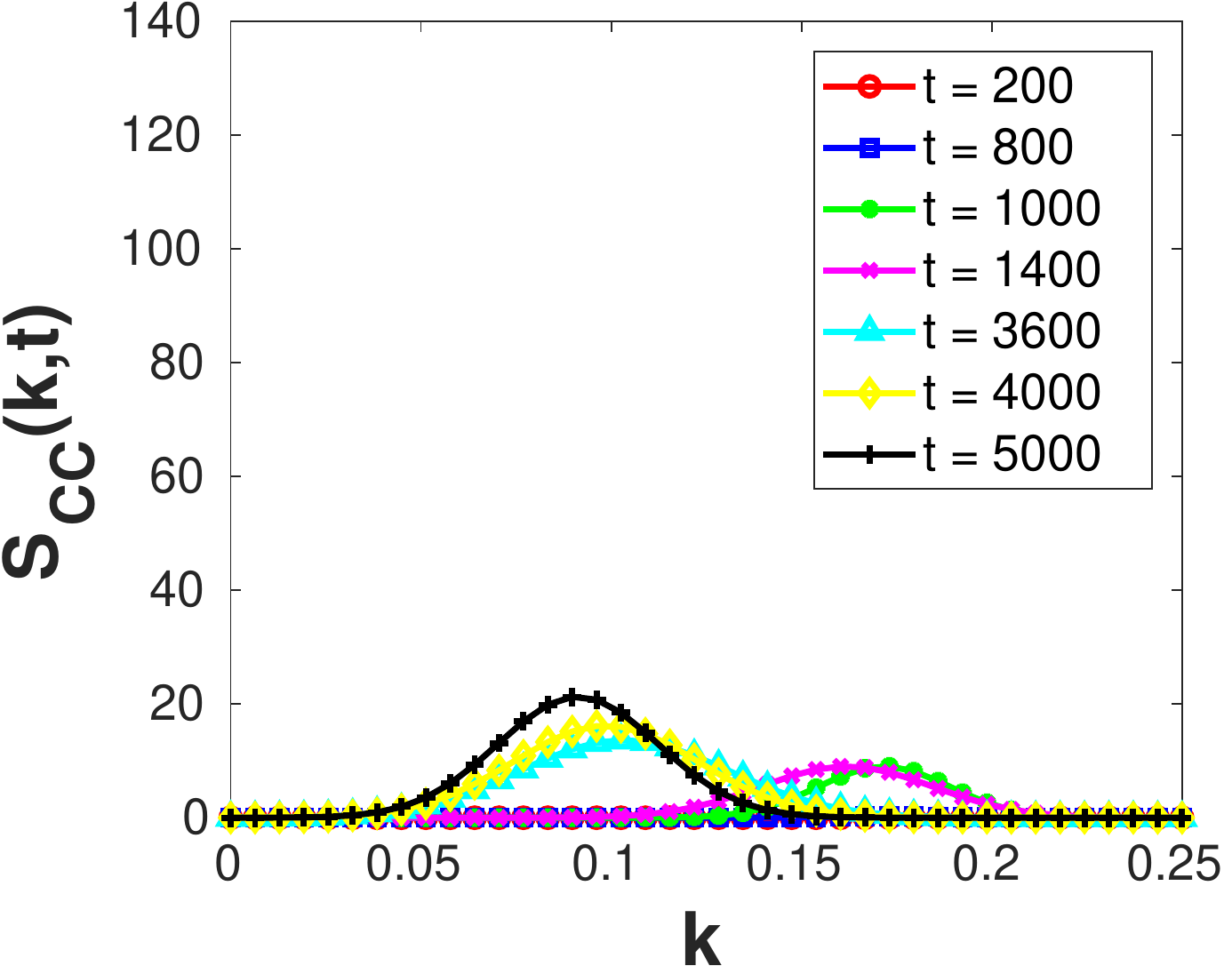}
  \caption{}
  \label{fig: P2_Scc}
\end{subfigure} \\
\begin{subfigure}{.32\textwidth}
  \centering
  \includegraphics[width=1.0\linewidth]{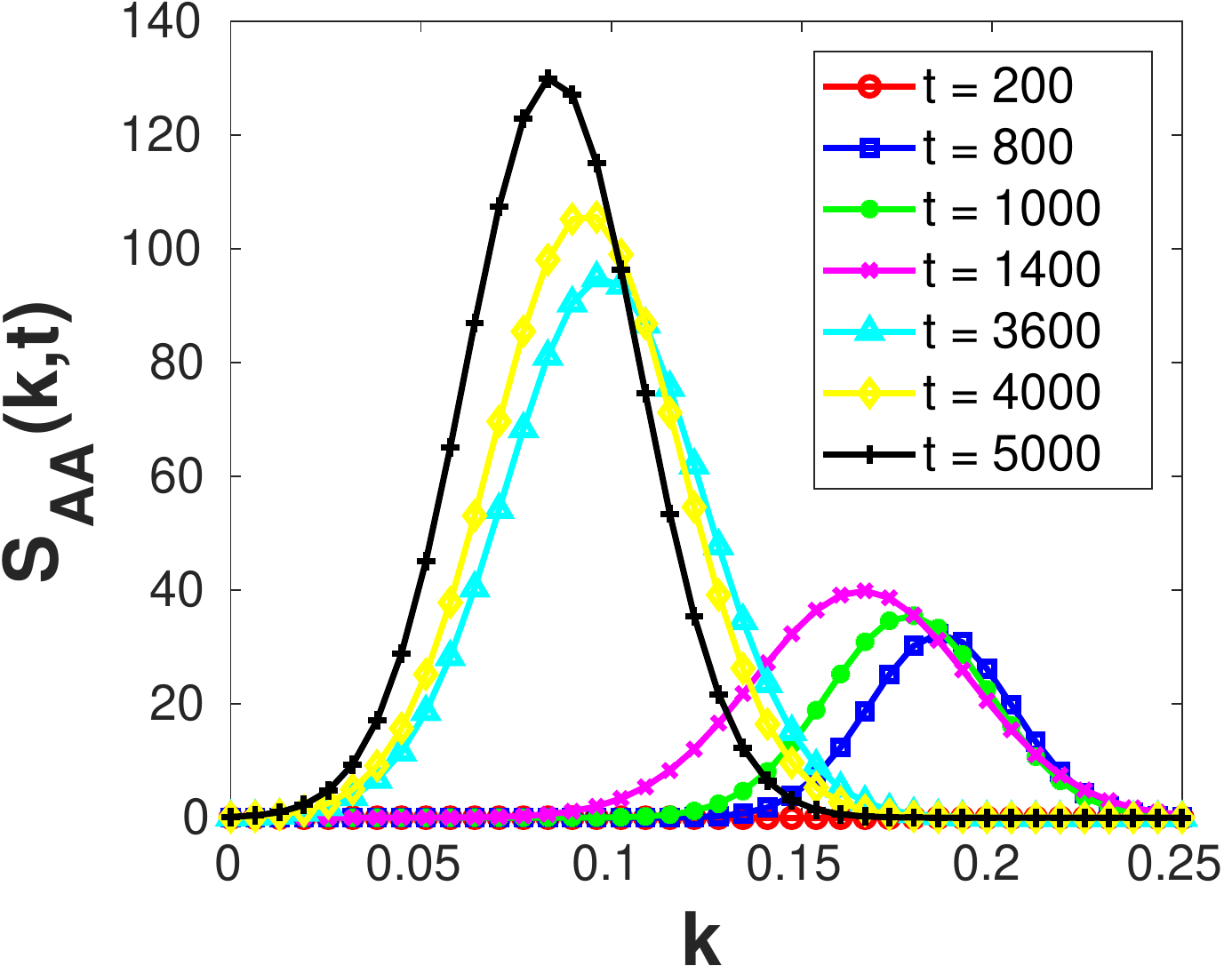}
  \caption{}
  \label{fig: P3_Saa}
\end{subfigure} 
\begin{subfigure}{.32\textwidth}
  \centering
  \includegraphics[width=1.0\linewidth]{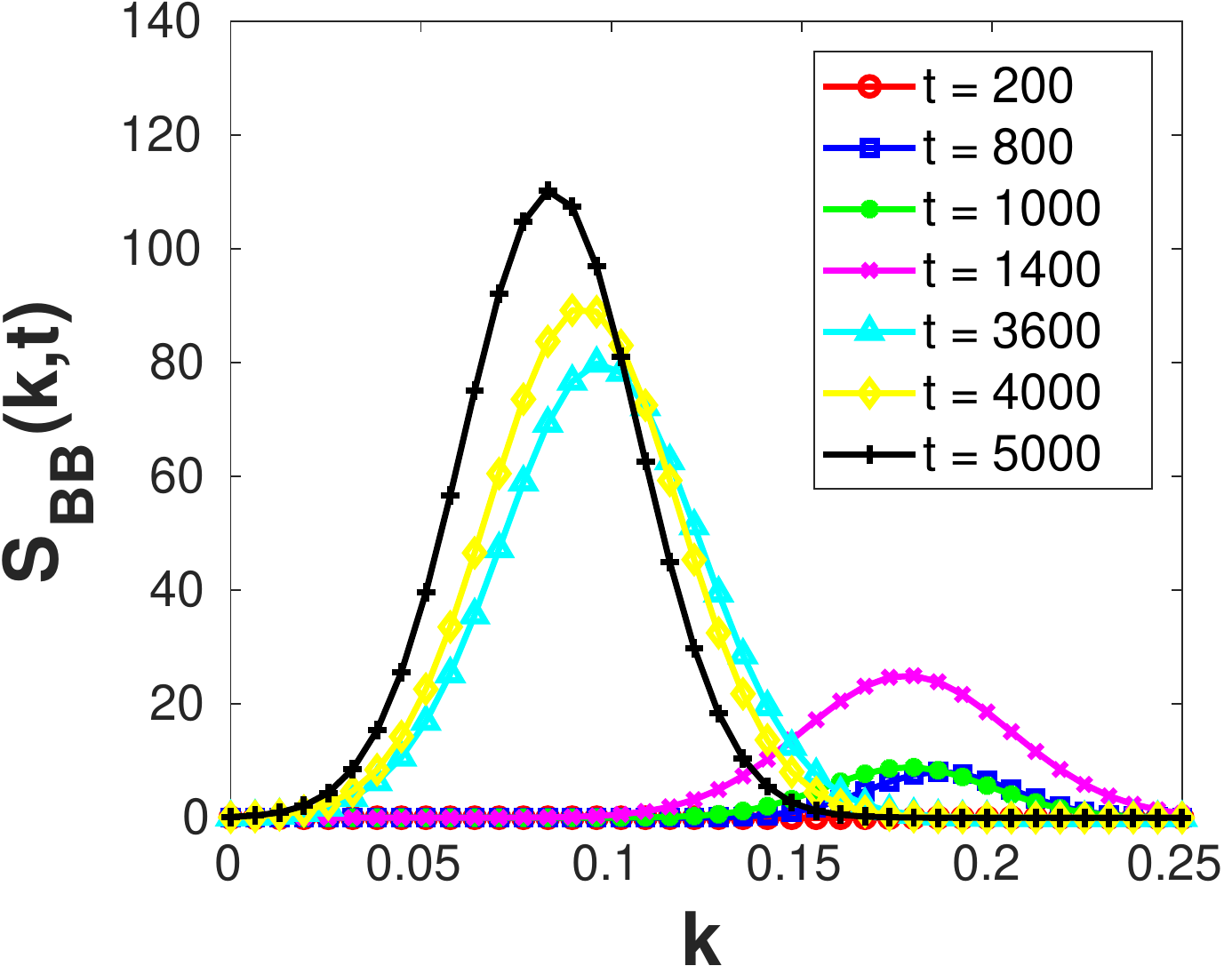}
  \caption{}
  \label{fig: P3_Sbb}
\end{subfigure} 
\begin{subfigure}{.32\textwidth}
  \centering
  \includegraphics[width=1.0\linewidth]{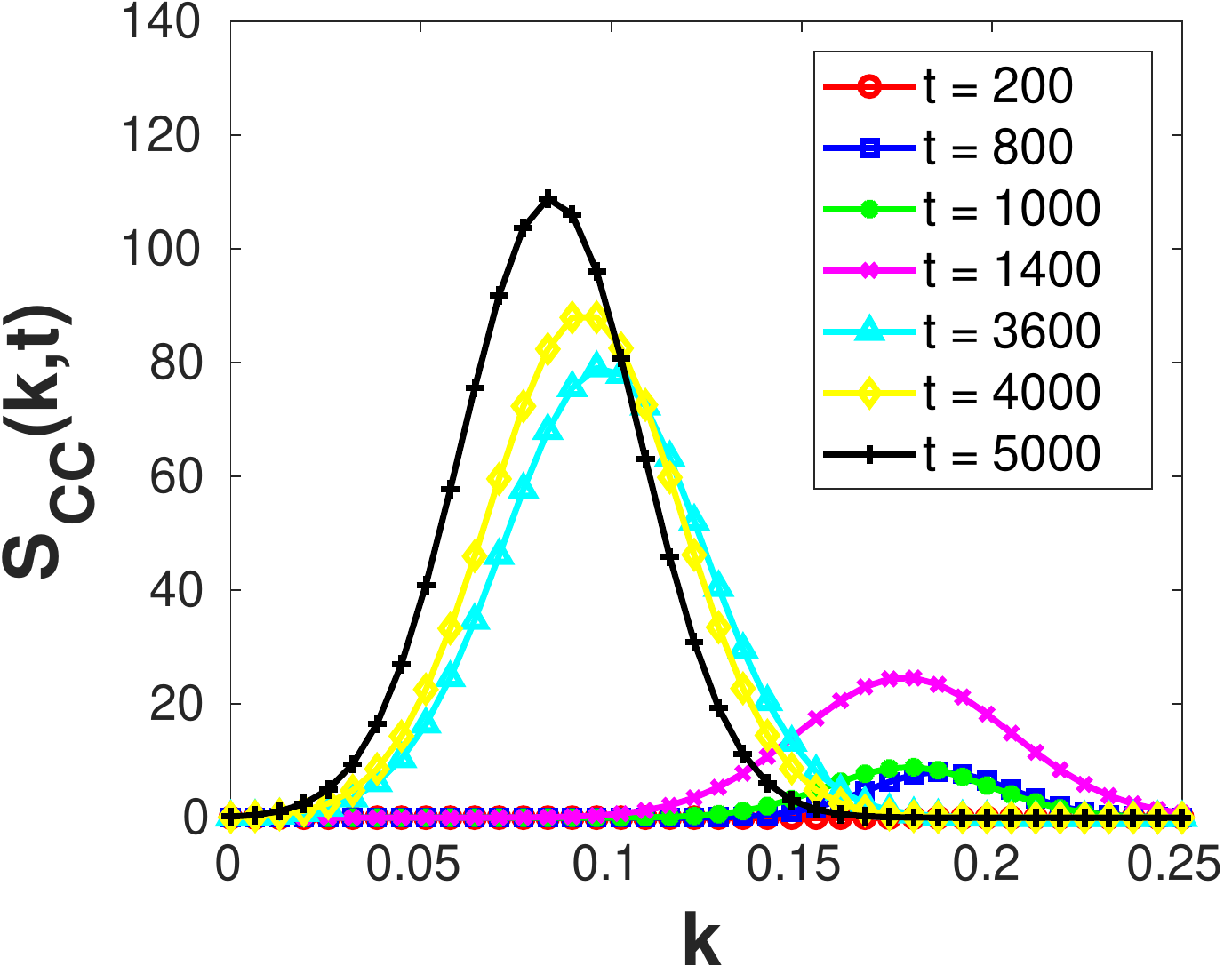}
  \caption{}
  \label{fig: P3_Scc}
\end{subfigure}
\caption{Structure functions for alloy systems (a-c)$P_1$, (d-f)$P_2$ and (g-i)$P_3$.}
\label{fig:P_Str}
\end{figure}
The structure function curves exhibit certain common characteristics 
for all alloy systems undergoing SD:
they are broad and shallow at the initial stages of SD, 
become narrower and sharper with increasing time, 
and the position of the peak $k_{max}$ shifts to lower values of $k$ at the later stages, 
indicating coarsening of domains. 
Moreover, the area under the curve denotes the extent of partitioning of species during SD. 
Note that $k_{max}$ can be used as a measure of the characteristic 
length scale of $A-$rich, $B$-rich and $C-$rich clusters. 
The evolution of structure functions $S_{ii}(k,t)\; i=A,B,C$ 
for alloy systems $P_1$, $P_2$, and $P_3$ are shown 
in Fig.~\ref{fig:P_Str}. 
For alloy system $P_1$, $A-$rich domains appear late ($t>1000$), 
as is evident from the evolution of $S_{AA}$ (Fig.~\ref{fig: P1_Saa}).
On the other hand, $S_{BB}$ and $S_{CC}$ show similar behavior with 
respect to the extent of decomposition and domain coarsening (Figs.~\ref{fig: P1_Sbb},~~\ref{fig: P1_Scc}). 
Comparison of areas under the $S_{AA}$, $S_{BB}$ and $S_{CC}$ curves
at $t=5000$ indicates that $A$, associated with larger misfit, partitions to 
a lesser extent.  

In alloy $P_2$, although the curves $S_{BB}$ and $S_{CC}$ 
develop peaks of low intensity during the evolution (Figs.~\ref{fig: P2_Sbb},~~\ref{fig: P2_Scc}), 
comparison of the areas enclosed by $S_{BB}$ and $S_{CC}$ 
with respect to that by $S_{AA}$ (Fig.~\ref{fig: P2_Saa}) at $t=5000$ points to a pseudo-binary phase 
separation producing $A$-rich and $A$-poor domains. The broad and shallow nature 
of $S_{BB}$ and $S_{CC}$ curves indicates negligibly small 
partitioning of $B$ and $C$ within the $A-$poor domains.  
 
In alloy system $P_3$, the peak in the $S_{AA}$ curve (Fig.~\ref{fig: P3_Saa}) appears earlier ($t=800$) 
than that corresponding to $P_2$. Thus, 
the decomposition of $P_3$ into $A-$rich and $A-$poor regions produce finer domains
than in $P_2$. Moreover, at $t=3600$, a noticeable increase in peak values of 
$S_{BB}$ and $S_{CC}$ indicates secondary SD of $A-$poor region producing $B-$rich and $C-$rich 
domains (Figs.~\ref{fig: P3_Sbb},~~\ref{fig: P3_Scc}).  
The behavior of the structure functions confirm our earlier 
findings on morphological evolution and compositional history for these three systems.  

\subsection{Effect of misfit strain on solute partitioning}
\label{effect:partitioning}

We demonstrate the effect of degree of misfit on solute partitioning using an $A-$rich asymmetric 
alloy $S$, where $B$ is a minority constituent ($c_A=0.6,c_B=0.1,c_C=0.3$). The alloy $S$
is categorized on the basis of misfit strains $\epsilon_{\alpha\beta}$ and $\epsilon_{\alpha\gamma}$
as given below:
\begin{itemize}
	\item
	$S_1:\epsilon_{\alpha\beta}=\epsilon_{\alpha\gamma}=0.01$
	\item
	$S_2:\epsilon_{\alpha\beta}=-\epsilon_{\alpha\gamma}=0.01$
	\item
	$S_3:\epsilon_{\alpha\beta}=0.014,\epsilon_{\alpha\gamma}=0.0$		
\end{itemize}
\begin{figure}[H]
  \centering
\begin{subfigure}{.3\textwidth}
  \centering
  \includegraphics[width=0.9\linewidth]{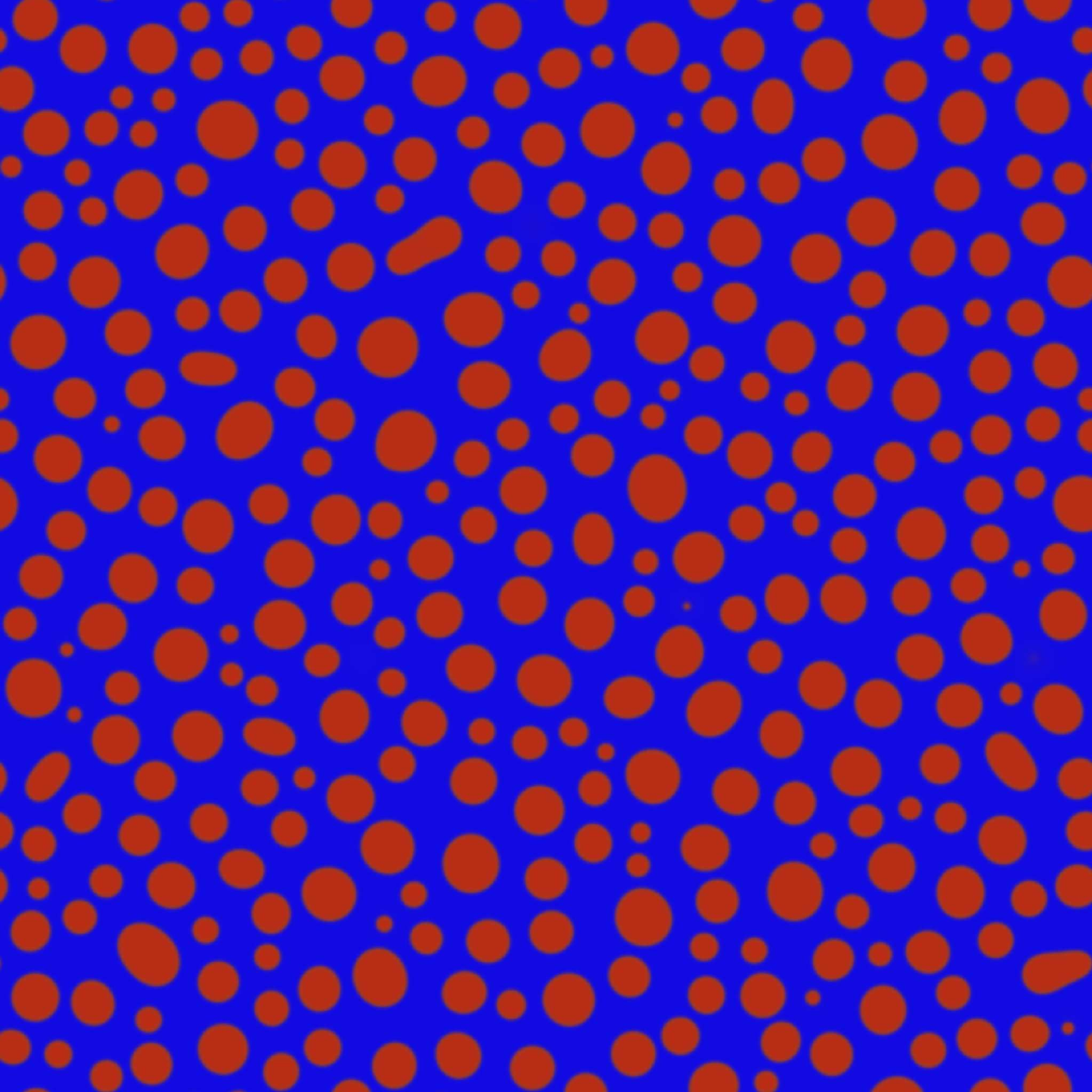}
  \caption{$S_1$}
  \label{S1_1}
\end{subfigure}
\begin{subfigure}{.3\textwidth}
  \centering
  \includegraphics[width=0.9\linewidth]{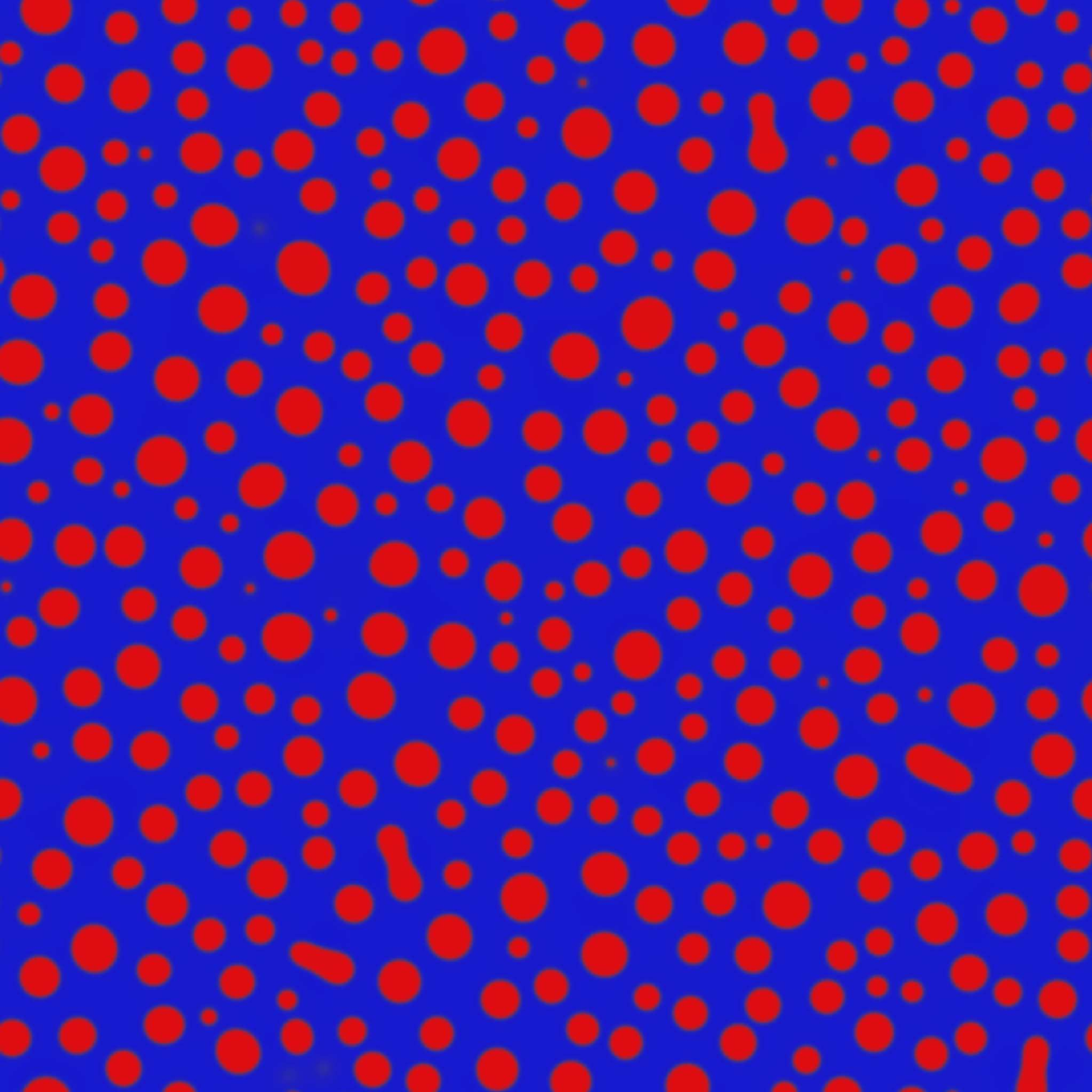}
  \caption{$S_2$}
  \label{S2_1}
\end{subfigure}
\begin{subfigure}{.3\textwidth}
  \centering
  \includegraphics[width=0.9\linewidth]{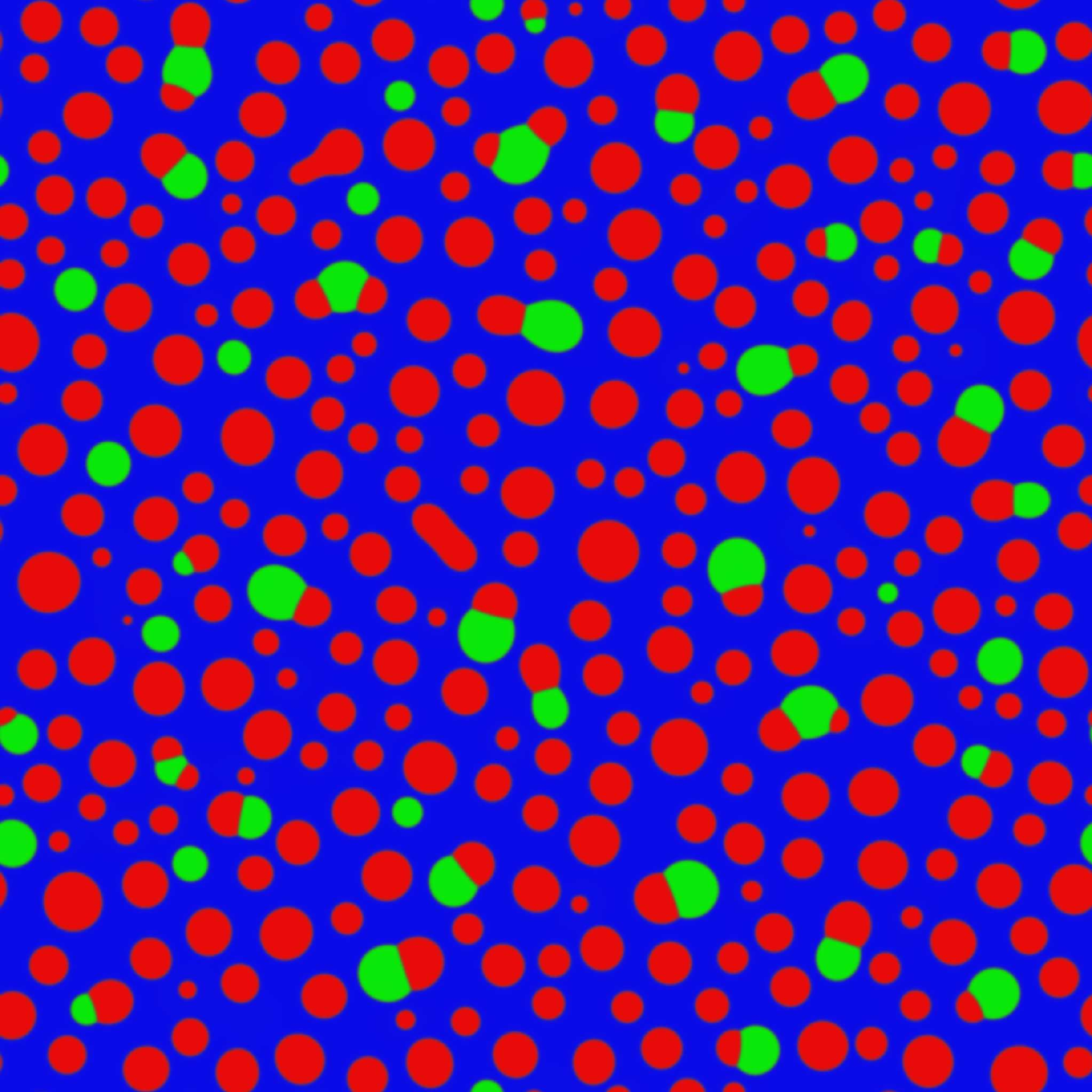}
  \caption{$S_3$}
  \label{S3_1}
\end{subfigure}\\
\caption{Time snapshots of microstructures of $A-$rich asymmetric alloy $S$ at $t=30000$. $S_1$, $S_2$ and $S_3$
correspond to misfit strains $\epsilon_{\alpha\beta}=\epsilon_{\alpha\gamma}$,
$\epsilon_{\alpha\beta}=-\epsilon_{\alpha\gamma}$, and $|\epsilon_{\alpha\beta}|\neq|\epsilon_{\alpha\gamma}|$,
respectively.}
\label{fig:S123}
\end{figure}
Figs.~\ref{S1_1},~\ref{S2_1}, and~\ref{S3_1} show the time snapshots of microstructures  
of alloys $S_1$, $S_2$, and $S_3$, respectively.
In these alloy systems, SD produces $C-$rich $\gamma$ domains in $\alpha$ matrix.
In alloy system $S_1$, $\alpha$ has misfit of same magnitude and sign (positive) with both $\beta$
and $\gamma$. Therefore, $\alpha$ has a larger lattice parameter than both $\beta$ and $\gamma$. 
Also, there is no misfit between $\beta$ and $\gamma$ ($\epsilon_{\beta\gamma}=0$). 
Large misfit between $\alpha$ matrix and $C-$rich particles impedes partitioning of $A$ at the early
stages of SD. Therefore, the decomposition into $A-$rich and $A-$poor regions 
require a mechanism to lower the coherency strain energy between them. Reduction in coherency strain 
involves partitioning of minority component $B$ into $\alpha$ matrix. Such selective partitioning of $B$
leads to the formation of $\gamma$ domains in $\alpha$ matrix. 

The decomposition in alloy $S_2$, yielding $\alpha$ and $\gamma$ phases, begins earlier than in $S_1$.
Therefore, the microstructure of $S_2$ contains finer $\gamma$ particles in $\alpha$
matrix (Fig.~\ref{S2_1}). In this system, the lattice parameter of $\alpha$ lies between those 
of $\beta$ and $\gamma$ ($\epsilon_{\alpha\beta}=-\epsilon_{\alpha\gamma}$). Preferential 
partitioning of $B$ to $C-$rich regions significantly reduces the effective misfit strain
between $\alpha$ and $\gamma$.

In system $S_3$, $\beta$ has the same misfit with
$\alpha$ and $\gamma$ ($\epsilon_{\alpha\beta}=\epsilon_{\beta\gamma}=0.014$).
Therefore, in this case, the interface between $\alpha$ and $\gamma$ gets 
enriched in $B$. At the intermediate stages, further decomposition
leads to the formation of $\beta$ particles at the $\alpha$-$\gamma$ boundaries (Fig.~\ref{S3_1}). 

\begin{figure}[H]
  \centering
  \begin{subfigure}{.3\textwidth}
  \raggedleft
\begin{overpic}[width=1.0\textwidth]{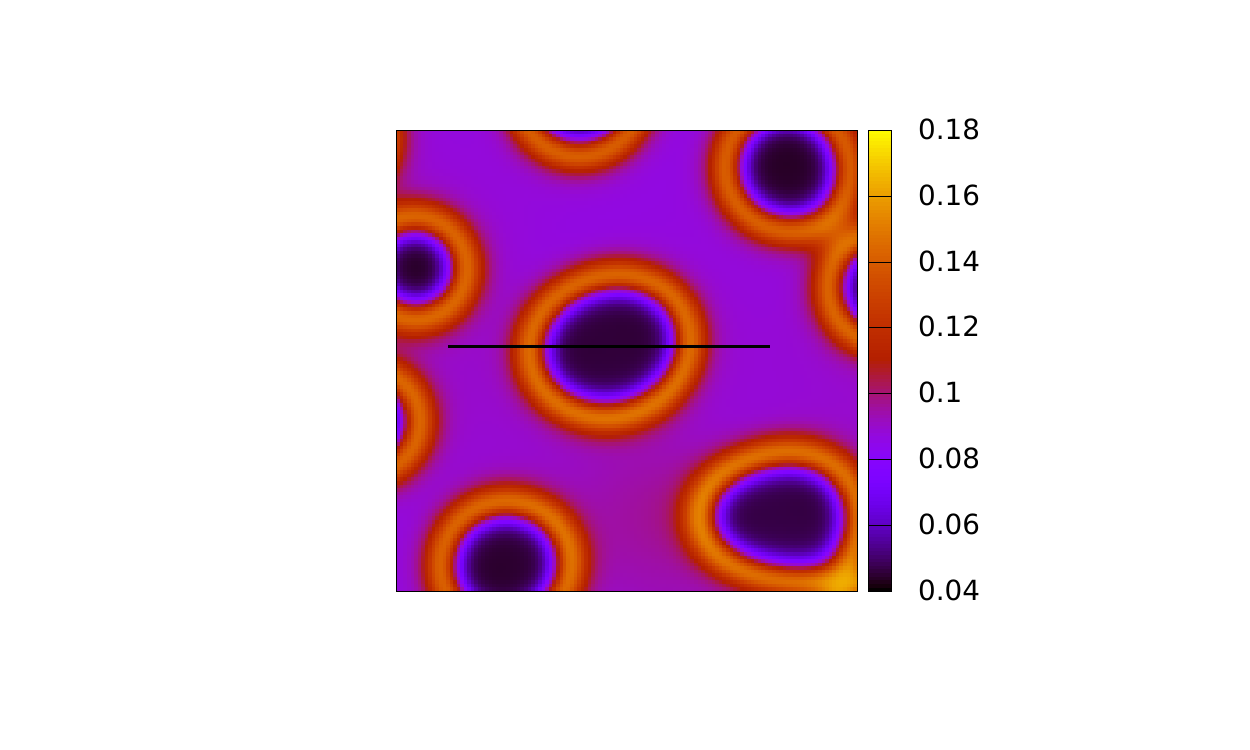}
 \put (68,7) {$c_B$}
\end{overpic}  
\caption{$S_1$}
  \label{S1_conc}
  \end{subfigure} 
\begin{subfigure}{.3\textwidth}
  \raggedleft  
\begin{overpic}[width=1.0\textwidth]{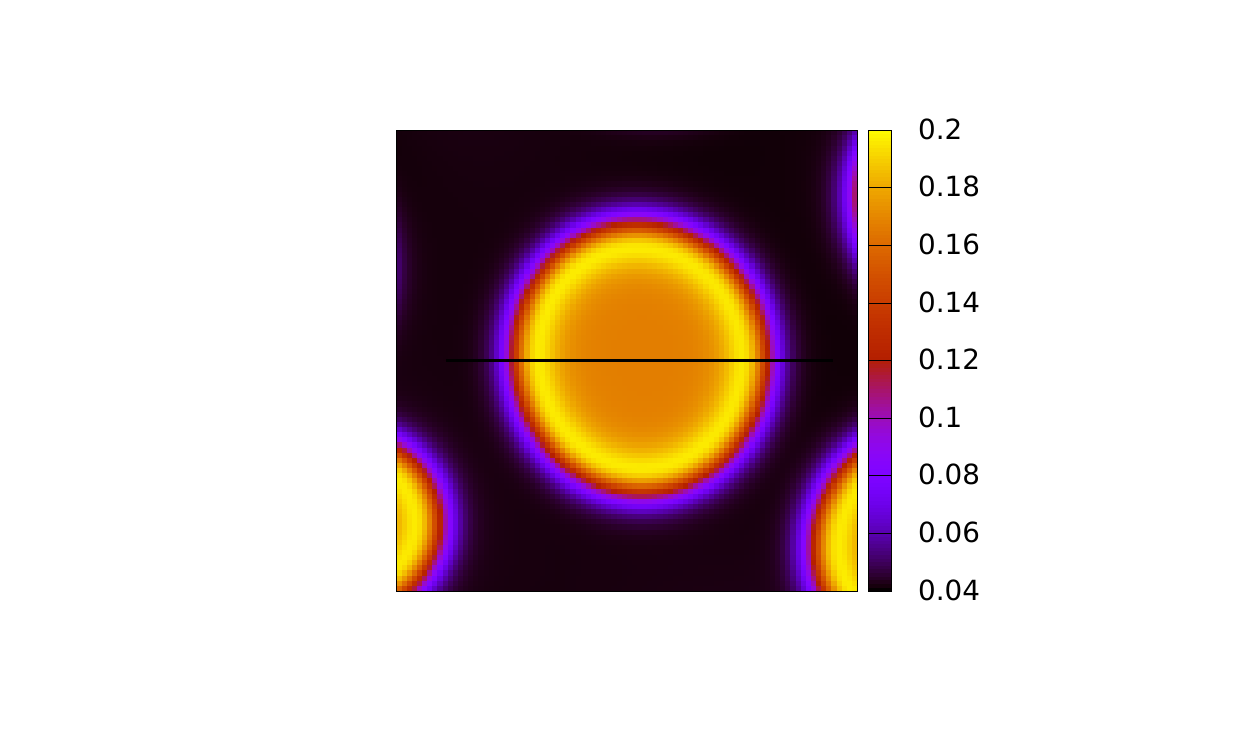}
 \put (68,7) {$c_B$}
\end{overpic} 
  \caption{$S_2$}
  \label{S2_conc}
\end{subfigure}%
\begin{subfigure}{.3\textwidth}
    \raggedleft
\begin{overpic}[width=1.0\textwidth]{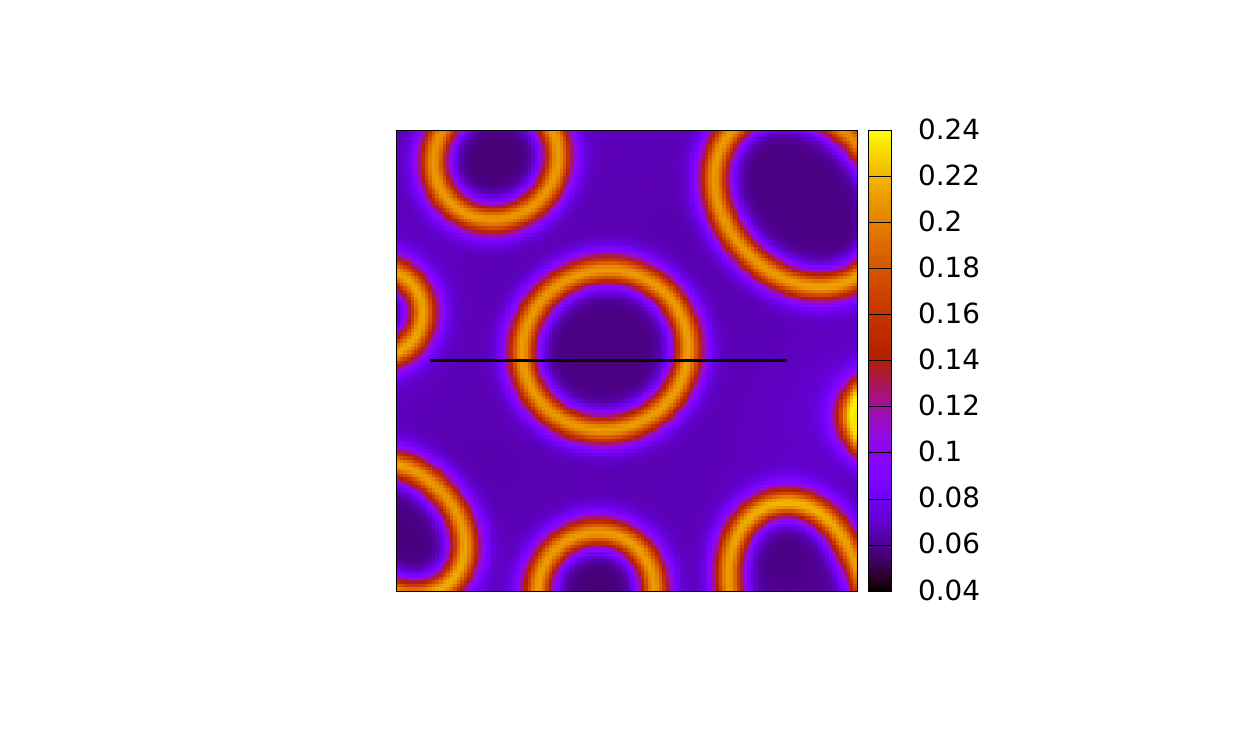}
 \put (68,7) {$c_B$}
\end{overpic}  
\caption{$S_3$}
  \label{S3_conc}
  \end{subfigure} \\
  \begin{subfigure}{.3\textwidth}
  \centering
\begin{overpic}[width=1.0\textwidth]{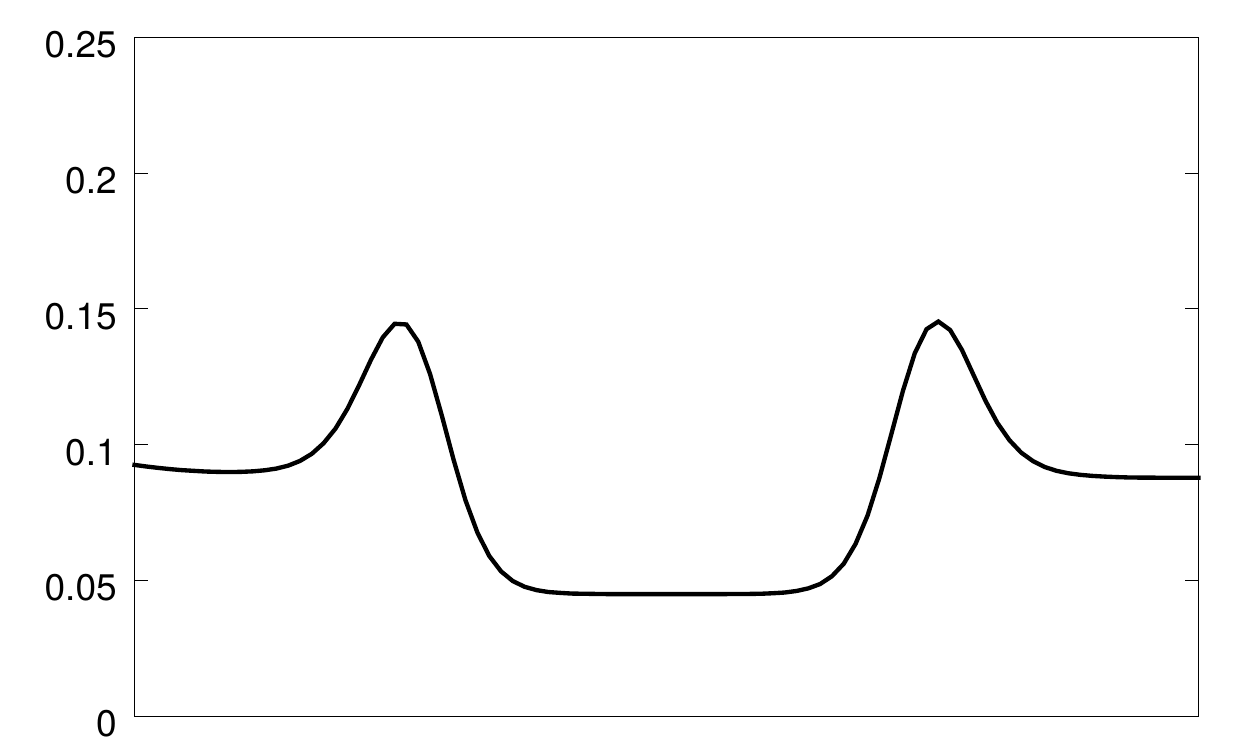}
 \put (-4,28) {\rotatebox{90}{$c_B$}}
\end{overpic}  
  \caption{}
  \label{S1_cprof}
  \end{subfigure} 
\begin{subfigure}{.3\textwidth}
  \centering
\begin{overpic}[width=1.0\textwidth]{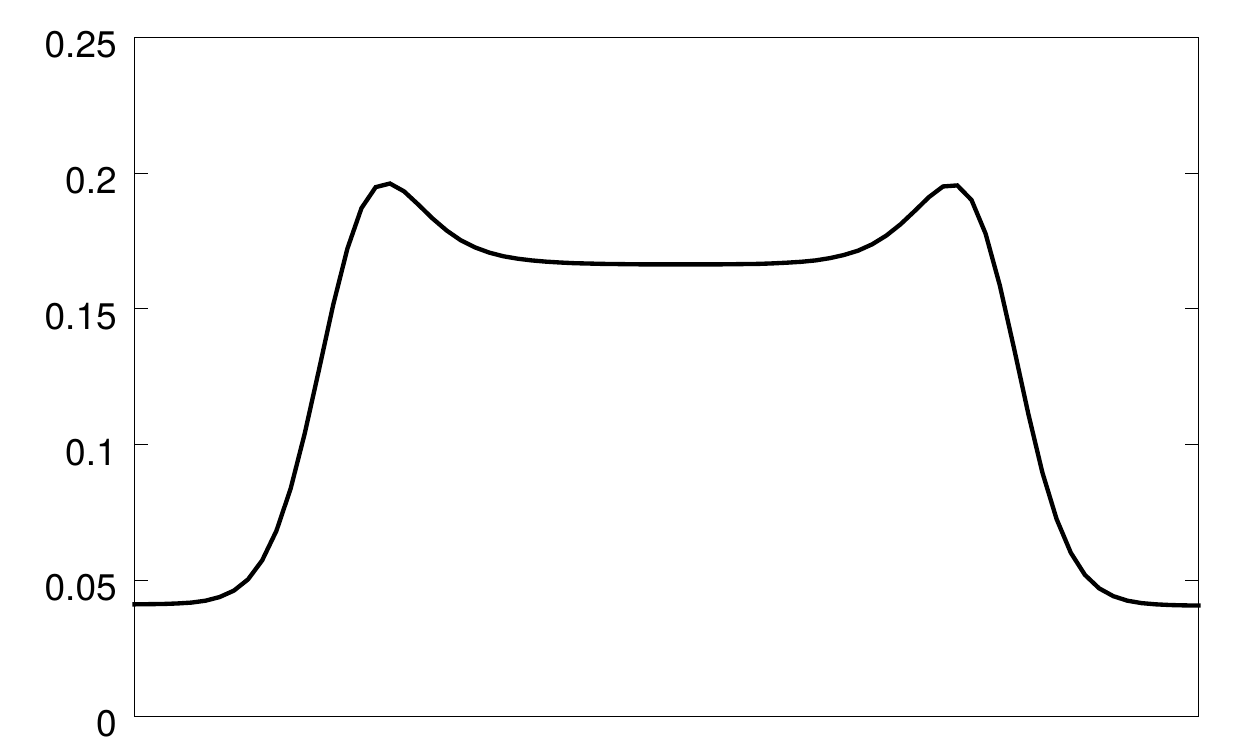}
 \put (-4,28) {\rotatebox{90}{$c_B$}}
\end{overpic} 
  \caption{}
  \label{S2_cprof}
\end{subfigure} 
\begin{subfigure}{.3\textwidth}
  \centering
\begin{overpic}[width=1.0\textwidth]{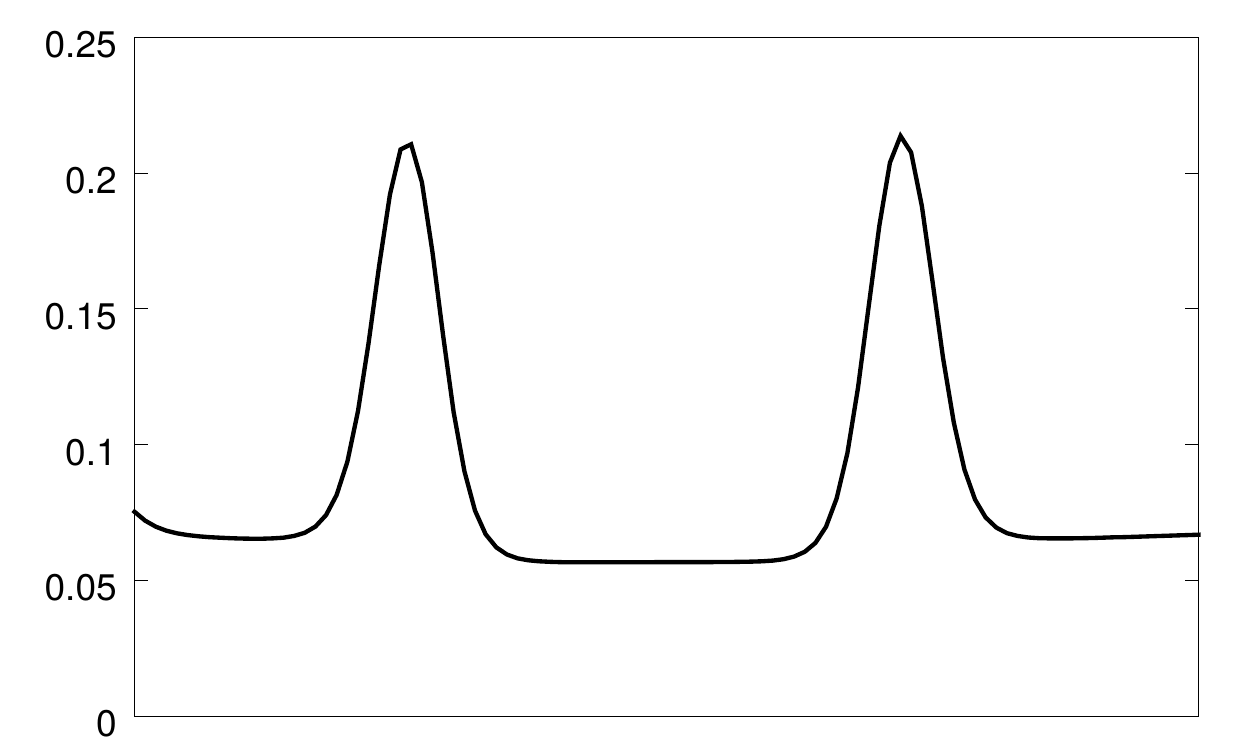}
 \put (-4,28) {\rotatebox{90}{$c_B$}}
\end{overpic}  
  \caption{}
  \label{S3_cprof}
  \end{subfigure}%
\caption{(a,b,c) Two-dimensional concentration maps of minority species $B$   
across a $\gamma$ particle in alloy systems $S_1$, $S_2$ and $S_3$. (d,e,f) 
Corresponding one-dimensional profiles drawn
across the interface illustrating the change in $B$ distribution in these systems.}
\label{fig:S_cprof}
\end{figure}

The composition profiles of $B$ across $\alpha$-$\gamma$ interfaces
in $S_1$, $S_2$ and $S_3$ alloy systems demonstrate the role of
degree and sign of misfit on solute partitioning (Fig.~\ref{fig:S_cprof}).
The relative magnitude of lattice expansion coefficients associated with B decides
the partitioning of $B$ to either $\alpha$ or $\gamma$. 
When the degree of misfit of $B$ with $\alpha$ is higher than that with $\gamma$ 
($|\epsilon_{\alpha\beta}|>|\epsilon_{\beta\gamma}|$), $B$ partitions to $\alpha$ phase ($S_1$). 
On the other hand, $B$ partitions to $\gamma$ particle 
when we reverse the condition on the misfit associated with $B$ ($S_2$). However, 
when $B$ has the same degree of misfit relative to $\alpha$ and $\gamma$, 
it segregates to the $\alpha\gamma$ boundary ($S_3$).
In all cases, strain energy minimization leads to preferential partitioning
of $B$.

The partition coefficient of $B$ ($K_p$), defined as the ratio of
concentration of $B$ in $\gamma$ particle to its concentration in the matrix, are
estimated for these alloy systems. $K_p$ for $S_1$, $S_2$ and $S_3$
are $0.51$, $4.09$ and $0.86$, respectively. A partition coefficient value close to unity indicates
preferential segregation of solute at the interface.    

\subsection{Effect of relative interfacial energies ($\Gamma_{\alpha\beta},\Gamma_{\alpha\gamma},\Gamma_{\beta\gamma}$)}
\label{effect:intfacenergy}

Additionally, we vary the interfacial energies between the phases to understand the interactions 
between interfacial energy and coherency strain energy on morphological development during ternary SD.
To this effect, we introduce a new system by modifying the bulk free energy coefficients 
($\chi_{AB}=3.5,\chi_{AC}=5,\chi_{BC}=2.5$) and gradient energy coefficients 
($\kappa_A=6,\kappa_B=2,\kappa_C=6$) such that the relative interfacial energies between the coexisting 
phases satisfy Cahn's spinodal wetting condition~\cite{cahn1977critical}: 
($\Gamma_{\alpha\gamma}\geq\Gamma_{\alpha\beta}+\Gamma_{\beta\gamma}$).

\begin{figure}[H]
\begin{subfigure}{.5\textwidth}
  \centering
  \includegraphics[width=0.55\linewidth]{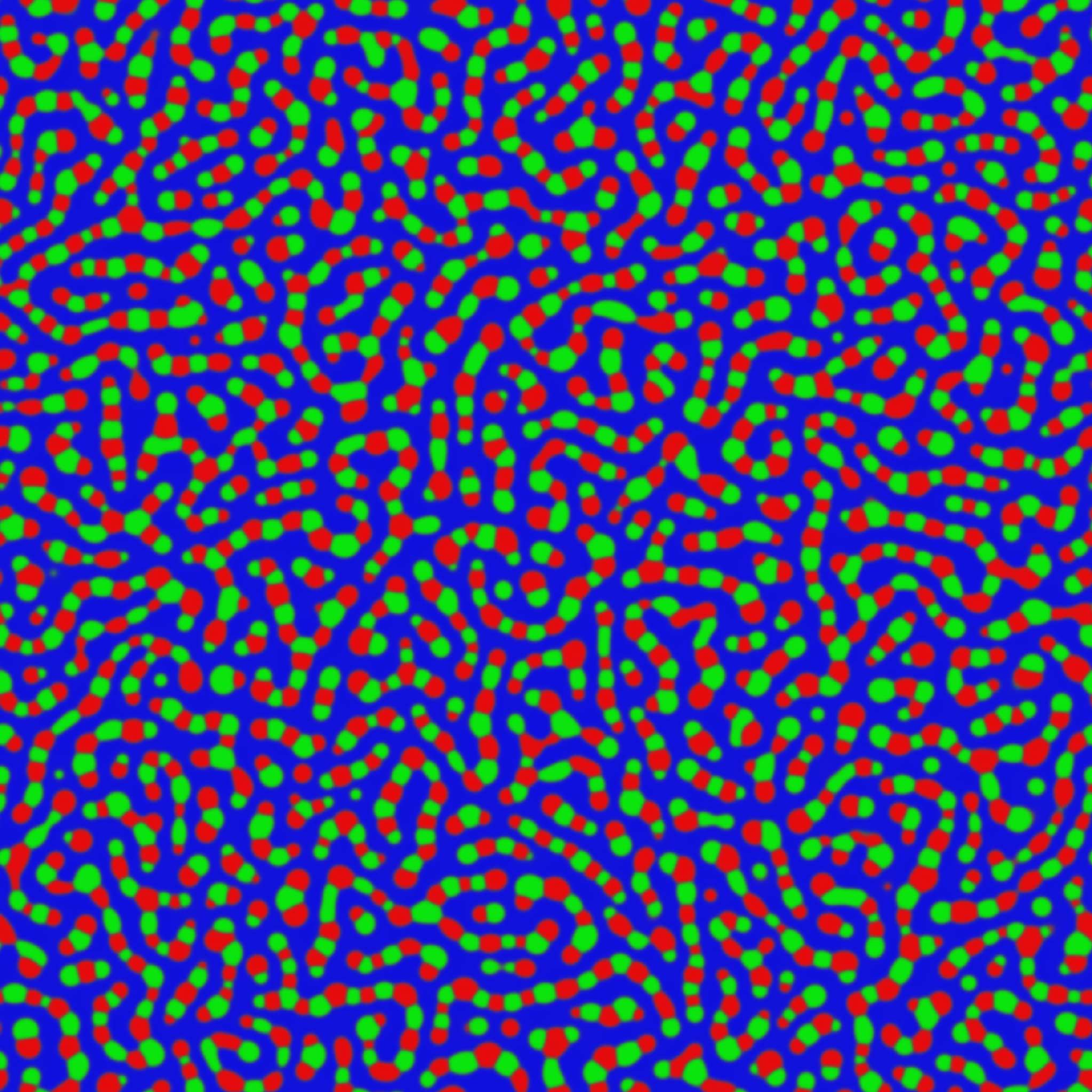}
  \caption{$Q_1$}
  \label{Q1_I}
\end{subfigure}%
\begin{subfigure}{.5\textwidth}
  \centering
  \includegraphics[width=0.55\linewidth]{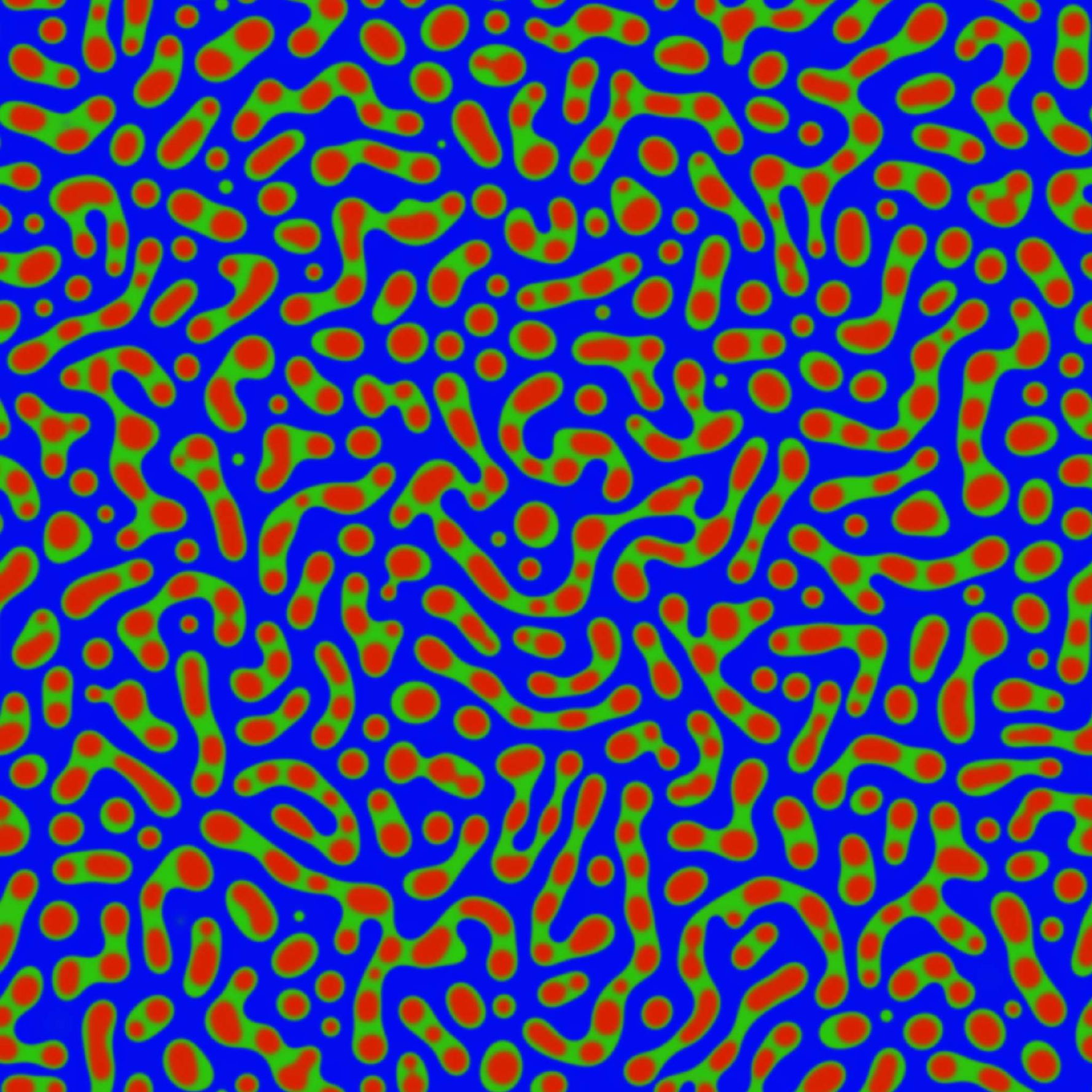}
  \caption{$Q_1^\prime$}
  \label{Q1_II}
\end{subfigure} \\
\caption{Comparison of microstructures of alloy systems $Q_1$ and $Q_1^\prime$ at $t=5000$. Interfacial energies
associated with $Q_1$ and $Q_1^\prime$ satisfy the conditions
$\Gamma_{\alpha\beta}=\Gamma_{\alpha\gamma}=\Gamma_{\beta\gamma}$ and 
$\Gamma_{\alpha\gamma}\geq\Gamma_{\alpha\beta}+\Gamma_{\beta\gamma}$, respectively. Both have same set of misfit 
strains ($\epsilon_{\alpha\beta}=\epsilon_{\alpha\gamma}=0.01$).}
\label{Q1_comp}
\end{figure}

We compare the microstructure of alloy $Q_1$ with another alloy $Q_1^\prime$ belonging to the new system where
interfacial energies obey perfect wetting condition (Fig.~\ref{Q1_comp}). We choose the alloy composition 
($c_A=0.5,c_B=0.25,c_C=0.25$) and the set of misfit strains ($\epsilon_{\alpha\beta}=\epsilon_{\alpha\gamma}=0.01$) 
of $Q_1^\prime$ to be the same as those of $Q_1$.   
Microstructure corresponding to $Q_1$ contains chains comprised of alternating beads
of $\beta$ and $\gamma$ embedded in the $\alpha$ matrix. On the other hand, microstructure of 
$Q_1^\prime$ shows discrete $\gamma$ particles completely wetted by $\beta$ films dispersed in
$\alpha$ matrix. In both cases, the misfit between $\beta$ and $\gamma$ is zero. This 
example clearly shows how the interaction between elastic and interfacial energies affects the 
morphology.

\subsection{Phase separation in alloys with compositions in the regions of absolute stability (Region III)}
\label{effect:abs_stability}

Further, we use this model to study phase separation in 
alloy systems whose compositions lie in the positive definite region (Region III).  
In this case, phase separation happens via nucleation and growth. 
Such alloy compositions can be used to study the development 
of compact core-shell morphology.
However, the stability of the compact morphology will 
depend on the relative interfacial energies 
and the sign and degree of misfit between the phases. 

We illustrate the development of core-shell morphology using an alloy 
$T$ ($c_B=0.10,c_C=0.05$) with perfect wetting condition imposed on interfacial 
energies between the coexisting phases. The alloy $T$ is categorized on the basis of 
misfit strains $\epsilon_{\alpha\beta}$ and $\epsilon_{\alpha\gamma}$ as given below:
\begin{itemize}
	\item
	$T_1^\prime:\epsilon_{\alpha\beta}=\epsilon_{\alpha\gamma}=0.01$
	\item
	$T_2^\prime:\epsilon_{\alpha\beta}=-\epsilon_{\alpha\gamma}=0.01$	
\end{itemize} 
\begin{figure}[H]
\begin{subfigure}{.5\textwidth}
  \centering
  \includegraphics[width=0.55\linewidth]{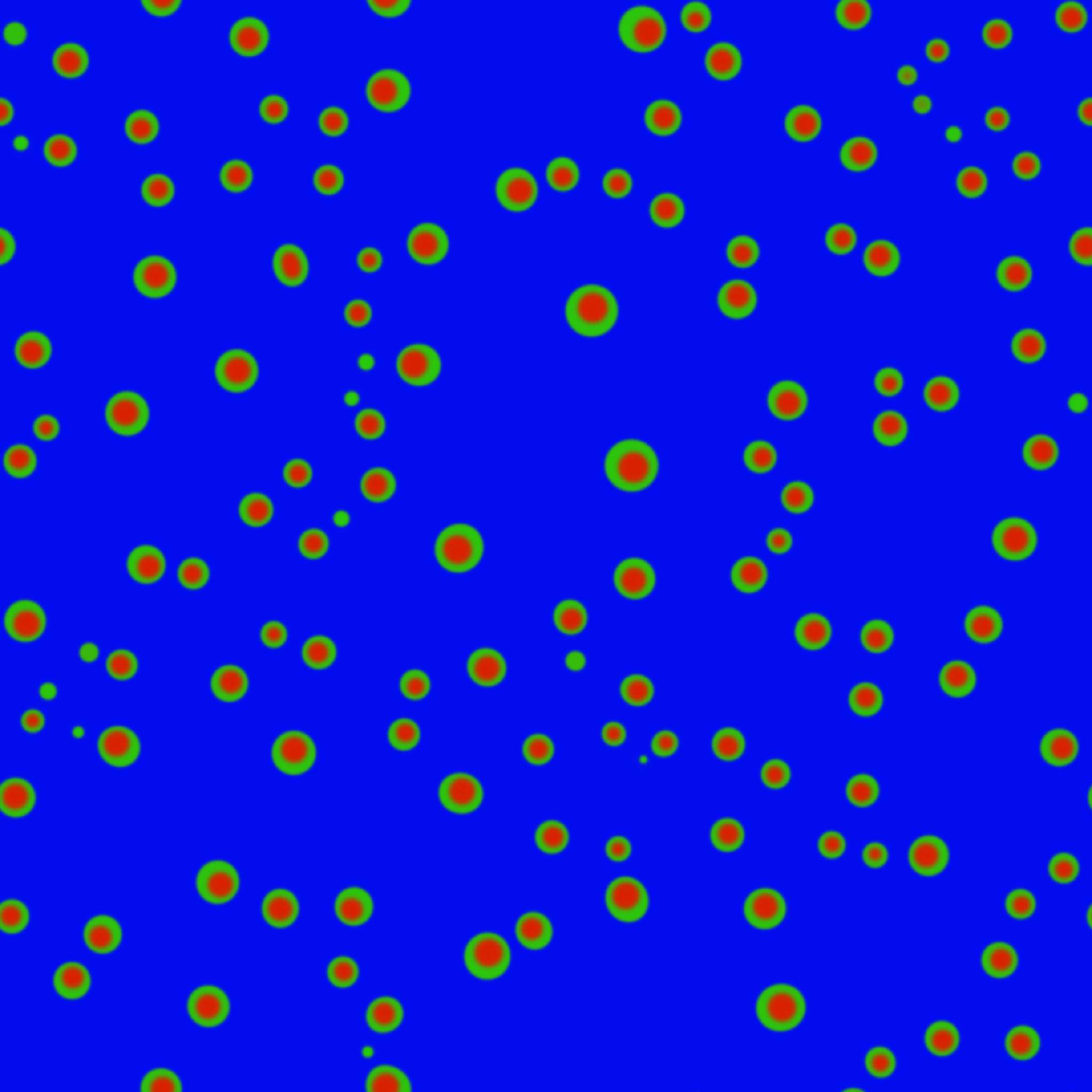}
  \caption{$T_1^\prime$}
  \label{Z1_4}
\end{subfigure}%
\begin{subfigure}{.5\textwidth}
  \centering
  \includegraphics[width=0.55\linewidth]{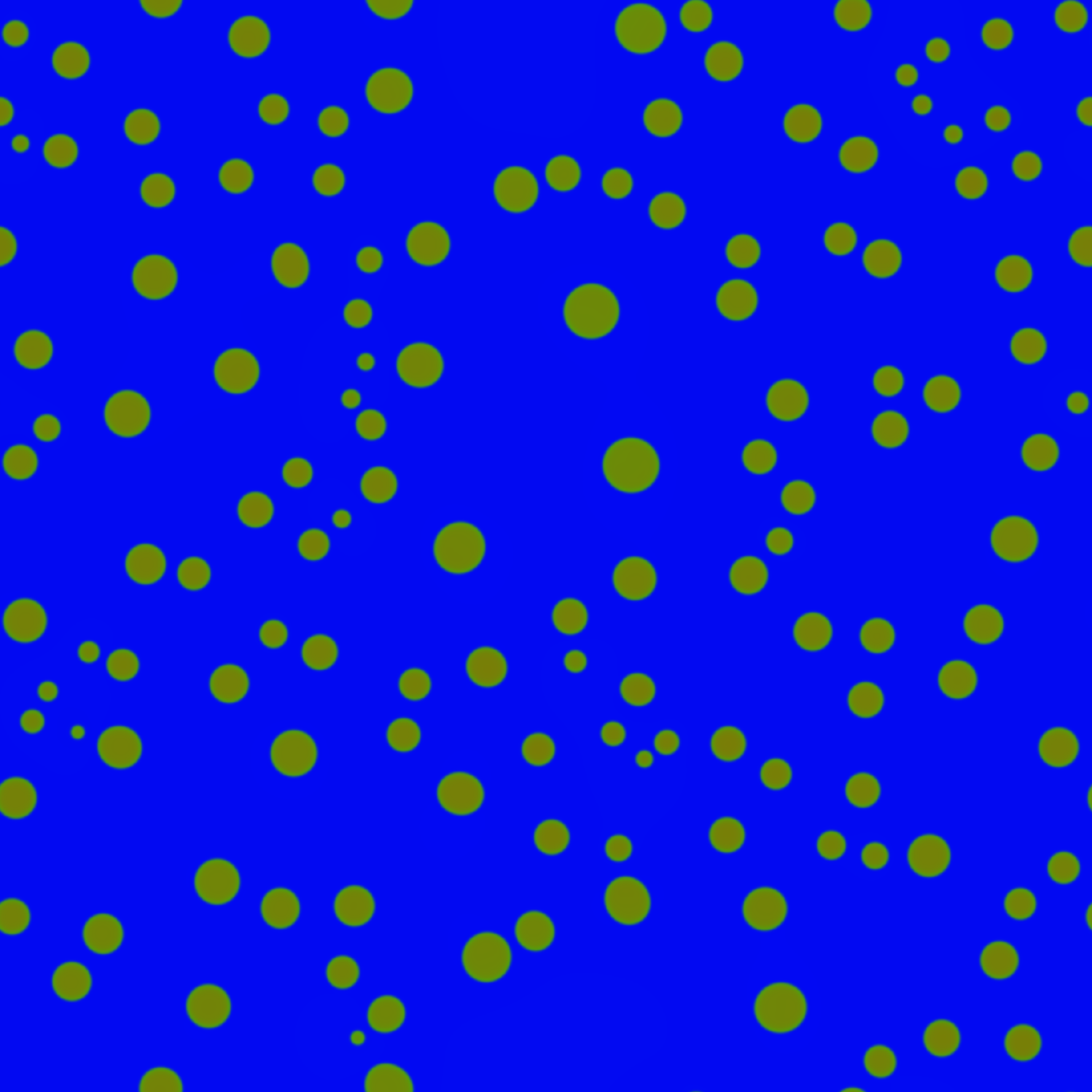}
  \caption{$T_2^\prime$}
  \label{Z2_4}
\end{subfigure} \\
\caption{Comparison of microstructures of alloy systems $T_1^\prime$ and $T_2^\prime$ at $t=6000$. The set of misfit
strains associated with $T_1^\prime$ and $T_2^\prime$ are $\epsilon_{\alpha\beta}=\epsilon_{\alpha\gamma}=0.01$
and $\epsilon_{\alpha\beta}=-\epsilon_{\alpha\gamma}=0.01$, respectively. The relative interfacial energies of
both systems satisfy spinodal wetting condition ($\Gamma_{\alpha\gamma}\geq\Gamma_{\alpha\beta}+\Gamma_{\beta\gamma}$).}
\label{Z1_comp}
\end{figure}
We incorporate nuclei with equilibrium 
$\gamma$ composition at random positions in a supersaturated matrix. 
We use classical nucleation theory to calculate the radius of the critical nucleus
of a given composition. 
Alloy system $T_1^\prime$ shows the development of stable 
core-shell microstructure at $t=6000$ 
($\gamma$ core enclosed by $\beta$ shell (Fig.~\ref{Z1_4})). 
Since $\epsilon_{\beta\gamma}=0$ in this case, 
the core-shell morphology remains stable till the late stages. In contrast,
$B-$rich shell grows inwards with the dissolution of core $\gamma$ 
in system $T_2^\prime$ (Fig.~\ref{Z2_4}). The destabilization of core-shell structure 
is attributed to large misfit strain between $\beta$
and $\gamma$ ($|\epsilon_{\beta\gamma}|=0.02$). This is yet another example 
describing the interplay between interfacial and elastic strain energies.   

\subsection{Coarsening kinetics of $\alpha$, $\beta$, and $\gamma$ domains in equiatomic $P$ alloys}
\label{coarsening}

In order to determine the dynamics of coarsening of multiple domains in  equiatomic alloys $P_1$, $P_2$ and $P_3$,
we need to ascertain the existence of a self-similar regime
where the three-phase microstructure can be dynamically scaled using 
a characteristic time-dependent length for each phase~\cite{sun2018self}.  
Thus, to study the dynamics of coarsening of $A-$rich, $B-$rich and $C-$rich domains,
we determine a characteristic length scale of the microstructure for each domain 
and record its temporal evolution. 
The characteristic length represents the time-dependent size 
of a domain.    

We can use different measures to define the characteristic length scale.
Here we use inverse of the first moment of $S_{ii}(\textbf{k},t)$ 
as the characteristic length~\cite{bhattacharyathesis}:  
\begin{equation}
R_i(t)=1/k_1^{(i)}(t)=\frac{\sum S_{ii}(\textbf{k},t)}{\sum k S_{ii}(\textbf{k},t)},
\end{equation}
where $i = A,B,C$. 

\begin{figure}[H]
  \centering
\begin{subfigure}{.32\textwidth}
  \centering
  \includegraphics[width=1.05\linewidth]{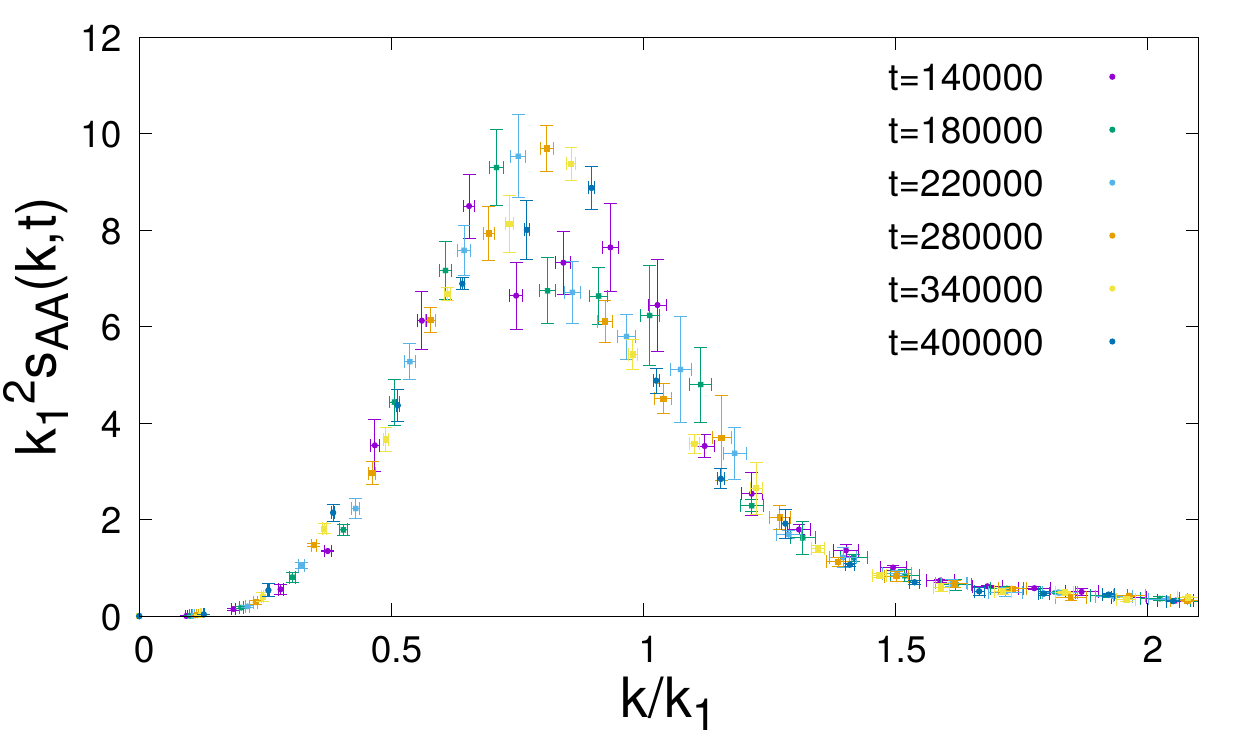}
  \caption{}
  \label{fig: P1_SFaa}
\end{subfigure} 
\begin{subfigure}{.32\textwidth}
  \centering
  \includegraphics[width=1.05\linewidth]{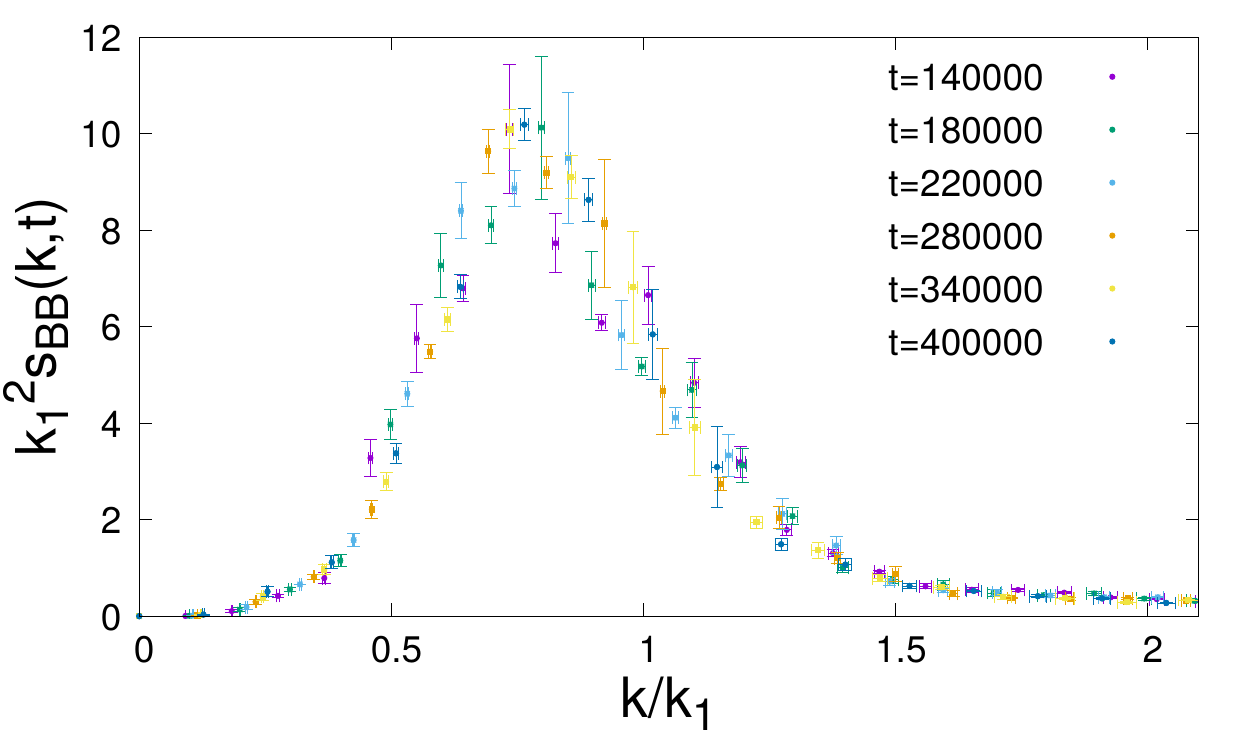}
  \caption{}
  \label{fig: P1_SFbb}
\end{subfigure} 
\begin{subfigure}{.32\textwidth}
  \centering
  \includegraphics[width=1.05\linewidth]{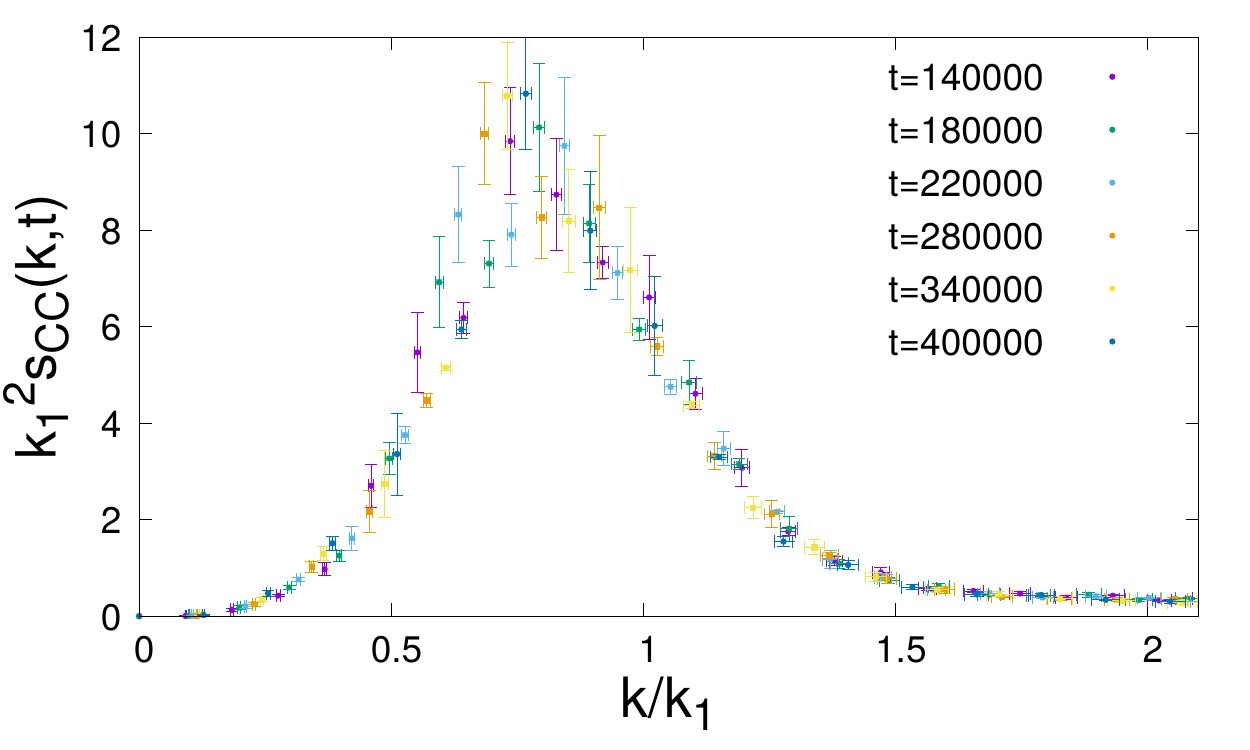}
  \caption{}
  \label{fig: P1_SFcc}
\end{subfigure} \\
\begin{subfigure}{.32\textwidth}
  \centering
  \includegraphics[width=1.05\linewidth]{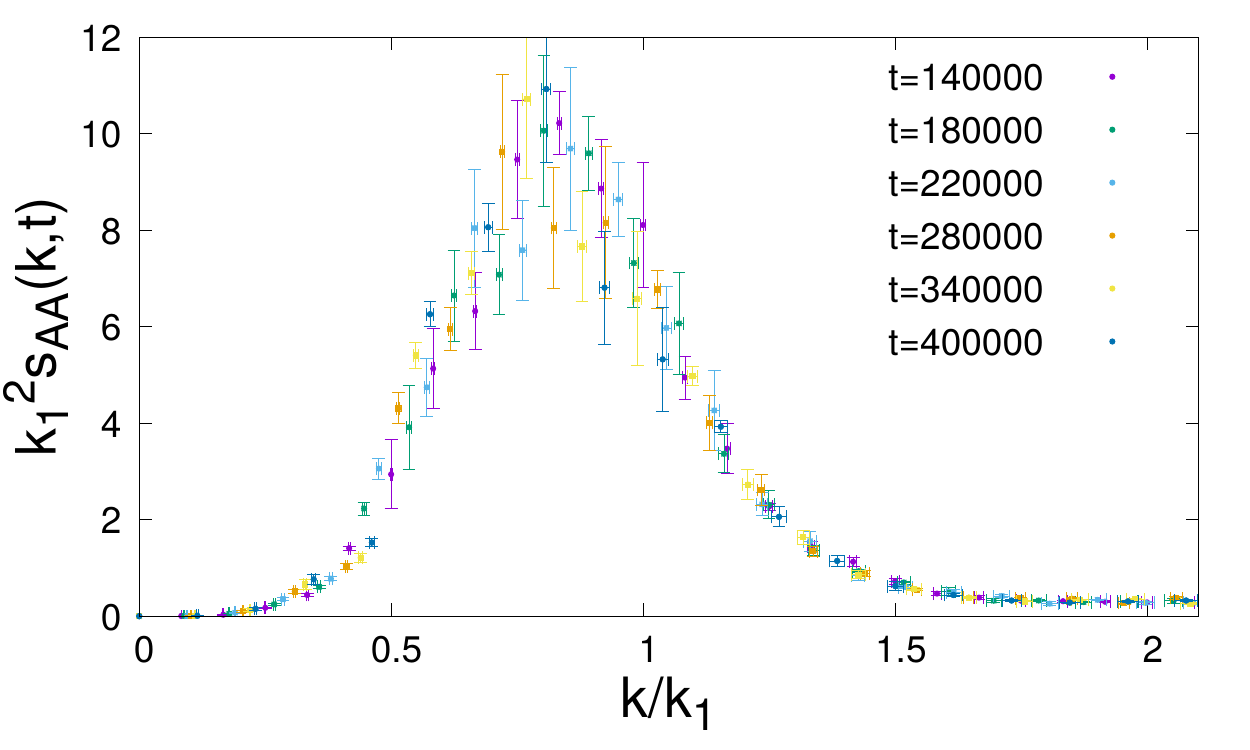}
  \caption{}
  \label{fig: P2_SFaa}
\end{subfigure} 
\begin{subfigure}{.32\textwidth}
  \centering
  \includegraphics[width=1.05\linewidth]{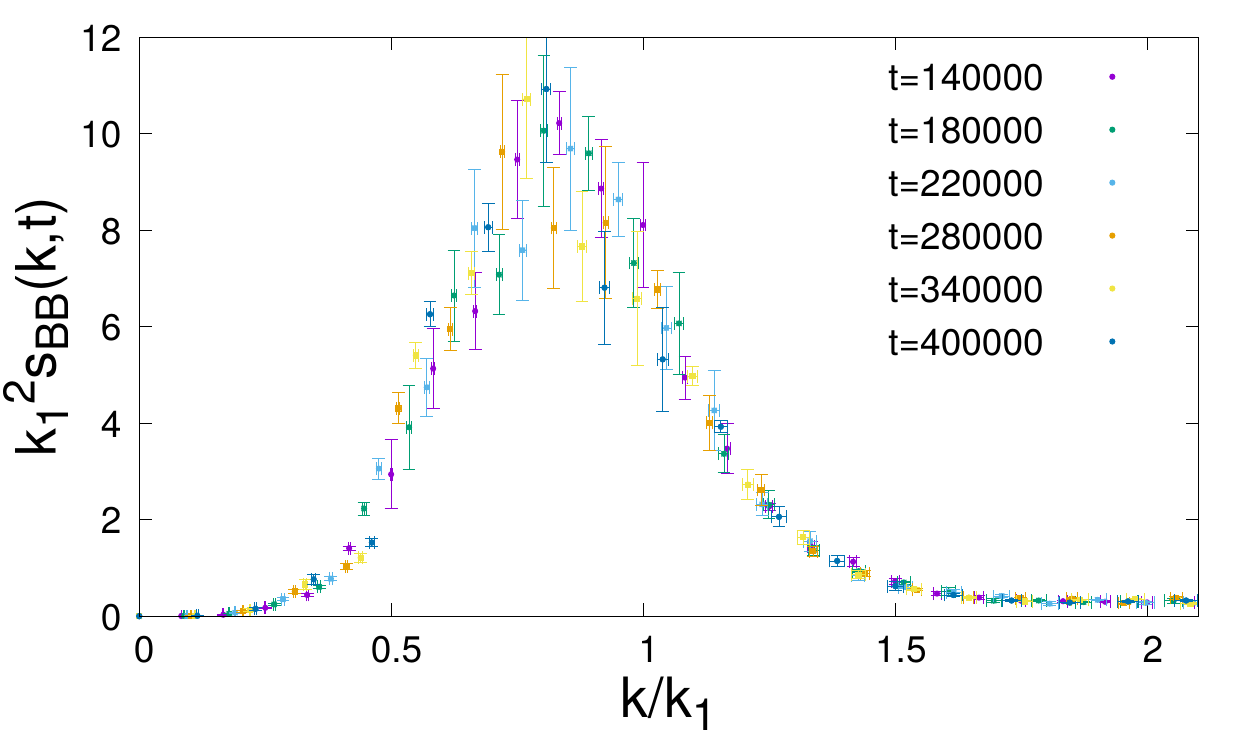}
  \caption{}
  \label{fig: P2_SFbb}
\end{subfigure} 
\begin{subfigure}{.32\textwidth}
  \centering
  \includegraphics[width=1.05\linewidth]{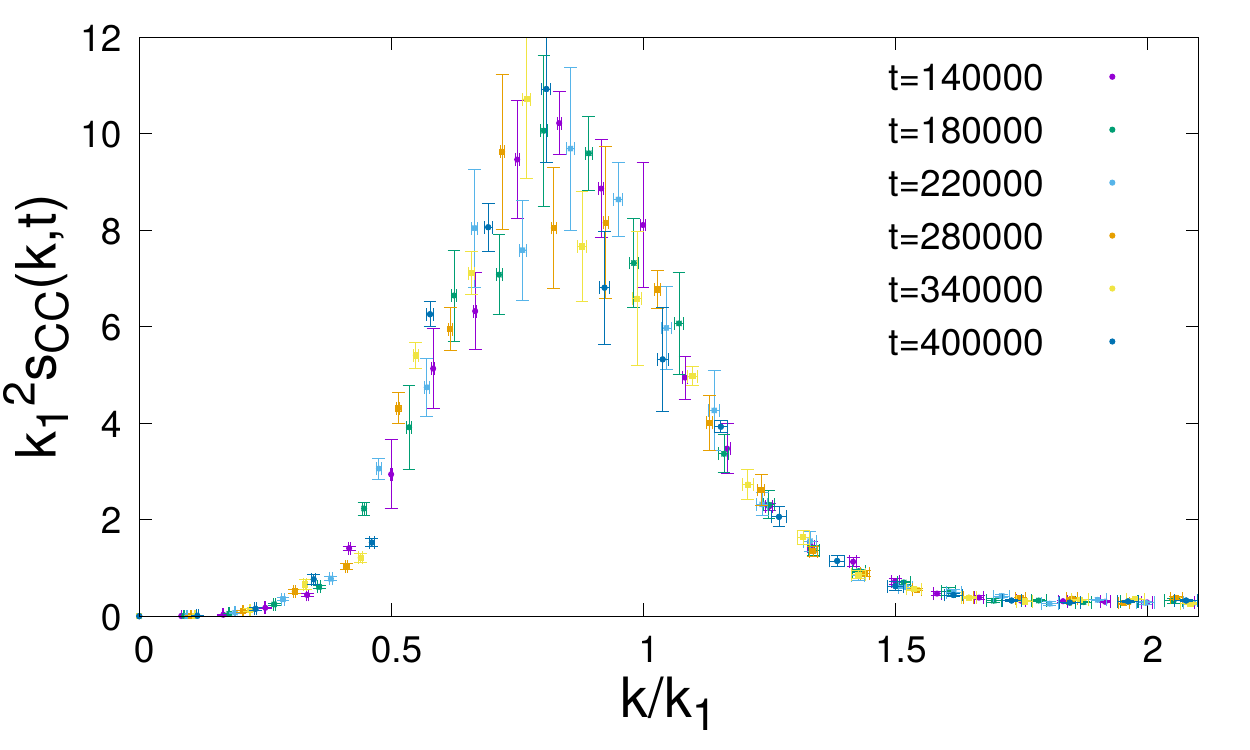}
  \caption{}
  \label{fig: P2_SFcc}
\end{subfigure} \\
\begin{subfigure}{.32\textwidth}
  \centering
  \includegraphics[width=1.05\linewidth]{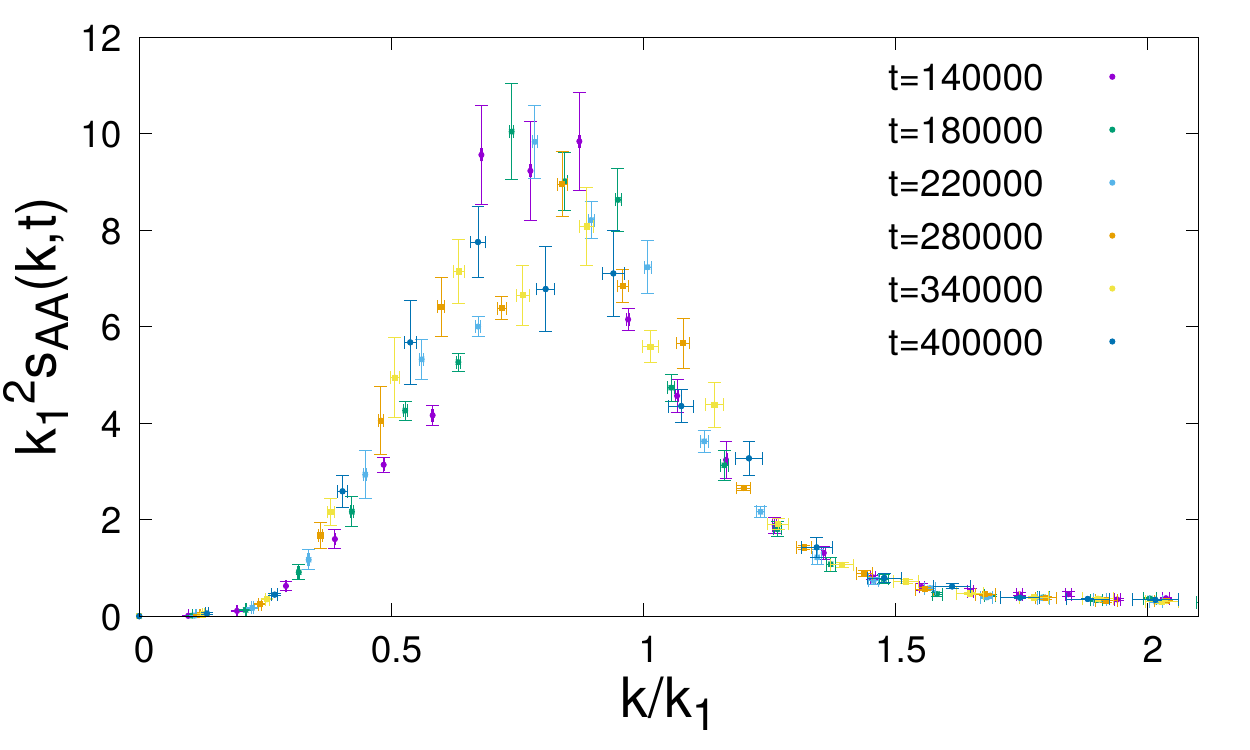}
  \caption{}
  \label{fig: P3_SFaa}
\end{subfigure} 
\begin{subfigure}{.32\textwidth}
  \centering
  \includegraphics[width=1.05\linewidth]{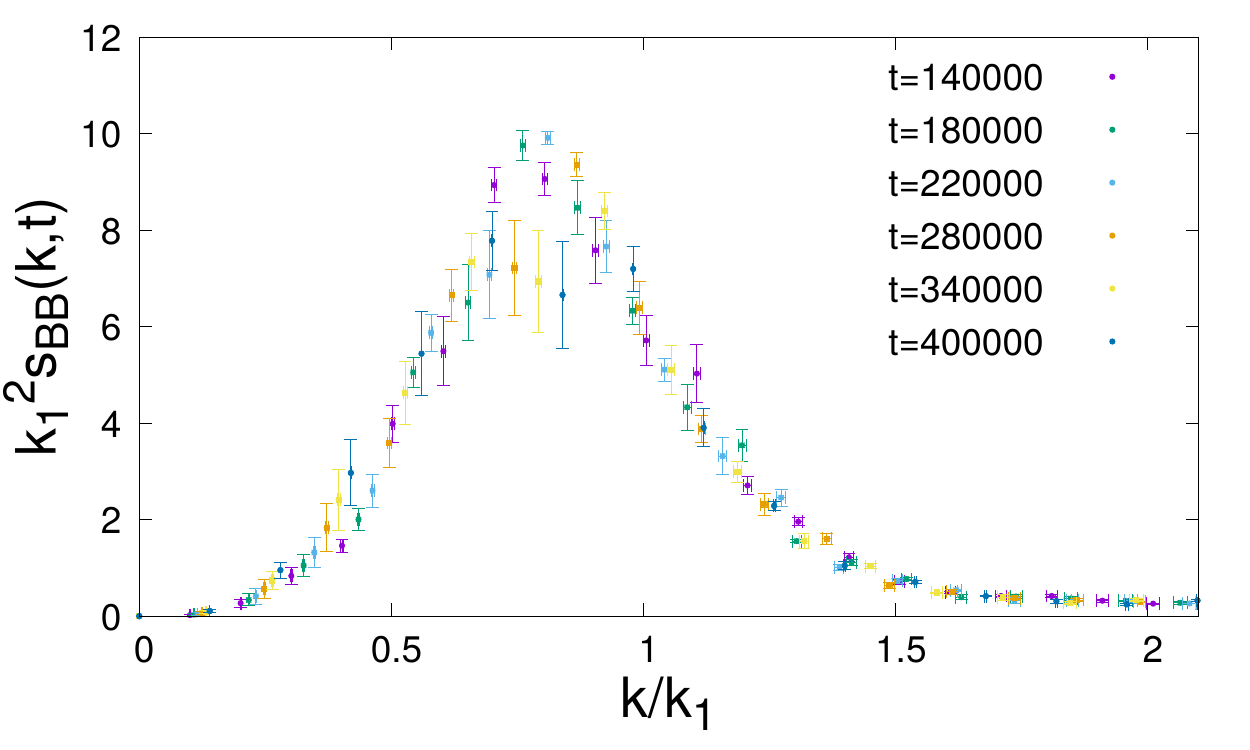}
  \caption{}
  \label{fig: P3_SFbb}
\end{subfigure} 
\begin{subfigure}{.32\textwidth}
  \centering
  \includegraphics[width=1.05\linewidth]{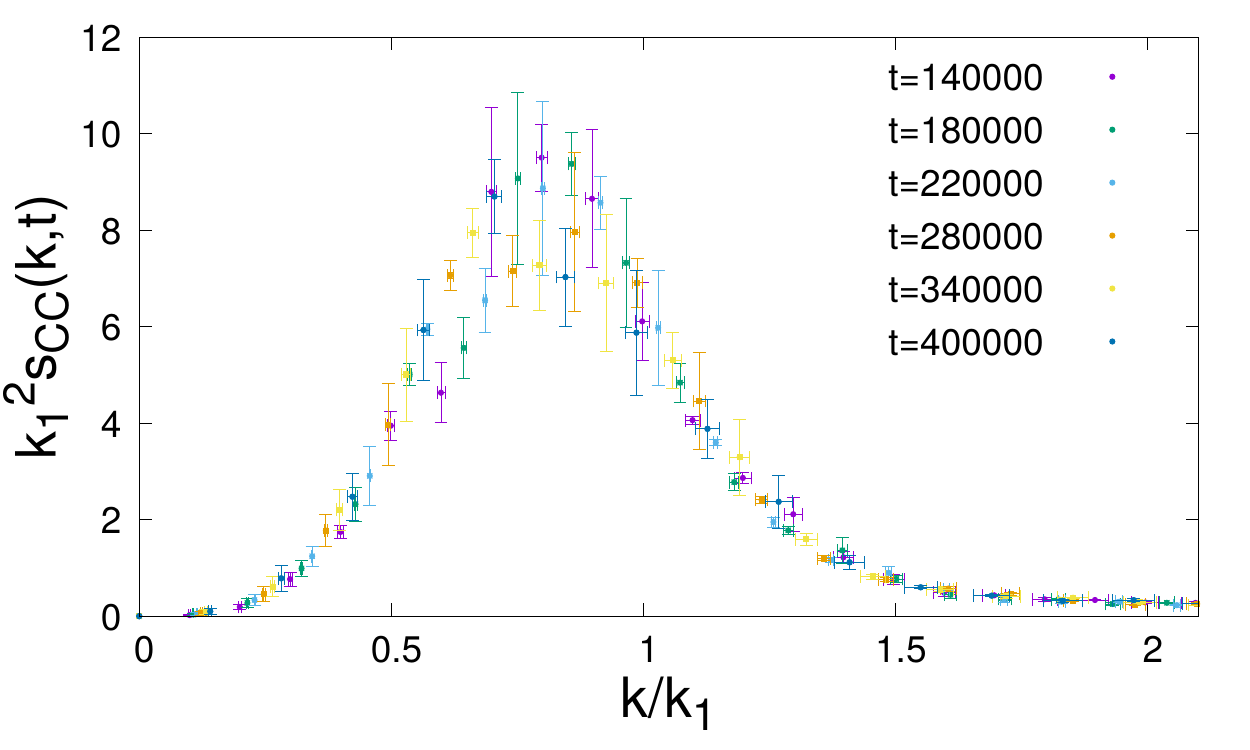}
  \caption{}
  \label{fig: P3_SFcc}
\end{subfigure}
\caption{Scaled structure functions for alloy systems (a-c) $P_1$, (d-f) $P_2$ and (g-i) $P_3$.}
\label{fig:P_Scal}
\end{figure}

\begin{figure}[H]
  \centering
\begin{subfigure}{.45\textwidth}
  \centering
  \includegraphics[width=0.8\linewidth]{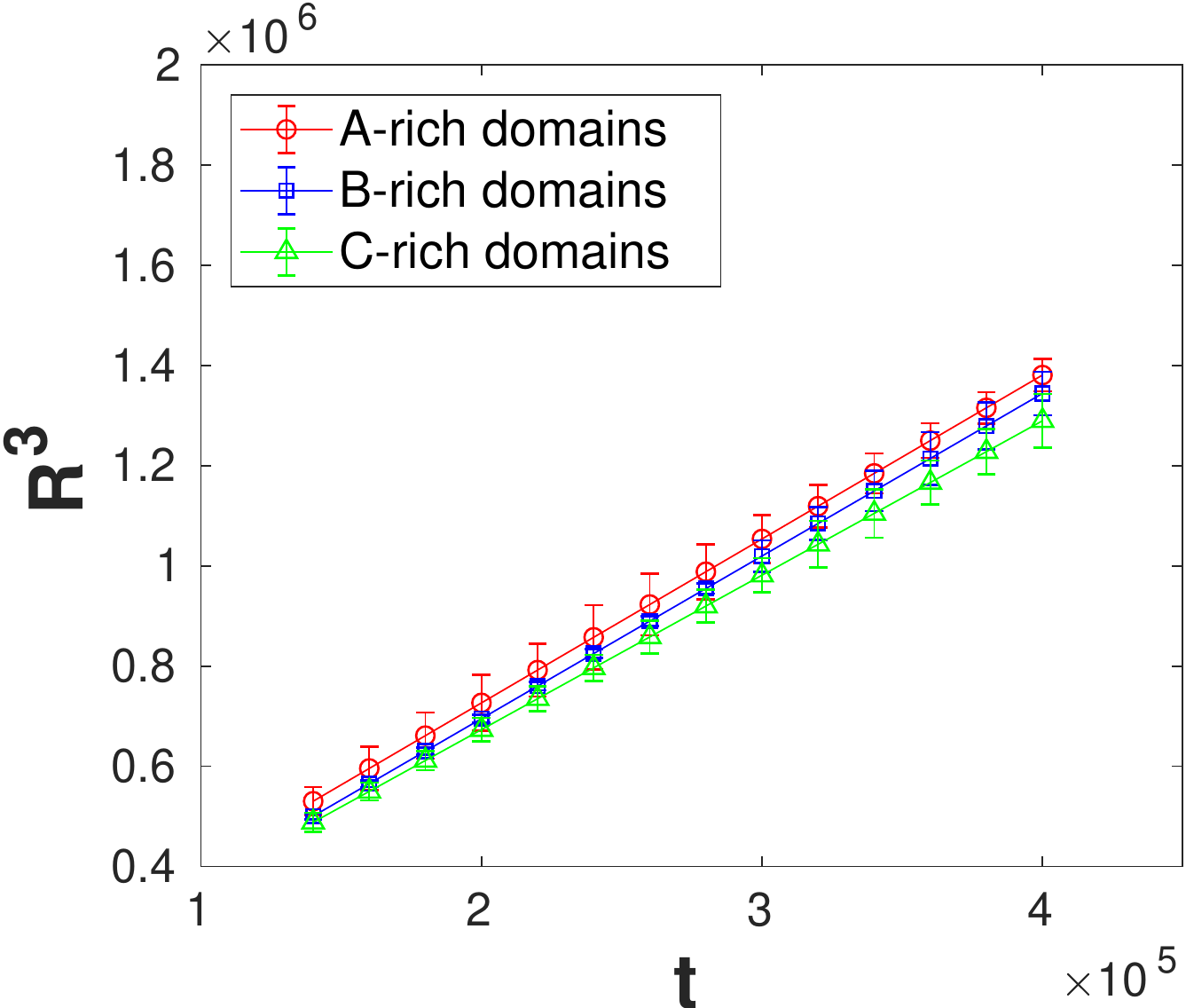}
  \caption{Alloy $P_1$}
  \label{fig: P1_rad}
\end{subfigure} 
\begin{subfigure}{.45\textwidth}
  \centering
  \includegraphics[width=0.8\linewidth]{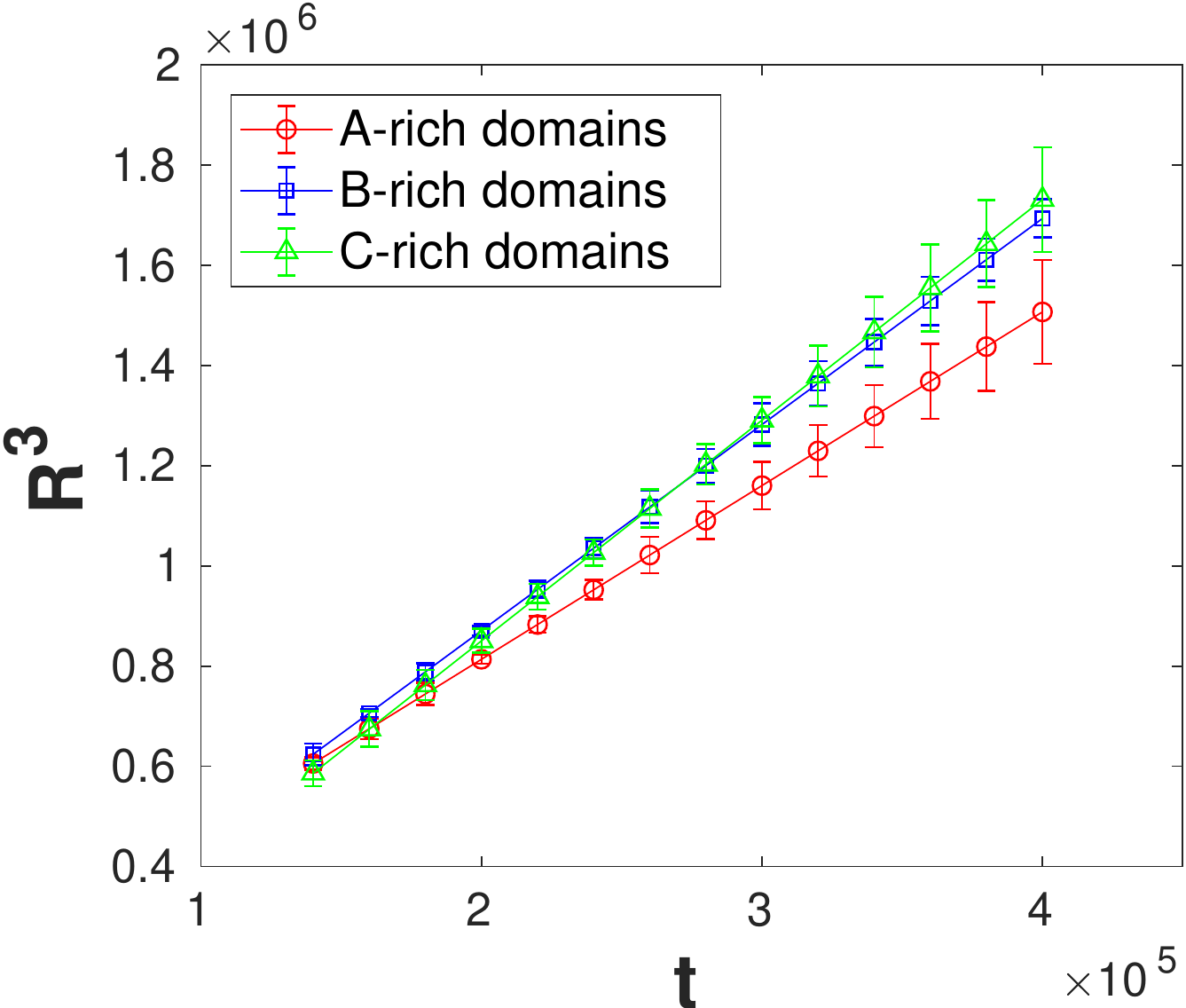}
  \caption{Alloy $P_3$}
  \label{fig: P3_rad}
\end{subfigure}
\caption{Comparison of coarsening rates of multiple domains ($\alpha$, $\beta$ and $\gamma$) 
in systems (a) $P_1$ and (b) $P_3$.}
\label{fig:P_rad}
\end{figure}

In all systems, the scaled structure functions for each phase fall on a single curve 
(accurate within numerical uncertainties) indicating self-similarity 
of the microstructures during coarsening and the existence of dynamic scaling regime
(Fig.~\ref{fig:P_Scal}). 
Therefore, we apply the classical LSW temporal power law $R^3 \propto t$
to analyze the coarsening rates of each phase in each 
system~\cite{lifshitz1961kinetics,wagner1961theorie}. 
$R=k_1^{-1}$ denotes the characteristic domain size at time $t$. 
However, in a recent study, Sun \emph{et al.} argue that the LSW power law holds true 
although the microstructure does not show self-similarity. Thus, the cubic power law
may be applied even in cases where the morphological evolution does not enter dynamic
scaling regime~\cite{sun2018self}. 
In equiatomic systems $P_1$ and $P_3$, we compare the coarsening rates of the coexisting
phases $\alpha$, $\beta$, and $\gamma$ in Figs.~\ref{fig: P1_rad} and~\ref{fig: P3_rad}, respectively. 
However, system $P_2$ contains $\alpha$ particles in a $`BC'-$rich matrix.
Since phase separation of $`BC'-$rich regions is completely arrested in this case,
a comparison of the coarsening kinetics of all phases is not possible in system $P_2$.
\begin{figure}[H]
  \centering
\begin{subfigure}{.45\textwidth}
  \centering
  \includegraphics[width=1.0\linewidth]{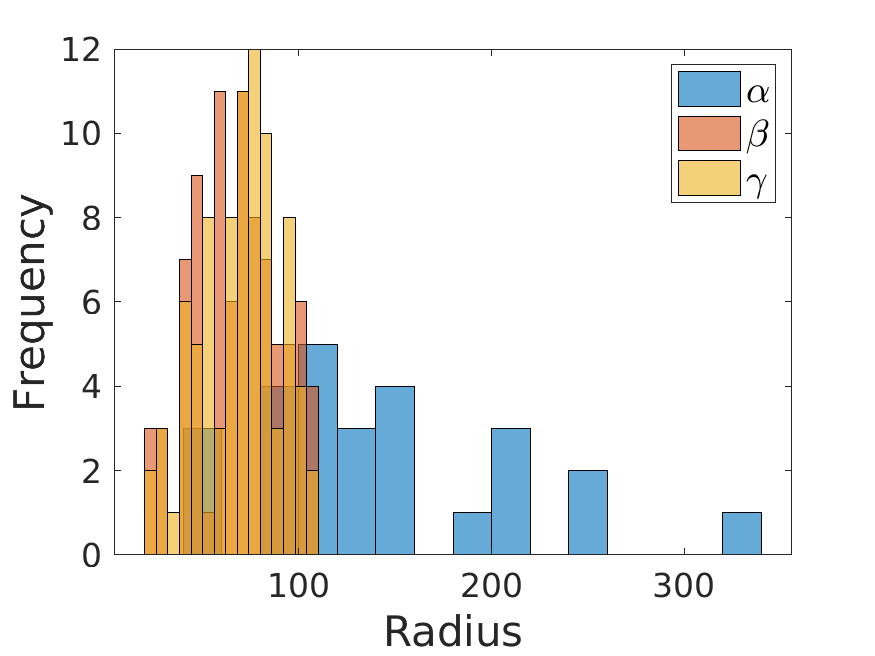}
  \caption{Alloy $P_1$}
  \label{fig: P1_psd}
\end{subfigure} 
\begin{subfigure}{.45\textwidth}
  \centering
  \includegraphics[width=1.0\linewidth]{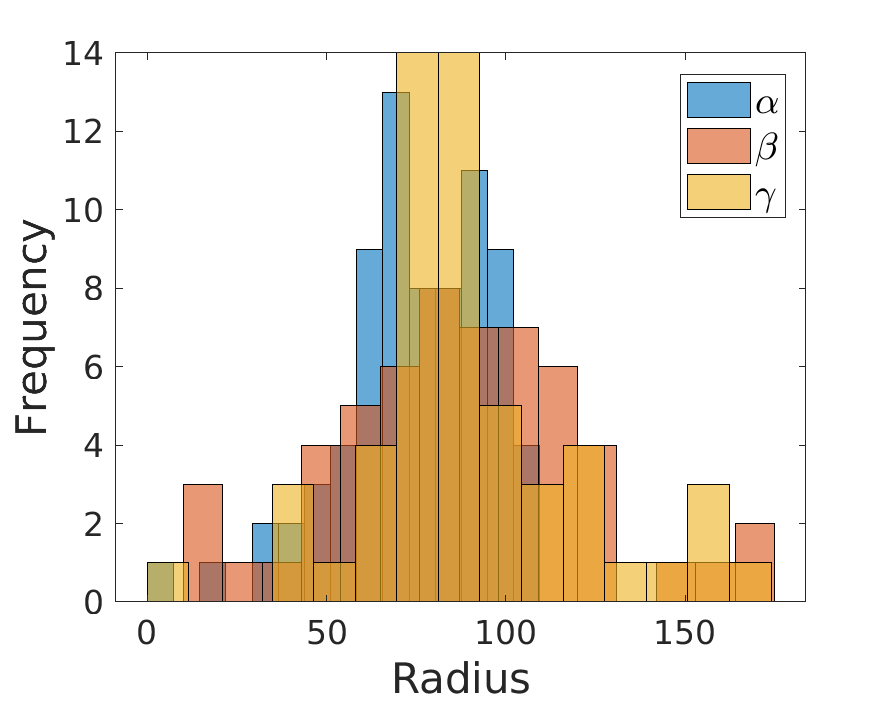}
  \caption{Alloy $P_3$}
  \label{fig: P3_psd}
\end{subfigure}
\caption{Comparison of particle size distributions of multiple domains ($\alpha$, $\beta$ and $\gamma$) 
in systems (a) $P_1$ and (b) $P_3$ at $t=200000$.}
\label{fig:P_psd}
\end{figure} 
In $P_1$, $\alpha$ domains form late and they appear coarser than $\beta$ and $\gamma$. 
All the three phases coarsen at a similar rate, although $\alpha$ 
domains are associated with larger coherency strain (Fig.~\ref{fig: P1_rad}). 
Since $\epsilon_{\beta\gamma}=0$ in this case, the system tries to maximize 
the number of $\beta\gamma$ interfaces. As a result, $\beta$ and $\gamma$ have a narrower size distribution. 
On the other hand, $\alpha$ shows a wider size distribution (see Fig~\ref{fig: P1_psd}). 
These competing factors lead to similar coarsening rates for all the domains in $P_1$.

However, in $P_3$, where the misfits between the coexisting phases are unequal, 
$\beta$ and $\gamma$ domains are coarser and they coarsen faster than $\alpha$ (Fig.~\ref{fig: P3_rad}).
Although the misfit between $\beta$ and $\gamma$ domains is high in this case ($|\epsilon_{\beta\gamma}|=0.017$, 
these domains are more extended and have wider size distribution than $\alpha$ 
(see Fig.~\ref{fig: P3_psd}). Thus, the $\beta$ and $\gamma$ domains coarsen faster in this case.   

\section{Conclusions}
\label{conclusions}

We developed a phase-field model to study phase separation in ternary alloys with coherent misfit. Our simulations
demonstrate the effects of sign and degree of misfit between the phases on their morphological evolution during 
ternary SD. Both magnitude and sign of relative misfit between the phases significantly alter the kinetic
paths during the early stages of phase separation and lead to morphological changes during subsequent 
growth and coarsening. 

Strain partitioning and, as a consequence, solute partitioning influence the morphology as well as the
coarsening kinetics of the phases. The phases associated with large misfits appears coarser and exhibit
wider particle size distributions, whereas those associated with low misfits coarsen faster 
and show narrower particle size distributions. 

Further, our simulations show the development of stable composite morphologies (core-shell) when the
interfacial energies as well as coherency strain energies between the phases favor selective 
wetting conditions ($\Gamma_{\alpha\gamma}>\Gamma_{\alpha\beta}+\Gamma_{\beta\gamma}$, 
$|\epsilon_{\beta\gamma}|<|\epsilon_{\alpha\beta}|+|\epsilon_{\alpha\gamma}|$).

\section*{Acknowledgments}
\label{acknowledgments}

The authors would like to acknowledge the financial support from R $\&$ D and SS, Tata Steel Ltd.


\section*{References}
\bibliographystyle{elsarticle-num} 
\bibliography{thesis_ref}

\begin{thebibliography}{10}
\expandafter\ifx\csname url\endcsname\relax
  \def\url#1{\texttt{#1}}\fi
\expandafter\ifx\csname urlprefix\endcsname\relax\def\urlprefix{URL }\fi
\expandafter\ifx\csname href\endcsname\relax
  \def\href#1#2{#2} \def\path#1{#1}\fi

\bibitem{voorhees2004thermodynamics}
P.~W. Voorhees, W.~C. Johnson, {The thermodynamics of elastically stressed
  crystals}, Solid State Physics-Advances in Research and Applications 59~(C)
  (2004) 1--201.

\bibitem{kostorz1995metals}
G.~Kostorz, Metals and alloys: Phase separation and defect agglomeration, in:
  Modern Aspects of Small-Angle Scattering, Springer, 1995, pp. 255--266.

\bibitem{doi1996elasticity}
M.~Doi, Elasticity effects on the microstructure of alloys containing coherent
  precipitates, Progress in Materials Science 40~(2) (1996) 79--180.

\bibitem{fratzl1999modeling}
P.~Fratzl, O.~Penrose, J.~L. Lebowitz, {Modeling of phase separation in alloys
  with coherent elastic misfit}, Journal of Statistical Physics 95~(5-6) (1999)
  1429--1503.

\bibitem{abinandanan1997computer}
T.~A. Abinandanan, {Computer modelling of elastic stress effects during
  precipitation}, Bulletin of Materials Science 20~(6) (1997) 901--908.

\bibitem{findik2002modulated}
F.~Findik, {Modulated structures in Cu-32Ni-3Cr and Cu-46Ni-17Cr alloys},
  Canadian metallurgical quarterly 41~(3) (2002) 337--347.

\bibitem{liu1985alpha}
T.~F. Liu, C.~M. Wan, $\alpha$-mn structure in a fealmncr alloy, Scripta
  metallurgica 19~(6) (1985) 727--732.

\bibitem{hanna2005new}
J.~A. Hanna, I.~Baker, M.~W. Wittmann, P.~R. Munroe, {A new high-strength
  spinodal alloy}, Journal of materials research 20~(4) (2005) 791--795.

\bibitem{Cozar_1973}
R.~Cozar, A.~Pineau, \href{https://doi.org/10.1007{\%}2Fbf02649604}{{Morphology
  of y' and y'' precipitates and thermal stability of inconel 718 type
  alloys}}, Metallurgical Transactions 4~(1) (1973) 47--59.
\newblock \href {http://dx.doi.org/10.1007/bf02649604}
  {\path{doi:10.1007/bf02649604}}.
\newline\urlprefix\url{https://doi.org/10.1007{\%}2Fbf02649604}

\bibitem{radmilovic2011highly}
V.~Radmilovic, C.~Ophus, E.~A. Marquis, M.~D. Rossell, A.~Tolley, A.~Gautam,
  M.~Asta, U.~Dahmen, {Highly monodisperse core--shell particles created by
  solid-state reactions}, Nature materials 10~(9) (2011) 710.

\bibitem{KumarMakineni2017}
S.~{Kumar Makineni}, S.~Sugathan, S.~Meher, R.~Banerjee, S.~Bhattacharya,
  S.~Kumar, K.~Chattopadhyay,
  \href{https://doi.org/10.1038/s41598-017-11540-2}{{Enhancing elevated
  temperature strength of copper containing aluminium alloys by forming L12
  Al3Zr precipitates and nucleating $\theta$″ precipitates on them}},
  Scientific Reports 7~(1) (2017) 11154+.
\newblock \href {http://dx.doi.org/10.1038/s41598-017-11540-2}
  {\path{doi:10.1038/s41598-017-11540-2}}.
\newline\urlprefix\url{https://doi.org/10.1038/s41598-017-11540-2}

\bibitem{pickering2015fine}
E.~J. Pickering, H.~J. Stone, N.~G. Jones, {Fine-scale precipitation in the
  high-entropy alloy Al0. 5CrFeCoNiCu}, Materials Science and Engineering: A
  645 (2015) 65--71.

\bibitem{santodonato2015deviation}
L.~J. Santodonato, Y.~Zhang, M.~Feygenson, C.~M. Parish, M.~C. Gao, R.~J.~K.
  Weber, J.~C. Neuefeind, Z.~Tang, P.~K. Liaw, {Deviation from high-entropy
  configurations in the atomic distributions of a multi-principal-element
  alloy}, Nature communications 6 (2015) 5964.

\bibitem{xiao2017microstructure}
D.~H. Xiao, P.~F. Zhou, W.~Q. Wu, H.~Y. Diao, M.~C. Gao, M.~Song, P.~K. Liaw,
  {Microstructure, mechanical and corrosion behaviors of AlCoCuFeNi-(Cr, Ti)
  high entropy alloys}, Materials {\&} Design 116 (2017) 438--447.

\bibitem{morral2017high}
J.~E. Morral, S.~L. Chen, {High entropy alloys, miscibility gaps and the Rose
  geometry}, Journal of Phase Equilibria and Diffusion 38~(3) (2017) 319--331.

\bibitem{johnson1989effects}
W.~C. Johnson, P.~W. Voorhees, D.~E. Zupon, {The effects of elastic stress on
  the kinetics of Ostwald ripening: the two-particle problem}, Metallurgical
  Transactions A 20~(7) (1989) 1175--1187.

\bibitem{abinandanan1993coarsening}
T.~A. Abinandanan, W.~C. Johnson, {Coarsening of elastically interacting
  coherent particles—I. Theoretical formulation}, Acta metallurgica et
  materialia 41~(1) (1993) 17--25.

\bibitem{Lee1995}
J.~K. Lee, J.~M.~T. Univ., Houghton, M.~U. S. D. o.~M. Engineering), Materials,
  \href{http://linkinghub.elsevier.com/retrieve/pii/0956716X9590837A}{{Coherency
  strain analyses via a discrete atom method}}, Scripta Metallurgica et
  Materialia 32~(4) (1995) 559--564.
\newblock \href {http://dx.doi.org/10.1016/0956-716X(95)90837-A}
  {\path{doi:10.1016/0956-716X(95)90837-A}}.
\newline\urlprefix\url{http://linkinghub.elsevier.com/retrieve/pii/0956716X9590837A}

\bibitem{nishimori1990pattern}
H.~Nishimori, A.~Onuki, {Pattern formation in phase-separating alloys with
  cubic symmetry}, Physical Review B 42~(1) (1990) 980.

\bibitem{onuki1991anomalously}
A.~Onuki, H.~Nishimori, {Anomalously slow domain growth due to a modulus
  inhomogeneity in phase-separating alloys}, Physical Review B 43~(16) (1991)
  13649.

\bibitem{zhou2014computer}
N.~Zhou, D.~C. Lv, H.~L. Zhang, D.~McAllister, F.~Zhang, M.~J. Mills, Y.~Wang,
  {Computer simulation of phase transformation and plastic deformation in IN718
  superalloy: microstructural evolution during precipitation}, Acta Materialia
  65 (2014) 270--286.

\bibitem{shi2019growth}
R.~Shi, D.~P. McAllister, N.~Zhou, A.~J. Detor, R.~DiDomizio, M.~J. Mills,
  Y.~Wang, Growth behavior of $\gamma$'/$\gamma$''coprecipitates in ni-base
  superalloys, Acta Materialia 164 (2019) 220--236.

\bibitem{de1972analysis}
D.~{De Fontaine}, {An analysis of clustering and ordering in multicomponent
  solid solutions—I. Stability criteria}, Journal of Physics and Chemistry of
  Solids 33~(2) (1972) 297--310.

\bibitem{morral1971spinodal}
J.~E. Morral, J.~W. Cahn, {Spinodal decomposition in ternary systems}, Acta
  metallurgica 19~(10) (1971) 1037--1045.

\bibitem{hoyt1989spinodal}
J.~J. Hoyt, {Spinodal decomposition in ternary alloys}, Acta metallurgica
  37~(9) (1989) 2489--2497.

\bibitem{lifshitz1961kinetics}
I.~M. Lifshitz, V.~V. Slyozov, The kinetics of precipitation from
  supersaturated solid solutions, Journal of physics and chemistry of solids
  19~(1-2) (1961) 35--50.

\bibitem{wagner1961theorie}
C.~Wagner, Theorie der alterung von niederschl{\"a}gen durch uml{\"o}sen
  (ostwald-reifung), Zeitschrift f{\"u}r Elektrochemie, Berichte der
  Bunsengesellschaft f{\"u}r physikalische Chemie 65~(7-8) (1961) 581--591.

\bibitem{bhattacharyya1972activation}
S.~K. Bhattacharyya, K.~C. Russell, {Activation energies for the coarsening of
  compound precipitates}, Metallurgical Transactions 3~(8) (1972) 2195--2199.

\bibitem{hoyt1998coarsening}
J.~Hoyt, Coarsening in multiphase multicomponent systems—i. the mean field
  limit, Acta materialia 47~(1) (1998) 345--351.

\bibitem{philippe2013ostwald}
T.~Philippe, P.~W. Voorhees, {Ostwald ripening in multicomponent alloys}, Acta
  Materialia 61~(11) (2013) 4237--4244.

\bibitem{wang2017phase}
K.~G. Wang, G.~Q. Wang, {Phase coarsening in multicomponent systems}, Physical
  Review E 95~(2) (2017) 22609.

\bibitem{chen1993computer}
L.~Q. Chen, {A computer simulation technique for spinodal decomposition and
  ordering in ternary systems}, Scripta Metallurgica et Materialia;(United
  States) 29~(5).

\bibitem{chen1994computer}
L.~Q. Chen, {Computer simulation of spinodal decomposition in ternary systems},
  Acta metallurgica et materialia 42~(10) (1994) 3503--3513.

\bibitem{eyre1993systems}
D.~J. Eyre, {Systems of Cahn--Hilliard equations}, SIAM Journal on Applied
  Mathematics 53~(6) (1993) 1686--1712.

\bibitem{BHATTACHARYYA2009646}
S.~Bhattacharyya, T.~A. Abinandanan,
  \href{http://www.sciencedirect.com/science/article/pii/S1359645408007209}{{Evolution
  of multivariant microstructures with anisotropic misfit: A phase field
  study}}, Acta Materialia 57~(3) (2009) 646--656.
\newblock \href
  {http://dx.doi.org/https://doi.org/10.1016/j.actamat.2008.10.008}
  {\path{doi:https://doi.org/10.1016/j.actamat.2008.10.008}}.
\newline\urlprefix\url{http://www.sciencedirect.com/science/article/pii/S1359645408007209}

\bibitem{ghosh2017particles}
S.~Ghosh, A.~Mukherjee, T.~A. Abinandanan, S.~Bose, {Particles with selective
  wetting affect spinodal decomposition microstructures}, Physical Chemistry
  Chemical Physics 19~(23) (2017) 15424--15432.

\bibitem{bhattacharyya2003study}
S.~Bhattacharyya, T.~A. Abinandanan, {A study of phase separation in ternary
  alloys}, Bull. Mater. Sci 26~(1) (2003) 193--197.

\bibitem{khachaturyan2013theory}
A.~G. Khachaturyan, {Theory of structural transformations in solids}, Courier
  Corporation, 2013.

\bibitem{kramer1984interdiffusion}
E.~J. Kramer, P.~Green, C.~J. Palmstr{\o}m, {Interdiffusion and marker
  movements in concentrated polymer-polymer diffusion couples}, polymer 25~(4)
  (1984) 473--480.

\bibitem{allnatt_lidiard_1993}
A.~R. Allnatt, A.~B. Lidiard, {Atomic Transport in Solids}, Cambridge
  University Press, 1993.
\newblock \href {http://dx.doi.org/10.1017/CBO9780511563904}
  {\path{doi:10.1017/CBO9780511563904}}.

\bibitem{chen1998applications}
L.~Q. Chen, J.~Shen, {Applications of semi-implicit Fourier-spectral method to
  phase field equations}, Computer Physics Communications 108~(2-3) (1998)
  147--158.

\bibitem{PhysRevE.60.3564}
J.~Zhu, L.-Q. Chen, J.~Shen, V.~Tikare,
  \href{https://link.aps.org/doi/10.1103/PhysRevE.60.3564}{{Coarsening kinetics
  from a variable-mobility Cahn-Hilliard equation: Application of a
  semi-implicit Fourier spectral method}}, Phys. Rev. E 60~(4) (1999)
  3564--3572.
\newblock \href {http://dx.doi.org/10.1103/PhysRevE.60.3564}
  {\path{doi:10.1103/PhysRevE.60.3564}}.
\newline\urlprefix\url{https://link.aps.org/doi/10.1103/PhysRevE.60.3564}

\bibitem{fialkowski2001scaling}
M.~Fia{\l}kowski, A.~Aksimentiev, R.~Ho{\l}yst, {Scaling of the Euler
  characteristic, surface area, and curvatures in the phase separating or
  ordering systems}, Physical review letters 86~(2) (2001) 240.

\bibitem{aksimentiev2002morphology}
A.~Aksimentiev, M.~Fialkowski, R.~Holyst, {Morphology of surfaces in mesoscopic
  polymers, surfactants, electrons, or reaction-diffusion systems: Methods,
  simulations, and measurements}, Advances in Chemical Physics 121 (2002)
  141--240.

\bibitem{fialkowski2002morphology}
M.~Fia{\l}kowski, R.~Ho{\l}yst, {Morphology from the maximum entropy principle:
  Domains in a phase ordering system and a crack pattern in broken glass},
  Physical Review E 65~(5) (2002) 57105.

\bibitem{bhattacharyathesis}
S.~Bhattacharya, {Ternary Spinodal Decomposition: Effect of Interfacial Energy
  (Doctoral dissertation)}, Indian Institute of Science, Bangalore.

\bibitem{chakrabarti1993late}
A.~Chakrabarti, R.~Toral, J.~D. Gunton, {Late-stage coarsening for off-critical
  quenches: Scaling functions and the growth law}, Physical Review E 47~(5)
  (1993) 3025.

\bibitem{cahn1977critical}
J.~W. Cahn, {Critical point wetting}, The Journal of Chemical Physics 66~(8)
  (1977) 3667--3672.

\bibitem{sun2018self}
Y.~Sun, W.~B. Andrews, K.~Thornton, P.~W. Voorhees, Self-similarity and the
  dynamics of coarsening in materials, Scientific reports 8~(1) (2018) 17940.

\end{thebibliography}





\end{document}